\documentclass[12pt]{report} \usepackage{amssymb,latexsym}
\usepackage{amsmath}

\columnwidth\textwidth

\begin{document}

\newtheorem{df}{Definition} \newtheorem{thm}{Theorem} \newtheorem{lem}{Lemma}
\newtheorem{prop}{Proposition} \newtheorem{assump}{Assumption}
\newtheorem{rl}{Rule}

\begin{titlepage}
 
\noindent
 
\begin{center} {\LARGE Survey of an approach to quantum measurement, \\
classical properties and realist interpretation problems} \vspace{1cm}

P. H\'{a}j\'{\i}\v{c}ek \\ Institute for Theoretical Physics \\ University of
Bern \\ Sidlerstrasse 5, CH-3012 Bern, Switzerland \\ hajicek@itp.unibe.ch \\
\vspace{.5cm} and \\ \vspace{.5cm} Ji\v{r}\'{\i} Tolar \\ Department of
Physics \\ Faculty of Nuclear Sciences and Physical Engineering \\ Czech
Technical University in Prague \\ B\v{r}ehov\'{a} 7 \\ 11519 Prague, Czech
Republic \vspace{1cm}
 
December 2010 \\ \vspace{1cm}
 
\end{center} \vspace*{2cm}
 
\nopagebreak[4]
 
\begin{abstract} The paper gives a systematic review of the basic ideas of
(non-relativistic) quantum mechanics including all improvements that result
from previous work of the authors. The aim is to show that the new theory is
self-consistent and that it reproduces all measurable predictions of standard
quantum mechanics. The most important improvements are: 1) A new realist
interpretation of quantum mechanics assumes that all properties that are
uniquely determined by preparations can be viewed as objective. These
properties are then sufficient to justify the notion of quantum object. 2)
Classical systems are defined as macroscopic quantum objects in states close
to maximum entropy. For classical mechanics, new states of such kind are
introduced, the so-called maximum-entropy packets, and shown to approximate
classical dynamics better than Gaussian wave packets. 3) A new solution of
quantum measurement problem is proposed for measurements that are performed on
microsystems. First, it is assumed that readings of registration apparatuses
are always signals from detectors. Second, an application of the cluster
separability principle leads to a locality requirement on observables and to
the key notion of separation status. Separation status of a microsystem is
shown to change in preparation and registration processes. This gives
preparation and registration new meaning and enhances the importance of these
notions. Changes of separation status are alterations of kinematic description
rather than some parts of dynamical trajectories and thus more radical than
'collapse of the wave function'. Standard quantum mechanics does not provide
any information of how separation status changes run, hence new rules must be
formulated. As an example of such a new rule, Beltrametti-Cassinelli-Lahti
model of measurement is modified and shown then to satisfy both the
probability-reproducibility and the objectification requirements.
\end{abstract}

\end{titlepage}

\tableofcontents

\section{Introduction} Since its beginnings, quantum mechanics has been
plagued by two often discussed paradoxes. The first one, let us call it
'non-objectivity-of-the measurable', is that the values obtained by most
measurements on microsystems cannot be assumed to exist before the
measurements, that is, to be objective properties of the microsystem on which
the measurement is done. An assumption of this kind of objectivity would lead
to contradictions under various circumstances (contextuality
\cite{bell4,kochen}, Bell inequalities \cite{bell1}, Hardy impossibilities
\cite{hardy}, Greenberger-Horne-Zeilinger equality \cite{GHZ}). This seduces
people into thinking that quantum systems are 'elusive' \cite{peres} or 'not
physical objects' \cite{ludwig1}.

The second one, let us call it 'Schr\"{o}dinger-cat' paradox, is that
application of quantum mechanics to interactions between microscopic and
macroscopic systems leads to results that contradict experience (if artificial
augmentations of the theory such as 'collapse of the wave function' are
forbidden). Indeed, the result often predicted by quantum mechanics for such
an interaction is that the macroscopic system evolves into a linear
superposition of macroscopically different states (dead and alive cat). This
also underlies the problem of quantum measurement: the state of a macroscopic
apparatus after a measurement as predicted by quantum mechanics does not
satisfy the objectification requirement \cite{BLM}. According to the
requirement, a well-defined value can be read off the apparatus after each
individual measurement.

There is, however, another paradox that has as yet been left rather in
shadow. Let us call it 'double-ruling' paradox: quantum mechanics has one set
of rules for dealing with a microscopic system as if it were alone in the
world (an isolated system) and another set of rules for it as a subsystem of a
family of microsystems that are identical to it. To avoid contradictions
resulting from the double ruling, the idea of cluster separability
\cite{peres} has been invented. However, only short and superficial discussion
of the idea within (non-relativistic) quantum mechanics seems to exist.

The present paper deals with the paradoxes. The
non-objectivity-of-the-measur\-able is a fact of life but it itself only
implies that some properties are not objective. Under certain conditions,
microsystems have sufficient number of other properties that can be viewed as
objective without contradictions \cite{PHJT}. Under these conditions, the
microsystems can be considered as physical objects. This also helps to explain
the emergence of classical properties \cite{hajicek}. The double-ruling and
the Schr\"{o}dinger-cat paradoxes are shown to be related. They can both be
solved if the nature of cluster separability is clarified
\cite{hajicek2,hajicek3}. The meaning and the role of preparations and
registrations appear then in a new light. In any case, the quantum mechanics
emerging on the basis of our ideas is at many points different from the usual
textbook version. This paper gives a short but systematic account of it in
order to show that it is a coherent theory free from internal contradictions.

Any physical theory can be divided into two parts. First, there are basic
notions and rules that form a kind of language. The language is general so
that it is, on the one hand, very broadly applicable, and, on the other,
unable to make any specific numerical predictions. Thus, it is difficult---may
be even impossible---to disprove by some observational facts. Second, there
are models formulated with the help of the language that are to explain and
predict properties of observed systems. Each model introduces assumptions
about the structure of the system and values of various parameters so that
quantitative results can be obtained. These can be disproved. For example, in
Newtonian mechanics, the notion of state as a point in phase space, of forces,
the law of gravitation and Newton's dynamical equations can form the
language. To describe the solar system, assumptions such as the number of
bodies, their masses and certain class of their trajectories define a
model. In the present paper, we attempt to separate quantum mechanics into the
language and the model parts.

The idea that the main activity of theoreticians is to construct models of
observed phenomena and that the general rules and notions must be adapted to
this aim plays an important role in our approach. For example, many textbooks
(see, e.g., Ref.\ \cite{peres}) maintain that any mathematically existing
state of any quantum system can be prepared by some physical device, and that
any mathematically existing observable of it can be registered by some
physical device, without specifying the devices. The fact that such claims
seem to be false for most macroscopic systems leads some authors (see, e.g.,
\cite{ludwig1,ludwig2}) to abandon the universality of quantum
mechanics. However, quantum mechanics would work perfectly well without any of
the two assumptions. On the one hand, one is free to use any mathematically
existing state or observable for a model construction and for prediction of
its properties. Mostly, only few are needed. On the other, it will turn out
(see Secs.\ 2.2.2 and 6.2) that quantum analysis of the model assumptions
itself implies that some of the mathematical observables cannot be registered.

\subsection{Examples of quantum systems} To explain what is quantum mechanics
about, this section mentions some known quantum systems. Thus, it belongs
mainly to the model part of theory. However, it also introduces some general
notions, such as microsystem, macrosystem, structural property or type of
system.

Quantum mechanics is a theory that describes certain class of properties of
certain class of systems in a similar way as any other physical theory
does. For example, Newtonian mechanics describes bodies that can be considered
as rigid in a good approximation and studies the motion of the bodies (we
consider 'mass point' as an idealisation of a small rigid body).

Quantum systems that we shall consider are photons\footnote{One can ask the
question whether there is a non-relativistic limit of photons. In one such
limit, photons may move with infinite velocity and their position need not
therefore be very well defined. In another, photons may be represented by a
classical electromagnetic wave.}, electrons, neutrons and nuclei, which we
call {\em particles}, and systems containing some number of particles, such as
atoms, molecules and macroscopic systems, which are called {\em composite}. Of
course, neutrons and nuclei themselves are composite of quarks and gluons, but
non-relativistic quantum mechanics can and must start from some
phenomenological description of neutrons and nuclei. We just view them as
structureless.

Let us call particles and quantum systems that are composite of small number
of particles {\em microsystems}. They are extremely tiny and they can never be
perceived directly by our senses. We can observe directly only macroscopic
quantum systems that are composed of very many particles (except for photons:
the eye can recognize signals of several photons). 'Very many' is: not too
different from $10^{23}$ (the Avogadro number). Let us call these {\em
macrosystems}. Some properties of most macroscopic systems obey classical
theories. For example, shape and position of my chair belong to Euclidean
geometry, chemical composition of its parts to classical chemistry and
thermodynamic properties of the parts such as phase or temperature to
phenomenological thermodynamics. Such properties are called 'classical'. Thus,
properties of microsystems can only be observed via classical properties of
macrosystems, if microsystems interact with them and this interaction leaves
classical traces.

Microsystems are divided into types, such as electrons or hydrogen
atoms. Systems of one type are not distinguishable from each other in a sense
not existing in classical physics. Systems of the same type are often called
{\em identical}. Microsystems exist always in a huge number of identical
copies. The two properties of microsystems, viz. 1) their inaccessibility to
direct observations and 2) utter lack of individuality that is connected with
the existence of a huge number of identical copies, make them rather different
from classical systems or 'things'. Each classical system can be observed
directly by humans (in principle: for example, the distant galaxies) and each
can be labelled and distinguished from other classical systems, because it is
a quantum system composite of a huge number of particles and hence it is
highly improbable that it has a relative of the same type in the world.

Objective properties that are common to all microsystems of the same type will
be called {\em structural}. Thus, each particle has a mass, spin and electric
charge. These are parameters that must be chosen in constructing models of
existing systems taking into account their empirical values; they are not
provided by the language part of theory. For example, each electron has a mass
about .5 MeV, a spin 1/2 and a charge about $10^{-19}$ C. Any composite system
consists of definite numbers of different particles with their masses, spins
and charges. E.g., a hydrogen atom contains one electron and one proton
(nucleus). The composition of a system is another structural property. The
structural properties influence the dynamics of quantum systems; the way they
do it and what dynamics is will be explained later. Only then, it will be
clear what the meaning of these parameters is and that the type of each system
can be recognized, if its dynamics is observed. When we shall know more about
dynamics, further structural properties will emerge.

Structural properties are objective in the sense that they can be assumed to
exist before and independently of any measurement. Such assumption does not
lead to any contradictions with standard quantum mechanics and is at least
tacitly made by most physicists. In fact, the ultimate goal of practically all
experiments is to find structural properties of quantum systems.

From the formally logical point of view, all possible objective properties of
given kind of objects ought to form a Boolean lattice. The structural
properties satisfy this condition: systems with a given structural property
form a subset of all systems. These subsets are always composed of whole
type-classes of quantum systems. Clearly, the intersection of two such subsets
and the complement of any such subset is again a structural property.

\subsection{Examples of quantum experiments} The topic of this section plays
an important role in understanding quantum mechanics. Specific examples of
typical experiments will be given in some detail. That belongs to the model
part of theory, but we gain access to some general notions of the language
part such as preparation and registration. The concept of individual quantum
object will be introduced and its relation to preparation and registration
mentioned.
 
Let us first consider experiments with microsystems that are carried out in
laboratories. Such an experiment starts at a source of microsystems that are
to be studied. Let us give examples of sources for each microsystem.
\begin{enumerate}
\item Electrons. One possible source (called field emission, see e.g.\ Ref.\
\cite{merton}, p.\ 38) consists of a cold cathode in the form of a sharp tip
and a flat anode with an aperture in the middle at some distance from the
cathode, in a vacuum tube. The electrostatic field of, say, few kV will enable
electrons to tunnel from the metal and form an electron beam of about $10^7$
electrons per second through the aperture, with a relatively well-defined
energy.
\item Neutrons can be obtained through nuclear reaction. This can be initiated
by charged particles or gamma rays that can be furnished by an accelerator or
a radioactive substance. For example the so-called Ra-Be source consists of
finely divided RaCl$_2$ mixed with powdered Be, contained in a small
capsule. The reaction is produced by the alpha particle of Ra. The yield for 1
mg Ra is about $10^4$ neutrons per second with broad energy spectrum from
small energies to about 13 MeV. The emission of neutrons is roughly
spherically symmetric centred at the capsule.
\item Atoms and molecules. A macroscopic specimen of the required substance in
gaseous phase at certain temperature can be produced, e.g. by an oven. The gas
is in a vessel with an aperture from which a beam of the atoms or molecules
emerges.
\end{enumerate} One can easily observe that each source is completely
determined by an arrangement of macroscopic bodies of different shapes,
chemical compositions and temperatures, and that the electric or magnetic
fields are determined by their macroscopic characteristics, such as field
intensities: that is, by their classical properties. These properties
determine uniquely what type of microsystem is produced. Let us call this
description {\em empirical}. It is important that the classical properties
defining a source do not include time and position so that the source can be
reproduced later and elsewhere. We call different sources that are defined by
the same classical properties {\em equivalent}. Empirical description is
sufficient for reproducibility of experiments but it is not sufficient for
understanding of how the sources work. If a source defined by an empirical
description is set into action, we have an instance of the so-called {\em
preparation}.

Quantum mechanics assumes that these are general features of all sources,
independently of whether they are arranged in a laboratory by humans or occur
spontaneously in nature. For example, some classical conditions in the centre
of the Sun lead to emission of neutrinos.

Often, a source yields very many microsystems of a given type that are emitted
in all possible directions, a kind of radiation. We stress that the detailed
structure of the radiation as it is understood in classical physics, that is
where each individual classical system exactly is at different times, is not
determined for quantum systems and the question has even no sense. Still, a
fixed source gives the microsystems that originate from it some properties. In
quantum mechanics, these properties are described on the one hand by the
structural properties that define the prepared type, on the other, e.g., by
the so-called {\em quantum state}. We call a prepared quantum state {\em
quantum object}. One preparation creates one individual quantum
object. However, a statement about a quantum object make sense only if it is
independent of the individuality of the quantum microsystem that occupies the
corresponding state.

To determine the quantum state that results from a preparation with a given
empirical description in each specific case requires the full formalism of
quantum mechanics. Hence, we must postpone this point.

After arranging the source, another stage of the experiment can
start. Generally, only a very small part of the radiation from a source has
the properties that are needed for the planned experiment. The next step is,
therefore, to select the part and to block off the rest. This is done by the
so-called collimator, mostly a set of macroscopic screens with apertures and
macroscopic electric or magnetic fields. For example, the electron radiation
can go through an electrostatic field that accelerates the electrons and
through electron-microscope 'lenses', each followed by a suitable screen. A
narrow part of the original radiation, a beam, remains. For example, a beam of
molecules obtained from an oven can also contain parts of broken molecules
including molecules with different degrees of ionisation. The part with
suitable composition can then be selected by a mass spectrometer and the rest
blocked off by a screen. Again, the beam resulting from a raw source and a
collimator consists of individual quantum objects with a well-defined type and
quantum state. The process of obtaining these individual quantum objects is
another example of preparation. Again, there is an empirical description that
defines an equivalence class of preparations and equivalent preparations can
be repeated.

The final beam can be characterized not only by the quantum state of
individual objects but also by its approximate current, that is how many
individual objects it yields per second. The beam can be made very thin. For
example, in the electron-diffraction experiment \cite{tono}, the beam that
emerges from the collimator represents an electric current of about $10^{-16}
A$, or $10^3$ electrons per second. As the approximate velocity of the
electrons and the distance between the collimator and the detector are known,
one can estimate the average simultaneous number of electrons there at each
time. In the experiment, it is less than one. One can understand in this way
that it is an experiment with individual electron objects.

Next, the beam can be lead through further arrangement of macroscopic bodies
and fields. For example, to study the phenomenon of diffraction of electrons,
each electron object can be scattered by a thin slab of crystalline graphite
or by an electrostatic biprism interference apparatus. The latter consists of
two parallel plates and a wire in between with a potential difference between
the wire on the one hand and the plates on the other. An electron object runs
through between the wire and both the left and right plate simultaneously and
interferes with itself afterwards (for details see \cite{tono}). Again, the
beam from the graphite or the biprism can be viewed as prepared by the whole
arrangement of the source, collimator and the interference apparatus. This is
another example of preparation procedure.

Finally, what results from the original beam must be made directly perceptible
by its interaction with another system of macroscopic bodies and fields. This
process is called {\em registration} and the system {\em registration
apparatus}. The division of an experimental arrangement into preparation and
registration parts is not unique. For example, in the electron-diffraction
experiment, one example of a registration apparatus begins after the biprism
interference, another one includes also the biprism interference
apparatus. Similarly to preparations, the registrations are defined by an
empirical description of their relevant classical properties in such a way
that equivalent registrations can be repeated.

An important definition property of a registration is that it is applicable to
an individual quantum object and that each empirical result of a registration
is caused by just one individual object. In the above experiment, this
assumption is made plausible by the extreme thinning of the beam, but it is
adopted in general even if the beam is not thin.

An empirical description of a registration apparatus can determine a quantum
mechanical {\em observable}. Again, more theory is needed for understanding of
what are observables and how they are related to registration devices. Each
individual registration performed by the apparatus, i.e., registration
performed on a single quantum object, then gives some value of the
observable. The registration is not considered to be finished without the
registration apparatus having given a definite, macroscopic and classical
response.

A part of registration apparatuses for microsystems are {\em detectors}. At
the empirical level, a detector is determined by an arrangement of macroscopic
fields and bodies, as well as by the chemical composition of its sensitive
matter \cite{leo}. For example, in the experiment \cite{tono} on electron
diffraction, the electrons coming from the biprism interference apparatus are
absorbed in a scintillation film placed transversally to the beam. An incoming
electron object is thus transformed into a light signal. The photons are
guided by parallel system of fibres to a photo-cathode. The resulting
(secondary) electrons are accelerated and lead to a micro-channel plate, which
is a system of parallel thin photo-multipliers. Finally, a system of tiny
anodes enables to record the time and the transversal position of the small
flash of light in the scintillation film. In this way, each individual
electron object coming from the biprism is detected at a position (two
transversal coordinates determined by the anodes and one longitudinal
coordinate determined by the position of the scintillation film). This triple
of numbers is the result of each registration and the value of the
corresponding observable, which is the position in this case. Also the time of
the arrival at each anode can be approximately determined. Thus, each position
obtains a certain time.

An important observation is the following. When an electron that has been
prepared by the source and the collimator hits the scintillation film, it is
lost as an object. Indeed, there is no property that would distinguish it from
other electrons in the scintillation matter. Thus, this particular
registration is a process inverse to preparation: while the preparation has
created a quantum object with certain individuality, the registration entails
a loss of the object. Our theory of quantum registration in Chap.\ 6 will
generalise this observation.

If we repeat the experiment with individual electron objects many times and
record the transversal position coordinates, the gradual formation of the
electron interference pattern can be observed. The pattern can also be
described by some numerical values. For example, the distance of adjacent
maxima and the direction of the interference fringes can be such
values. Still, the interference pattern is {\em not} a result of one but of a
whole large set of individual registrations.

In some sense, each electron object must be spread out over the whole plane of
the scintillation detector after coming from the biprism but the excitation of
the molecules in the detector matter happens always only within a tiny
well-localized piece of it, which is different for different objects. Thus,
one can say that the interference pattern must be encoded in each individual
electron object, even if it is not possible to obtain the property by a single
registration. The interference pattern can be considered as an objective
property of the individual electron objects prepared by the source, the
collimator and the biprism interference apparatus. On the other hand, the
hitting position of each individual electron object cannot be considered as
its objective property. Such an assumption would lead to contradictions with
results of other experiments. The position must be regarded as created in the
detection process. The interference pattern is not a structural property:
preparations that differ in the voltage at some stage of the experiment (e.g.,
the accelerating field in the collimator or the field between the wire and the
side electrodes in the biprism interference apparatus, etc.) will give
different interference patterns. We call such objective properties {\em
dynamical}.

Some structural properties can be measured directly by a registration (on
individual quantum objects) and their values are real numbers. For example,
mass can be measured by a mass spectrometer. Such structural properties can be
described by quantum observables (see Sec.\ 1.2)\footnote{These observables
must commute with all other observables (\cite{ludwig1}, IV.8), and can be
associated with the so-called superselection rules, see Sec. 3.3.}. However,
there are also structural properties that cannot be directly measured on
individual objects similarly to the interference pattern, such as cross
sections. They cannot be described by observables: the value of a cross
section is not a value of an observable.

\subsection{Realism} The fact that the description at empirical level is
sufficient for reproducibility of experiments and for the account of their
results has been used to construct a kind of realist interpretation of quantum
mechanics. This interpretation maintains that the domain of real world to
which quantum mechanics is applicable are exactly the classical properties of
macroscopic devices described at the empirical level. A rigorous formulation
is Ref.\ \cite{ludwig1}. The macroscopic systems are not considered as quantum
systems. Such interpretation does not even need to assume that there are any
microsystems \cite{ludwig2}. This is in agreement with statements such as
Bohr's 'there is no quantum concept', see Ref.\ \cite{cheval}, pp.\ 33--57.

There are two reasons for such a cautious approach. First, the values of
observables concerning a quantum object $\mathcal S$ that are read off as the
classical properties of a macroscopic device $\mathcal A$ after its
interaction with the quantum object are not objective properties of the
quantum object $\mathcal S$. If one assumes that such values are the only
'properties' any quantum systems can exhibit or that each objective property
must be accessible by an individual registration, then quantum systems do not
possess enough objective properties to be physical objects (for a mathematical
proof, see Ref.\ \cite{ludwig1}, p. 60). Second, there is neither a completely
satisfactory quantum theory of all classical properties of macrosystems nor a
generally accepted understanding of quantum measurement process as yet.

The cautious approach has its drawbacks. The question naturally arises, what
is the physics that prevents macroscopic systems to be quantum systems. After
all, they are composite of quantum particles. The explanation offered in
Refs.\ \cite{ludwig1,ludwig2} by G. Ludwig is as follows. Ludwig's method is
to postulate certain statements on preparation and registration first and then
to derive quantum mechanics from it. For such method to be successful, some
completeness of the starting set of preparation and registration properties is
needed. This, however, leads to problems: the macroscopic systems described by
classical theories yield examples that do not admit such completeness (see
also Chap. 5 of the present paper). Thus, if preparations and registrations
are to form the domain of reality of a macroscopic system that is to be
described by quantum mechanics, then quantum mechanics is not applicable (see
Ref.\ \cite{ludwig2}, Vol.\ 2, p. 16). On the one hand, the argument underpins
the idea that the cautious approach is self-consistent, on the other, it only
shows that the approach itself is inapplicable to macroscopic systems and
leaves the question open whether there is another more comprehensive approach.

These and other problems motivate our attempt at an alternative
theory. Surprisingly enough, another realist interpretation that is compatible
with all physical principles of standard quantum mechanics is possible. The
only difficulty are certain prejudices and restrictions motivated, in fact, by
classical mechanics. Thus, our deviation from classical thinking (with the
exception of realism) is even greater than that of other interpretations of
quantum mechanics.

The general hypothesis of realism says that the world is not just an idea in
our mind but does really exist\footnote{Of course, we are a part of the world
and can influence it. Thus, the world is objective but not completely
independent of us.}. This is an utterly metaphysical hypothesis, which means
that it cannot be disproved by any observational facts. Only if some further
assumptions about the world are involved then it is indeed possible to prove
in some cases that such a world does not exist.

Our approach to properties of quantum systems is ne\-cessarily different from
those that can be found in literature in order that it can circumvent some
no-go theorems concerning objective properties (see, e.g., Ref.\
\cite{ludwig1}). First, we extend the notion of properties to include complex
ones in the following sense:
\begin{enumerate}
\item Their values may be arbitrary mathematical objects (sets, maps between
sets, etc.). For example, the Hamiltonian of a quantum system involves a
relation between energy and some other observables of the system. This
relation is an example of such a complex property.
\item Their values need not be directly obtained by individual
registrations. For example, to measure a cross-section a whole series of
scattering experiments must be done. Thus, their values need not possess
probability distribution but may be equivalent to, or derivable from,
probability distributions.
\end{enumerate} Second, we weaken the relation of objective properties to the
basic quantum-mecha\-nical concepts of preparation and registration. What is
objective? A popular opinion is (\cite{ludwig1}, p. 60):
\begin{quote} An objective property should refer directly to the microsystem
itself and be independent of preparation and registration process. By this we
mean that a set $a$ [of systems] which is selected by a preparation (and
similarly a set $b$ which is selected by a registration) may be divided
according to objective properties into subsets...
\end{quote} The 'independence' does not mean that properties are not
manipulable or observable. In fact, if we look at Newton mechanics, a mass
point can be put into a space point $\vec{x}$ with a momentum $\vec{p}$ so
that it is then objectively in the state $(\vec{x}, \vec{p})$, and measurement
procedures exist that can inform us in which state it objectively is in the
moment of measurement. One can say that {\em all classical objective
properties can be prepared and can be obtained as results of
measurements}. The problem in quantum mechanics is that a registration in most
cases disturbs the microsystem so that the result of the registration is only
created during the registration process. The result of the registration cannot
thus be assumed to refer directly to the microsystem and to be independent of
the registration process. Hence, a set of prepared microsystem objects cannot
already be subdivided into subsets according to the properties that will only
be registered later.

It is clear then that the objective properties of microsystems, if there are
any, cannot be directly related to registrations. To confirm an objective
property, many registration procedures often carried out by several
registration devices must be done on many copies of equally prepared objects.

Any criterion of objectivity must therefore ignore registration. We are thus
led to the following:\par \vspace{.5cm} \noindent {\bf Basic Ontological
Hypothesis of Quantum Mechanics} {\it A property is objective if its value is
uniquely determined by a preparation according to the rules of standard
quantum mechanics. The 'value' is the value of the mathematical expression
that describes the property and it may be more general than just a real
number. No registration is necessary to establish such a property but a
correct registration cannot disprove its value; in many cases, registrations
can confirm the value.}\par \vspace{.5cm}

The motivation for the hypothesis is that such a property can be assumed to be
possessed by the prepared object without either violating any rules of
standard quantum mechanics or contradicting possible results of any
registration performed on the prepared object. Any set of equally prepared
objects can, therefore, always be subdivided into subsets according to such
objective properties. It follows that the properties satisfy Boolean lattice
rules. The relation of registrations to such objective properties must be
studied for each property separately.

We shall discus the objective properties in more detail in Chap.\ 1 and see
that they can be divided into structural and dynamical. Already now, we
stress: with the help of the above criterion, we obtain enough objective
properties to characterize quantum systems completely (at least from the
standpoint of standard quantum mechanics). Moreover, there will be objective
properties of quantum systems that can play the role of classical properties
of macroscopic quantum systems.

A few words have to be said on ontological hypotheses. As it is well-known,
the objective existence of anything cannot be proved (even that of the chair
on which I am now sitting, see, e.g., Ref.\ \cite{d'Espagnat}, where this old
philosophical tenet is explained from the point of view of a physicist). Thus,
all such statements are only hypotheses, called ontological. However, an
ontological hypothesis may lead to contradictions with some observations;
exactly that happens if one tries to claim objectivity of quantum
observables. Hypotheses that do not lead to contradictions may be useful. For
example, the objective existence of the chair nicely explains why we all agree
on its properties. Similarly, the assumption that quantum systems can possess
some objective properties can be useful for the quantum theory of classical
properties or for a solution of the problem of quantum measurement. The fact
that a realist version of quantum mechanics appeals to our intuition is not
essential, but it is a bonus.

\part{General notions and rules} As announced in the Introduction, we are
going to divide quantum mechanics into two parts. The first one contains
general notions and rules as well as the corresponding mathematical formalism
and its physical interpretation, while the second is a menagerie of models.

\chapter{States and observables} This chapter describes the first part of the
general language of quantum mechanics concerning the notions and most
important properties of states, observables and their relation to space-time
transformations. We shall use some of the nomenclature introduced in Refs.\
\cite{ludwig1,BLM} but no previous knowledge of these books is assumed. Ref.\
\cite{ludwig1} lists certain axioms first that must be satisfied by
preparations and registrations and then derives quantum mechanics from these
axioms. Our way will be the opposite one, similar to that of Ref.\ \cite{BLM}:
we assume the basic notions and rules of quantum mechanics first and derive
general consequences for preparations and registrations later.

\section{States} A preparation yields quantum object of certain type in a {\em
state} while a registration results in a value of an {\em observable}. What is
the type and what is the state is uniquely determined by the classical
properties of the preparation and what is the observable is uniquely
determined by the classical properties of the registration.

In quantum mechanics, the states and observables are described by specific
mathe\-matical entities. If we say 'state' or 'observable', we always mean
these entities. It is important to distinguish between the fact of being
uniquely determined and the way from the empirical description to a particular
mathematical state or observable. The calculation of a state from classical
conditions defined by the preparation needs full quantum mechanics. Similarly,
to calculate an observable from classical properties of the registration
device, quantum mechanics must be used. At this stage of exposition, we just
assume that the classical properties determine states and observables and
consider the way of how they do it as belonging to the model part of the
theory.

\subsection{Mathematical preliminaries} We assume that the reader is already
more or less acquainted with most of the necessary mathematics and the purpose
of the following text is mainly to settle the notation.

Let ${\mathbf H}$ be a complex separable Hilbert space with inner product
$\langle\cdot|\cdot\rangle$ satisfying $\langle a\phi|b\psi\rangle =
a^*b\langle\phi|\psi\rangle$, where '$*$' denotes complex conjugation. An
element $\phi\in{\mathbf H}$ is a unit vector if its norm defined by
$$\| \phi \| = \langle\phi|\phi\rangle$$
equals one,
$$\| \phi \| = 1\ ,$$
and the non-zero vectors $\phi,\ \psi\in{\mathbf H}$ are orthogonal if
$$\langle\phi|\psi\rangle = 0\ .$$
A set $\{\phi_k\}\subset{\mathbf H}$ is orthonormal if the vectors $\phi_k$
are mutually orthogonal unit vectors. $\{\phi_k\}\subset{\mathbf H}$,
$k=1,2,\cdots$, is an orthomormal basis of ${\mathbf H}$ if any
$\psi\in{\mathbf H}$ can be expressed as a series
$$
\psi = \sum_k\langle\phi_k|\psi\rangle\phi_k
$$
with
$$
\| \psi\|^2 = \sum_k|\langle\phi_k|\psi\rangle|^2\ .
$$
Separability means that there is at least one countable basis.

Let ${\mathbf H}$ and ${\mathbf H}'$ be two separable Hilbert spaces,
$\{\phi_k\}$ and $\{\phi'_k\}$ two orthonormal bases, $\{\phi_k\} \subset
{\mathbf H}$ and $\{\phi'_k\} \subset {\mathbf H}'$. Let us define map
${\mathsf U} : \{\phi_k\} \mapsto \{\phi'_k\}$ by
$$
{\mathsf U}\phi_k = \phi'_k
$$
for each $k$. Then, ${\mathsf U}$ can be extended by linearity and continuity
to the whole of ${\mathbf H}$ and it maps ${\mathbf H}$ onto ${\mathbf
H}'$. The map ${\mathsf U}$ is called {\em unitary}. Unitary maps preserve
linear superposition,
$$
{\mathsf U}(a\psi + b\phi) = a{\mathsf U}\psi + b{\mathsf U}\phi
$$
and inner product,
\begin{equation}\label{sp} \langle {\mathsf U}\psi|{\mathsf U}\phi\rangle =
\langle \psi|\phi\rangle\ .
\end{equation} They can be defined by these properties for general Hilbert
spaces and used as equivalence morphisms in the theory of Hilbert spaces. Each
two separable Hilbert spaces are thus unitarily equivalent.

Any unit vector $\phi\in{\mathbf H}$ determines a one-dimensional (orthogonal)
projection operator ${\mathsf P}[\phi]$ by the formula
$${\mathsf P}[\phi]\psi = \langle\phi|\psi\rangle\phi$$
for all $\psi\in{\mathbf H}$. We also use the Dirac notation
$|\phi\rangle\langle\phi|$ for this projector. If $\{\phi_k\}$ is an
orthonormal basis of ${\mathbf H}$, then the projection operators ${\mathsf
P}[\phi_k]$ satisfy
$$
{\mathsf P}[\phi_k]{\mathsf P}[\phi_l] = 0
$$
for all $k\neq l$--we say they are mutually orthogonal--and
$$
\sum_k {\mathsf P}[\phi_k] = {\mathsf 1}\ ,
$$
where ${\mathsf 1}$ is the identity operator on ${\mathbf H}$.

A linear operator ${\mathsf A} : {\mathbf H} \mapsto {\mathbf H}$ is defined
by the property
$$
{\mathsf A}(a\phi + b\psi) = a{\mathsf A}\phi + b{\mathsf A}\psi
$$
for all $\phi, \psi \in {\mathbf H}$ for which the right-hand side makes
sense. A linear operator ${\mathsf A}$ is called {\em bounded} if its norm
\begin{equation}\label{opnorm} \|{\mathsf A}\| = \sup_{\|\psi\| = 1}\|{\mathsf
A}\psi\|\
\end{equation} is finite. Let us denote the set of all bounded linear
operators by ${\mathbf L}({\mathbf H})$. Bounded linear operators can be
multiplied,
$$
({\mathsf A}{\mathsf B})\phi = {\mathsf A}({\mathsf B}\phi)\ ,
$$
and linearly combined,
$$
(a{\mathsf A} + b{\mathsf B})\phi = a({\mathsf A}\phi) + b({\mathsf B}\phi)
$$
for any $a,b \in {\mathbb C}$. ${\mathbf L}({\mathbf H})$ is a Banach algebra
with the norm (\ref{opnorm}).

The adjoint ${\mathsf A}^\dagger$ of bounded operator ${\mathsf A}$ is defined
by
$$
\langle {\mathsf A}^\dagger \phi|\psi\rangle = \langle \phi| {\mathsf A}
\psi\rangle
$$
for all $\phi,\ \psi\in{\mathbf H}$, and ${\mathsf A}$ is self-adjoint (s.a.)
if
$$
{\mathsf A}^\dagger = {\mathsf A}\ .
$$
Linear combinations of bounded s.a.\ operators with real coefficients are
again s.a., but their products are not, in general. Hence, the set of all
bounded s.a.\ operators forms a real linear space. If completed with respect
to the norm (\ref{opnorm}), it is a Banach space denoted by ${\mathbf
L}_r({\mathbf H})$.

Unitary maps ${\mathsf U} : {\mathbf H} \mapsto {\mathbf H}$ are bounded
operators and we obtain from Eq.\ (\ref{sp})
$$
{\mathsf U}^\dagger\cdot{\mathsf U} = {\mathsf U}\cdot{\mathsf U}^\dagger =
{\mathsf 1}\ .
$$

Let ${\mathbf H}$ and ${\mathbf H}'$ be two separable Hilbert spaces and
${\mathsf U} : {\mathbf H} \mapsto {\mathbf H}'$ be a unitary map. Then
${\mathsf U}$ defines a map of ${\mathbf L}({\mathbf H})$ onto ${\mathbf
L}({\mathbf H'})$ by ${\mathsf A} \mapsto {\mathsf U}{\mathsf A}{\mathsf
U}^\dagger$. This map preserves operator action,
$$
({\mathsf U}{\mathsf A}{\mathsf U}^\dagger)({\mathsf U}\phi) = {\mathsf
U}({\mathsf A}\phi)\ ,
$$
linear relation,
$$
{\mathsf U}(a{\mathsf A} + b{\mathsf B}){\mathsf U}^\dagger = a{\mathsf
U}{\mathsf A}{\mathsf U}^\dagger + b{\mathsf U}{\mathsf B}{\mathsf U}^\dagger\
,
$$
operator product
$$
({\mathsf U}{\mathsf A}{\mathsf U}^\dagger)({\mathsf U}{\mathsf B}{\mathsf
U}^\dagger) = {\mathsf U}({\mathsf A}{\mathsf B}){\mathsf U}^\dagger
$$
and norm,
$$
\|{\mathsf U}{\mathsf A}\| = \|{\mathsf A}\|
$$
on ${\mathbf L}({\mathbf H})$. The Banach-space structure of ${\mathbf
L}_r({\mathbf H})$ is also preserved by unitary maps.

An operator ${\mathsf A}\in{\mathbf L}_r({\mathbf H})$ is positive, ${\mathsf
A}\geq {\mathsf 0}$, where ${\mathsf 0}$ is the null operator, if
$$
\langle\phi|{\mathsf A}\phi\rangle \geq 0
$$
for all vectors $\phi\in{\mathbf H}$. The relation ${\mathsf A}\geq {\mathsf
B}$ defined by
$$
{\mathsf A}-{\mathsf B}\geq {\mathsf 0}
$$
is an ordering on this space. With this (partial) order relation, ${\mathbf
L}_r({\mathbf H})$ is an ordered Banach space. The order relation is preserved
by unitary maps,
$$
{\mathsf U}{\mathsf A}{\mathsf U}^\dagger \geq {\mathsf U}{\mathsf B}{\mathsf
U}^\dagger \quad \text{if} \quad {\mathsf A} \geq {\mathsf B}\ .
$$

Let $\{\phi_k\}$ be any orthonormal basis of ${\mathbf H}$. For any ${\mathsf
A} \in{\mathbf L}({\mathbf H})$, we define the trace
$$
tr[{\mathsf A}] = \sum_k\langle\phi_k|{\mathsf A}\phi_k\rangle\ .
$$
Trace is independent of basis and invariant with respect to unitary maps,
$$
tr[{\mathsf U}{\mathsf A}{\mathsf U}^\dagger] = tr[{\mathsf A}]\ .
$$.

\begin{prop} Trace defines the norm $\|{\mathsf A}\|_s$ on ${\mathbf
L}_r({\mathbf H})$ by
\begin{equation}\label{trnorm} \|{\mathsf A}\|_s = tr\left[\sqrt{{\mathsf
A}^2}\right]
\end{equation} satisfying
$$
\|{\mathsf A}\|_s \geq \|{\mathsf A}\|
$$
for all ${\mathsf A} \in {\mathbf L}_r({\mathbf H})$
\end{prop} For proof, see Ref.\ \cite{ludwig1}, Appendix IV.11.
\begin{df} The norm (\ref{trnorm}) is called trace norm and all s.a.\
operators on ${\mathbf H}$ with finite trace norm are called trace-class. The
set of all trace-class operators is denoted by ${\mathbf T}({\mathbf H})$.
\end{df} Trace norm is preserved by unitary maps.
\begin{prop} ${\mathbf T}({\mathbf H})$ completed with respect to the norm
(\ref{trnorm}) is an ordered Banach space. A trace-class operator is bounded,
its trace is finite and its spectrum is discrete.
\end{prop} For proof, see Ref.\ \cite{ludwig1}, Appendix IV.11.

Let ${\mathbf T}({\mathbf H})^+_1$ be the set of all positive self-adjoint
trace-class operators on ${\mathbf H}$ with trace 1. As these operators are
positive, their trace is equal to their trace norm and they lie on the unit
sphere in ${\mathbf T}({\mathbf H})$. ${\mathbf T}({\mathbf H})^+_1$ is a
convex set: let ${\mathsf T}_1, {\mathsf T}_2 \in {\mathbf T}({\mathbf
H})^+_1$, then
$$
w{\mathsf T}_1 + (1-w){\mathsf T}_2 \in {\mathbf T}({\mathbf H})^+_1
$$
for all $0<w<1$. It follows that any convex combination
\begin{equation}\label{properm'} {\mathsf T} = \sum_k w_k{\mathsf T}_k
\end{equation} of at most countable set of states ${\mathsf T}_k\in {\mathbf
T}({\mathbf H})^+_1$ with weights $w_k$ satisfying
\begin{equation}\label{weight} 0\leq w_k \leq 1\ ,\quad \sum w_k = 1
\end{equation} and the series converging in the trace-norm topology also lies
in ${\mathbf T}({\mathbf H})^+_1$. In general, a given state ${\mathbf
T}({\mathbf H})^+_1$ can be written in (infinitely) many ways as a convex
combination. All possible components of such convex combinations form a
so-called 'face' in the convex set ${\mathbf T}({\mathbf H}_\tau)^+_1$ (cf.\
\cite{ludwig1}, p. 75).

An element ${\mathsf T}$ is called extremal element of ${\mathbf T}({\mathbf
H})^+_1$ if it lies in a zero-dimensional face, i.e., if the condition
$$
{\mathsf T} = w{\mathsf T}_1 + (1-w){\mathsf T}_2
$$
with ${\mathsf T}_1, {\mathsf T}_2 \in {\mathbf T}({\mathbf H})^+_1$ and
$0\leq w \leq 1$, implies that ${\mathsf T} = {\mathsf T}_1 = {\mathsf
T}_2$. Any extremal element has the form ${\mathsf P}[\phi]$, where $\phi$ is
a unit vector of ${\mathbf H}$ and, conversely, any projection onto a
one-dimensional sub-space of ${\mathbf H}$ is extremal. The set of all
extremal elements of ${\mathbf T}({\mathbf H})^+_1$ generates ${\mathbf
T}({\mathbf H})^+_1$ in the sense that any ${\mathsf T} \in {\mathbf
T}({\mathbf H})^+_1$ can be expressed as countable convex combination of some
extremal elements,
$$
{\mathsf T} = \sum_k w_k{\mathsf P}[\phi_k]\ .
$$
Such a decomposition can be obtained, in particular, from the spectral
decomposition
$$
{\mathsf T} = \sum_k t_k {\mathsf P}_k\ .
$$
In that case ${\mathsf T}$ is decomposed into mutually orthogonal one
dimensional projectors ${\mathsf P}[\phi_l]$, with the eigenvalues
representing the weights $w_l$, each term appearing as many times as given by
the dimension of the eigenspace.

A unit vector $\phi$ of ${\mathbf H}$ defines a unique extremal element
${\mathsf P}[\phi] \in {\mathbf T}({\mathbf H})^+_1$ but ${\mathsf P}[\phi]$
determines $\phi$ only up to a phase factor $e^{i\alpha}$. Due to the
(complex) linear structure of ${\mathbf H}$, there is an operation on vectors
called linear superposition. Linear superposition $\psi = \sum c_k\phi_k$ of
unit vectors with complex coefficients satisfying
$$
\sum_k |c_k|^2 = 1
$$
is another unit vector and the resulting projector ${\mathsf P}[\psi]$,
\begin{equation}\label{pures} {\mathsf P}[\psi] = \left|\sum_k
c_k\phi_k\right\rangle\left\langle\sum_l c_l\phi_l\right| =
\sum_{kl}c_l^*c_k|\phi_k\rangle\langle\phi_l| \neq \sum_k
|c_k|^2|\phi_k\rangle\langle\phi_k|\ .
\end{equation} is different from the corresponding convex combination,
\begin{equation}\label{npures} \sum_k |c_k|^2|\phi_k\rangle\langle \phi_k|\ .
\end{equation} Observe that ${\mathsf P}[\psi]$ is not determined by the
projectors ${\mathsf P}[\phi_k]$ because it depends on the relative phases of
vectors $\phi_k$.

\subsection{General rules} With each quantum system ${\mathcal S}$ of type
$\tau$, a complex separable Hilbert space ${\mathbf H}_\tau$ is
associated. ${\mathbf H}_\tau$ carries a further structure dependent on
$\tau$, such as a representation of Galilean group (see Sec.\ 1.3) and a
tensor-product structure (see Secs.\ 2.1 and 2.2).
\begin{rl} The states of a quantum system ${\mathcal S}$ of type $\tau$ are
elements of ${\mathbf T}({\mathbf H}_\tau)^+_1$. Every preparation creates a
quantum object of a definite type and a definite state. The state is an
objective property of the prepared system. In principle, each element of
${\mathbf T}({\mathbf H}_\tau)^+_1$ can be prepared as a state of a system of
type $\tau$.
\end{rl} Textbooks of quantum mechanics do not consider states as objective
but this assumption is crucial for our interpretation. The objectivity of
states however does not mean that each system of type $\tau$ is in some state
from ${\mathbf T}({\mathbf H}_\tau)^+_1$ even if it does not been so prepared
(see Sec.\ 2.2.1). Rule 1 does not exclude that one and the same state can be
obtained by different preparations. Methods of how states prepared by two
different preparations can be compared will be described Sec.\ 1.2.
\begin{df} Let ${\mathsf T}$ be a finite convex combination
(\ref{properm'}). In the case that the preparation ${\mathcal P}[{\mathsf T}]$
of ${\mathsf T}$ includes sub-preparations ${\mathcal P}[{\mathsf T}_k]$ of
each ${\mathsf T}_k$ with relative weights $w_k$ (independently of whether the
mixing is done by humans or by nature; in practice, it is achieved by mixing
beams), the decomposition (\ref{properm'}) is selected by the preparation of
${\mathsf T}$. Let us call the resulting state {\em gemenge}, the states
${\mathsf T}_k$ {\em components} of the gemenge.
\end{df} A gemenge is also called {\em direct mixture} \cite{ludwig1} or {\em
proper mixture} \cite{d'Espagnat}; the word gemenge is taken from Ref.\
\cite{BLM}. Gemenge concerns a physical property of preparation rather than
any mathematical one of the right-hand side of Eq.\ (\ref{properm'}) (such as
${\mathsf T}_k$ being vector states or being mutually orthogonal, etc). From
the mathematical point of view, many different convex decompositions of a
general state ${\mathsf T}$ may exist. A preparation of ${\mathsf T}$ selects
one of the mathematically possible convex decompositions of ${\mathsf T}$.

A random mixture of preparations is not uniquely determined by the preparation
process. It can be coarsened or refined, i.e., some of ${\mathcal P}[{\mathsf
T}_k]$ can be combined into one preparation procedure or ${\mathcal
P}[{\mathsf T}_k]$ for some $k$ can itself be a random mixture of other
preparations ${\mathcal P}[{\mathsf T}_{kl}]$.
\begin{df} The finest convex decomposition of state ${\mathsf T}$ defined by
its preparation as gemenge is called {\em gemenge structure} of ${\mathsf T}$.
\end{df} Thus, gemenge structure of ${\mathsf T}$ is uniquely determined by
its preparation. For extremal states, there is always only one gemenge
structure, the trivial one, independently of how it was prepared.

It may be advantageous to distinguish the mathematical convex combination of
states from their gemenge by writing the sum in Eq.\ (\ref{properm'}) as
follows
\begin{equation}\label{properm} {\mathsf T} = \left(\sum_{k=1}^n\right) w_k
{\mathsf T}_k
\end{equation} in the case that the right-hand side is a gemenge structure of
${\mathsf T}$.

A gemenge contains more information about an object than its state operator
does and it is, therefore, viewed as a refinement of the state concept.
\begin{prop} The individual objects prepared in a state with gemenge structure
(\ref{properm}) can be assumed always to objectively be in either of its
component states ${\mathsf T}_k$ with respective probabilities $w_k$.
\end{prop}

It also follows that a state itself even if it is prepared as a gemenge with
non-trivial structure (\ref{properm}) is an objective property of any
individual object prepared in this way. On the one hand, this is in agreement
with our objectivity criterion, on the other it is completely analogous to
classical properties which need not determine the state of a classical system
uniquely, either. The states and their gemenge structures are objective
properties (being determined by preparations) that we have called {\em
dynamical}\footnote{In Ref.\ \cite{PHJT}, they were called 'conditional'.}. We
shall later encounter more examples of dynamical properties, but many will be
derived from, or determined by, states.

Set ${\mathbf T}({\mathbf H}_\tau)^+_1$ of states can be viewed as analogous
to phase space of Newtonian mechanics. Each point of ${\mathbf T}({\mathbf
H}_\tau)^+_1$ is analogous to a point in the phase space if it is prepared so
that it has a trivial gemenge structure. The gemenges are apparently analogous
to the probability distributions over the phase space.

Let a classical system ${\mathcal S}_c$ be subject to physical conditions
$\mathcal C$ that do not uniquely determine a point $Z$ in its phase space but
a whole probability distribution $\rho(Z)$ on the phase space. Then, on the
one hand, $\rho(Z)$ is an objective property of each individual system
${\mathcal S}_c$ subject to $\mathcal C$ and, on the other, each individual
${\mathcal S}_c$ is usually assumed also to be objectively at a particular
point $Z$ (but cf.\ Chap. 5), which is, therefore, also its objective
property. Thus, the two different states, $\rho(Z)$ and $Z$, are
simultaneously objective properties of the particular system ${\mathcal
S}_c$. This is, of course, possible, because the two properties are logically
compatible.

The situation in quantum mechanics is completely analogous, but there is one
difference. In Newtonian mechanics, point $Z$ and non-trivial distribution
$\rho(Z)$ are mathematically different. In quantum mechanics, one and the same
element ${\mathsf T} \in {\mathbf T}({\mathbf H}_\tau)^+_1$ can have both a
non-trivial and a trivial gemenge structure. This depends on its
preparation. Mathematical properties of ${\mathsf T}$ can restrict its
possible gemenge structures, e.g.\ extremal states allow only the trivial one.

Any unit vector $\phi \in {\mathbf H}_\tau$ defines state ${\mathsf
P}[\phi]$. Such a state is called {\em vector state} (often, such states are
called 'pure'). The states spoken about in Rule 1 are not necessarily vector
states. However, vector states have a number of very interesting
properties. In some special cases, vector states ${\mathsf P}[\phi_k]$,
$k=1,\cdots, N$, can be prepared in such a way that their relative phases are
also fixed. Such set of states is called {\em coherent}. Then, the linear
combination of the vectors defines the state (\ref{pures}), ${\mathsf
P}[\psi]$, which is called {\em linear superposition} of the vector states
${\mathsf P}[\phi_k]$, $k=1,\cdots, N$. This leads to {\em interference}
phenomena that are purely quantum and unknown in the Newtonian mechanics such
as the electron interference in the Tonomura \cite{tono} experiment (see Sec.\
1.2.2).

The dynamical properties of quantum objects form a Boolean lattice similarly
to the structural properties. This is to be understood in the following sense
(cf.\ Ref.\ \cite{ludwig1}). There is a set ${\mathbf M}$ of entities and a
class ${\mathbf N}$ of possible properties of the entities. All entities
having a given property form a subset that can be identified with the
property. The family of all subsets form a Boolean lattice with the order
relation, disjunction, conjunction and complement being the set-theoretical
inclusion, intersection, union and complement in ${\mathbf M}$. This is
analogous to the properties of a system in Newtonian mechanics: All systems of
a given type form ${\mathbf M}$. Each element of ${\mathbf M}$ is objectively
in one point of the phase space of the system. Then, properties can be
identified with subsets of the phase space.

The set of entities ${\mathbf M}$ that are considered in the present case
consists of all quantum objects of the same type $\tau$. (Thus, they all have
the same structural properties.) All states of any such object ${\mathcal S}$
that can be prepared form set ${\mathbf N} = {\mathbf T}({\mathbf
H}_\tau)^+_1$. Any ${\mathbf A} \subset {\mathbf T}({\mathbf H}_\tau)^+_1$
represents all copies of ${\mathcal S}$ that are prepared in the states
contained in ${\mathbf A}$\footnote{Observe that the Boolean lattice of
subsets of $\mathbf N$ has nothing to do with the Boolean lattice of
preparation procedures as defined in Ref.\ \cite{ludwig1}: the lattice
operations are defined in a different way.}. Another example of ${\mathbf N}$
is the set of all gemenge structures (\ref{properm}) of a given state.

There is no finer or more complete description of a quantum system than by its
gemenge structure. This implies that a prepared quantum system can be
considered as a physical object in our theory\footnote{In Ref.\
\cite{ludwig1}, p. 60, the problem is discussed whether a quantum system can
be considered as a physical object. The criterion is that each state of the
system is uniquely determined by its objective properties. Of course, this
depends on what is a property. In Ref.\ \cite{ludwig1}, only values of
observables are admitted as properties. Then, quantum systems are not physical
objects. In our interpretation, the class of properties is much broader and
certain quantum systems are, then, physical objects.}.

Some of the above statements seem to contradict the well-known fact that the
subset of states consisting of all possible projections (quantum logic) does
not form a Boolean lattice. In Sec.\ 3.2.2, the lattice structure of the set
of projections will be described. It is clear that there is no contradiction
because the lattice operations that are used in quantum logic are different
from those introduced above. The deep reason is that the properties of quantum
logic are associated with registrations and are, therefore, not objective,
while our properties are determined by preparations and can be considered as
objective.

\section{Observables} Once an object in a state has been prepared, one can
carry out a registration on it. Mathematically, a registration is described by
an observable. A better understanding of the notion of state can be achieved
if results of all possible registrations applied to the state can be
calculated.

\subsection{Mathematical preliminaries}
\begin{df} Let ${\mathbf F}$ be the Boolean lattice of all Borel subsets of
${\mathbb R}^n$. A {\em positive operator valued (POV) measure}
$$
{\mathsf E} : {\mathbf F} \mapsto {\mathbf L}_r({\mathbf H})
$$
is defined by the properties
\begin{enumerate}
\item positivity: ${\mathsf E}(X) \geq 0$ for all $X\in {\mathbf F}\ ,$
\item $\sigma$-additivity: if $\{X_k\}$ is a countable collection of disjoint
sets in ${\mathbf F}$ then
$$
{\mathsf E}(\cup_k X_k) = \sum_k {\mathsf E}(X_k)\ ,
$$
where the series converges in weak operator topology, i.e., averages in any
state converge to an average in the state.
\item normalisation:
$$
{\mathsf E}({\mathbb R}^n) = {\mathsf 1}\ .
$$
\end{enumerate} The number $n$ is called {\em dimension} of ${\mathsf E}$. Let
us denote the support of the measure ${\mathsf E}$ by ${\mathbf \Omega}$. The
set ${\mathbf \Omega}$ is called the {\em value space} of ${\mathsf E}$. The
operators ${\mathsf E}(X)$ for $X\in {\mathbf F}$ are called {\em effects}.
\end{df} The support ${\mathbf \Omega}$ of measure ${\mathsf E}$ is defined as
follows
$$
{\mathbf \Omega} = \{\vec{x}\in {\mathbb R}^n| \mathsf
E(\{\vec{y}||\vec{x}-\vec{y}| < \epsilon\}) \neq {\mathsf 0}\quad \forall
\epsilon > 0\}\ .
$$

Let ${\mathbf H}$ and ${\mathbf H}'$ be two separable Hilbert spaces and
${\mathsf U} : {\mathbf H} \mapsto {\mathbf H}'$ a unitary map. Then ${\mathsf
U}{\mathsf E}{\mathsf U}^\dagger : {\mathbf F} \mapsto {\mathbf L}({\mathbf
H}')$ is a POV measure on ${\mathbf H}'$.

We denote by ${\mathbf L}_r({\mathbf H})^+_{\leq 1}$ the set of all effects.
\begin{prop} ${\mathbf L}_r({\mathbf H})^+_{\leq 1}$ is the set of elements of
${\mathbf L}_r({\mathbf H})$ satisfying the inequality
\begin{equation}\label{effect} {\mathsf 0}\leq {\mathsf E}(X)\leq {\mathsf 1}\
.
\end{equation}
\end{prop} That effects must satisfy (\ref{effect}) follows from the
positivity and normalisation of a POV measure. On the other hand, each element
of the set defined in Proposition 4 is an effect. For example, if ${\mathsf
E}$ is such an element, we can define ${\mathsf E}(\{1\}) = {\mathsf E}$ and
${\mathsf E}(\{-1\}) = {\mathsf 1}-{\mathsf E}$. The two operators satisfy
Eq.\ (\ref{effect}) and they sum to ${\mathsf 1}$, so they determine a POV
measure with the value set $\Omega = \{-1,+1\}$.

Proposition 4 implies that the spectrum of each effect is a subset of
$[0,1]$. An effect is a projection operator (${\mathsf E}(X)^2 = {\mathsf
E}(X)$) if an only if its spectrum is the two-point set $\{0,1\}$.

\begin{prop}For any POV measure ${\mathsf E} : {\mathbf F} \mapsto {\mathbf
L}({\mathbf H})$ the following two conditions are equivalent
$$
{\mathsf E}(X)^2 = {\mathsf E}(X)
$$
for all $X \in {\mathbf F}$ and
$$
{\mathsf E}(X\cap Y) = {\mathsf E}(X){\mathsf E}(Y)
$$
for all $X, Y\in {\mathbf F}$\ .
\end{prop} Thus, a POV measure is a {\em projection valued} (PV) measure
exactly when it is multiplicative. In this case, all effects commute with each
other.

PV measures for $n=1$ are equivalent to s.a.\ operators that need not be
bounded. Sec.\ 1.1.1 introduced only bounded s.a.\ operators and we need a
more general definition.

Let ${\mathsf A}$ be an operator on Hilbert space ${\mathbf H}$ that is
not necessarily bounded. Then it is not defined on the whole of ${\mathbf H}$
but only on linear subspace ${\mathbf D}_{\mathsf A}$ that is dense in
${\mathbf H}$ and is called {\em domain} of ${\mathsf A}$. The definition of
adjoint has two steps: first, ${\mathsf A}^\dagger$ is defined on ${\mathbf
D}_{\mathsf A}$ and then extended. Similarly the definition of the
self-adjoint operator, see e.g.\ Ref.\ \cite{RS}.

To define a sum and product of two unbounded operators ${\mathsf A}$ and
${\mathsf B}$, their domains are used as follows. If
$$
{\mathbf D}_{\mathsf A} = {\mathbf D}_{\mathsf B}
$$
then we say that ${\mathsf A}$ and ${\mathsf B}$ have a common domain. If,
moreover, the common domain ${\mathbf D}$ is invariant with respect to both
operators, i.e.,
$$
{\mathsf A}{\mathbf D}\subset {\mathbf D}\ ,\quad {\mathsf B}{\mathbf
D}\subset {\mathbf D}
$$
then the sum ${\mathsf A} + {\mathsf B}$ and product ${\mathsf A}{\mathsf B}$
are well defined on ${\mathbf D}$ and can be possibly extended.

For s.a.\ operators, bounded or unbounded, the so-called spectral theorem
holds, see Ref.\ \cite{RS}. This says that any s.a.\ operator ${\mathsf A}$ is
equivalent to a PV measure, which is called, in this case, the spectral
measure of ${\mathsf A}$. Let ${\mathsf E}$ be a PV measure, then ${\mathsf
E}$ determines a unique self-adjoint operator $\int_{\mathbb R}\iota d{\mathsf
E}$, where $\iota$ denotes the identity function on ${\mathbb R}$. Conversely,
each s.a.\ operator $\mathsf A$ on ${\mathbf H}$ determines a unique PV
measure ${\mathsf E} : {\mathbf F}\mapsto {\mathbf L}({\mathbf H})$ such that
\begin{equation}\label{sharp} {\mathsf A} = \int_{\mathbb R}\iota d{\mathsf
E}\ .
\end{equation} If PV measure ${\mathsf E}$ is equivalent to a s.a.\ operator
${\mathsf A}$, we shall denote it ${\mathsf E}^{\mathsf A}$. Thus, POV measure
is a generalisation of a self-adjoint operator.

Another important special case is a discrete POV measure. A POV measure
${\mathsf E}$ is {\em discrete} if its value set ${\mathbf \Omega}$ is an at
most countable subset of $\mathbb R$. Let ${\mathsf E}$ be a discrete POV
measure and let ${\mathbf \Omega} = \{a_k, k\in {\mathbb N}\}$, where
${\mathbb N}$ is the set of positive integers. Then ${\mathsf E}$ is a PV
measure, ${\mathsf E} = {\mathsf E}^{\mathsf A}$, ${\mathsf A} = \sum_k a_k
{\mathsf E}_k$, $a_k$ are eigenvalues of ${\mathsf A}$ and
$$
{\mathsf E}_k = {\mathsf E}(\{a_k\})
$$
are projections on the corresponding eigenspaces of the s.a.\ operator
${\mathsf A}$ that is defined in this way.

For any POV measure ${\mathsf E} : {\mathbf F} \mapsto {\mathbf L}({\mathbf
H})$ and any ${\mathsf T}\in {\mathbf T}({\mathbf H})^+_1$, the mapping
$$
p^{\mathsf E}_{\mathsf T} : {\mathbf F} \mapsto [0,1]
$$
defined by
\begin{equation}\label{probab} p^{\mathsf E}_{\mathsf T}(X) = tr[{\mathsf
T}{\mathsf E}(X)]
\end{equation} for all $X\in {\mathbf F}$ is a real $\sigma$-additive
probability measure with values in $[0,1]$. This follows from the defining
properties of ${\mathsf E}$ and the continuity and linearity of the trace. The
measure is preserved by unitary maps,
$$
tr[({\mathsf U}{\mathsf T}{\mathsf U}^\dagger)({\mathsf U}{\mathsf
E}(X){\mathsf U}^\dagger)] = tr[{\mathsf T}{\mathsf E}(X)]\ .
$$

We note that a convex combination of states induces a convex combination of
measures,
$$
{\mathsf T} = \sum_k w_k {\mathsf T}_k \mapsto p^{\mathsf E}_{\mathsf T} =
\sum_k w_k p^{\mathsf E}_{{\mathsf T}_k}\ .
$$

More about mathematical properties of states and effects and the corresponding
spaces ${\mathbf T}(\mathbf H)^+_1$ and ${\mathbf L}_r(\mathbf H)^+_{\leq 1}$
can be found in Ref.\ \cite{ludwig1}.

\subsection{General rules} In this subsection, we shall again work only with
systems of definite type $\tau$.
\begin{rl} Any quantum mechanical observable for system ${\mathcal S}$ of type
$\tau$ is mathematically described by some POV measure ${\mathsf E} : {\mathbf
F} \mapsto {\mathbf L}_r({\mathbf H}_\tau)$. Each outcome of an individual
registration of the observable ${\mathsf E}$ performed on an object ${\mathcal
S}$ yields an element of ${\mathbf F}$. Each registration apparatus determines
a unique observable.
\end{rl} Often, a stronger assumption is made, namely that each element of
${\mathbf L}_r({\mathbf H}_\tau)$ is a sharp observable of system ${\mathcal
S}$ of type $\tau$. Such systems are called {\em proper} quantum systems. In
Secs.\ 2.2.2 and 6.2, we shall show that, in strict sense, no quantum system
is proper. Usually, the construction of a model needs only few observables
and, for a given model, our theory of quantum measurement will itself
determine which POV measures cannot be observables.

An important assumption of quantum mechanics is the following generalisation
of Born's rule
\begin{rl} The number $p^{\mathsf E}_{\mathsf T}(X)$ defined by Eq.\
(\ref{probab}) is the probability that a registration of the observable
${\mathsf E}$ performed on object ${\mathcal S}$ in the state ${\mathsf T}$
leads to a result in the set $X$.
\end{rl} Using the same preparation $N$ times gives always the same state
${\mathsf T}$. Performing in each case the registration measuring ${\mathsf
E}$ results in a value (or in an approximate value) $y\in {\mathbf
\Omega}$. The relative frequency of finding $y\in X$ approaches $p^{\mathsf
E}_ {\mathsf T}(X)$ if $N \rightarrow\infty$. Hence, the observable is
associated with an apparatus and the corresponding effects with subsets of
values that can be obtained by registration processes.

\begin{prop} If ${\mathsf T}_1$ and ${\mathsf T}_2$ from ${\mathbf T}({\mathbf
H}_\tau)^+_1$ satisfy
$$
tr[{\mathsf T}_1{\mathsf E}] = tr[{\mathsf T}_2{\mathsf E}]
$$
for all ${\mathsf E} \in {\mathbf L}_r({\mathbf H}_\tau)^+_{\leq 1}$ then
${\mathsf T}_1 = {\mathsf T}_2$; if ${\mathsf E}_1$ and ${\mathsf E}_2$ from
${\mathbf L}_r({\mathbf H}_\tau)^+_{\leq 1}$ satisfy
$$
tr[{\mathsf T}{\mathsf E}_1] = tr[{\mathsf T}{\mathsf E}_2]
$$
for all ${\mathsf T} \in {\mathbf T}({\mathbf H}_\tau)^+_1$ then ${\mathsf
E}_1 = {\mathsf E}_2$.
\end{prop} Thus, a state ${\mathsf T}$ can be determined uniquely, if it is
prepared many times and a sufficient number of different registrations can be
performed on it. As $tr[{\mathsf E}{\mathsf T}]$ is independent of the gemenge
structure of $\mathsf T$, the question of what is its gemenge structure cannot
be decided in this way with one exception: vector states. A vector state is
extremal so that it cannot be a non-trivial gemenge.

In Rule 3, the probability $p^{\mathsf E}_{\mathsf T}(X)$ refers to
measurements on an object and not to the object itself. In general, the
property that value $x\in {\mathbf \Omega}$ lies within subset $X\in {\mathbf
F}$ is not an objective property of objects in the sense that it could be
attributed to an object itself and that the measurement would just reveal
it. Such assumption would lead to contradictions, see Sec.\ 3.2.2. The
property can be called a relational in the sense that it does not concern the
quantum object alone but only in interaction with a definite registration
device\footnote{There also are relational properties that are
objective. Consider a point particle in Newtonian mechanics. Its angular
momentum with respect to the origin of coordinates is an objective property of
the particle if the origin is uniquely determined. It is relational, because
it depends on the choice of origin, not only on the particle.}. It can also be
called incidental, because it is not predictable, except in special cases, see
Sec.\ 3.2.2.

Thus, there is an asymmetry between states and observables: states are
objective properties but observables are not; all elements of ${\mathbf
T}({\mathbf H}_\tau)^+_1$ can be prepared but not all POV measures are
observables.

Special cases in which the outcome of a registration is predictable are
described as follows. Let ${\mathsf E} : {\mathbf F} \mapsto {\mathbf
L}_r({\mathbf H}_\tau)$ and let ${\mathsf T} \in {\mathbf T}({\mathbf
H}_\tau)^+_1$ be such that
$$
tr[{\mathsf T}{\mathsf E}(X)] = 1
$$
for some $X\in {\mathbf F}$. Then the probability that the registration of
${\mathsf E}$ on ${\mathsf T}$ will give a value in $X$ is 1 and we can say
that the prepared object in the state ${\mathsf T}$ possesses the property
independently of any measurement.
\begin{prop} The condition $tr[{\mathsf T}{\mathsf E}(X)] = 1$ is equivalent
to
$$
{\mathsf E}(X){\mathsf T} = {\mathsf T}\ .
$$
\end{prop} In general, we call state ${\mathsf T}$ satisfying
$$
{\mathsf E}(X){\mathsf T} = a{\mathsf T}
$$
with some real $a$ {\em eigenstate} of ${\mathsf E}(X)$ to {\em eigenvalue}
$a$. If ${\mathsf T} = {\mathsf P}[\psi]$, then $\psi$ is called {\em
eigenvector}. Thus, ${\mathsf T}$ in the Proposition 7 is an eigenstate of
${\mathsf E}(X)$ to eigenvalue 1.

If a PV measure is an observable, we call the observable {\em sharp}. Let us
consider a discrete observable ${\mathsf A}$ (which is always sharp) with
spectrum $\{a_k\}$ that is non-degenerate, ${\mathsf P}_k = {\mathsf
P}[\psi_k]$ for all $k$. $\psi_k$ are eigenvectors of ${\mathsf A}$. If we
prepare the state ${\mathsf P}[\psi_k]$ and then register ${\mathsf A}$, the
result must be $a_k$ with probability 1. Next suppose that we prepare a linear
superposition ${\mathsf P}[\Psi]$,
$$
\Psi = \sum_1^\infty c_k\psi_k\ ,
$$
with $\sum |c_k|^2 = 1$. The probability $p_k$ that the registration of
${\mathsf A}$ will give the result $a_k$ is
\begin{equation}\label{probrep} p_k = tr[{\mathsf P}[\psi_k]{\mathsf P}[\Psi]]
= |c_k|^2\ .
\end{equation} Eq.\ (\ref{probrep}) gives the physical meaning to the
coefficients in the linear superposition and is called Born rule. Observe that
the probability is only meaningful if it concerns really existing (but may-be
unknown) outcomes, see Sec.\ 3.1. Thus, it can be only associated with the
registration, not with the object in state ${\mathsf P}[\Psi]$. One can say
that, before the registration, the object in this state possesses all values
$a_k$ for which $c_k \neq 0$ simultaneously.

This can be seen experimentally by means of the interference phenomena: the
average in state ${\mathsf P}[\Psi]$ of observable ${\mathsf B}$ that does not
commute with ${\mathsf A}$ does depend on matrix elements of ${\mathsf B}$
between different states $\psi_k$:
$$
tr[{\mathsf B}{\mathsf P}[\Psi]] = \sum_{kl}c^*_kc_l\langle \psi_k |{\mathsf
B}| \psi_l \rangle\ .
$$

Rule 3 implies useful formulae for averages and higher moments of ${\mathsf
E}$ that generalise the well-known formulae for sharp observables. For the
average $\langle{{\mathsf E}}\rangle$ of observable ${\mathsf E}$ with $n=1$
in the state ${\mathsf T}$, we have
$$
\langle{{\mathsf E}}\rangle= \int_{\mathbb R} \iota dp^{\mathsf E}_{\mathsf
T}\ .
$$
Using Eq.\ (\ref{probab}), we obtain
$$
\langle{{\mathsf E}}\rangle= \int_{\mathbb R} \iota\,d\left(tr[{\mathsf
T}{\mathsf E}]\right) = tr\left[{\mathsf T} \int_{\mathbb R} \iota d{\mathsf
E}\right]\ .
$$
For a sharp observable ${\mathsf E}^{\mathsf A}$, Eq. (\ref{sharp}) yields the
usual relation
$$
\langle{{\mathsf A}}\rangle= tr[{\mathsf T}{\mathsf A}]\ .
$$
In the case of vector state, ${\mathsf T} = {\mathsf P}[\phi]$, we obtain
$\langle{{\mathsf A}}\rangle = \langle \phi|{\mathsf A}\phi \rangle$. For
higher moments, we have just to substitute suitable power of $\iota$ in the
integral and the proof is analogous.

The values of an observable are not objective properties of an object but its
states are. As averages and moments of an observable are uniquely determined
by states, we have
\begin{prop} The average $\langle{{\mathsf E}}\rangle$ of any observable
${\mathsf E}$ in state ${\mathsf T}$ of object $\mathcal S$ is an objective
property of $\mathcal S$ that has been prepared in state ${\mathsf T}$.
\end{prop} One may wonder how an average can be objective if the values from
which the average is calculated are not. However, the value $\langle{{\mathsf
E}}\rangle$ of the average is uniquely determined by the preparation and
$\langle{{\mathsf E}}\rangle$ is fixed before any registration, whereas the
outcomes of individual registrations are not. In fact, the values that are
obtained by registrations must satisfy the condition that they sum to a given
average.

\section{Galilean group} In quantum mechanics, {\em Galilean relativity
principle} holds in the following form:
\begin{rl} The same experiments performed in two different Newtonian inertial
frames have the same results, i.e. give the same probability distributions.
\end{rl} 'The same experiment' means that all conditions of the first
experiment have the same description with respect to the first inertial frame
as those of the second experiment have with respect to the second frame. In
the present section, we restrict ourselves to the proper Galilean group and
work out some consequences of the principle. For example, the most important
PV measures of quantum mechanics will be defined and the time evolution
equation formulated. We just list the relevant facts; for motivation and
technical detail, see, e.g., Ref.\ \cite{peres}.

The group of transformations that leave the geometric structure of Newtonian
spacetime invariant\footnote{We understand symmetry as transformation that
leaves some well-defined structure invariant. For example, Galilean group
contains all transformations that leave Newtonian spacetime geometry invariant
and a symmetry of a quantum system leaves its Hamiltonian invariant.} (see,
e.g., Ref.\ \cite{MTW}, p. 296) is called Galilean group ${\mathbf G}$. It is
also the group of transformations between inertial frames. Let
$(x_1,x_2,x_3,t)$ be a Newtonian inertial frame so that $x_1,x_2,x_3$ are
Cartesian coordinates. A general element $g(\lambda)$ of the component of
unity of ${\mathbf G}$ (called proper Galilean group, ${\mathbf G}^+$) can be
written in the form
\begin{equation}\label{actiong1} \vec{x}^\top \mapsto {\mathsf O}\vec{x}^\top
+ \vec{a}^\top + \vec{v}^\top t
\end{equation} and
\begin{equation}\label{actiong2} t \mapsto t + \lambda_{10}\ ,
\end{equation} where ${\mathsf O}$ is a proper orthogonal matrix determined by
three parameters $\lambda_1,\lambda_2,\lambda_3$, $\vec{a}$ is a vector of
space shift with components $\lambda_4,\lambda_5,\lambda_6$, $\vec{v}$ is a
relative velocity of the frames with components
$\lambda_7,\lambda_8,\lambda_9$ and $\lambda_{10}$ is a time shift. The
relations are written in matrix notation so that, e.g., $\vec{x}^\top$ is a
column matrix with components of vector $\vec{x}$.

The group product $g(\lambda^3) = g(\lambda^2)g(\lambda^1)$ is defined as the
composition of the transformations $g(\lambda^2)$ and $g(\lambda^1)$ so that
$g(\lambda^1)$ is performed first and $g(\lambda^2)$ second. Then,
$$
{\mathsf O}_3 = {\mathsf O}_2{\mathsf O}_1\ ,
$$
$$
\vec{a}_3^\top = \vec{a}_2^\top + {\mathsf O}_2\vec{a}_1^\top +
\vec{v}_2^\top\lambda_{10}^1\ ,
$$
$$
\vec{v}_3^\top = \vec{v}_2^\top + {\mathsf O}_2\vec{v}_1^\top\ ,
$$
and
$$
\lambda_{10}^3 = \lambda_{10}^2 + \lambda_{10}^1\ .
$$

Given a quantum system ${\mathcal S}$ of type $\tau$, classical apparatuses
that are supposed to prepare and register ${\mathcal S}$ have well-defined
Galilean transformations. There are then some conventions on the corresponding
action on states and effects. Often, ${\mathcal S}$ is subject to influences
external to ${\mathcal S}$ such as classical external fields. These also
possess non-trivial transformation laws with respect to Galilean
transformations. We call system ${\mathcal S}$ {\em isolated}, if the
approximation that it is alone in the universe is a good one. Thus, there are
no external influences.

Let us consider a measurement that consists of a preparation and a
registration of system $\mathcal S$ of type $\tau$ by apparatuses ${\mathcal
A}_p$ and ${\mathcal A}_r$, respectively, and let there be some external
fields $f$. Let ${\mathcal A}_p$ prepare object ${\mathcal S}$ in state
${\mathsf T}$ and let ${\mathcal A}_r$ register effect ${\mathsf E}(X)$. Next,
we can transport both ${\mathcal A}_p$ and ${\mathcal A}_r$ by $g \in {\mathbf
G}$ to $g({\mathcal A}_p)$ and $g({\mathcal A}_r)$. Let ${\mathsf T}_{g,f}$ be
the state prepared by $g({\mathcal A}_p)$ and ${\mathsf E}_{g,f}(X)$ the
effect registered by $g({\mathcal A}_r)$. If the external influences are also
transformed by $g : f \mapsto g(f)$, we obtain state ${\mathsf T}_{g,g(f)}$
and effect ${\mathsf E}_{g,g(f)}(X)$. The experiment has then been completely
transferred by $g$ and, therefore, Galilean relativity principle implies
\begin{prop} State ${\mathsf T}_{g,g(f)}$ and effect ${\mathsf E}_{g,g(f)}(X)$
satisfy
\begin{equation}\label{consprob} tr[{\mathsf T}{\mathsf E}(X)] = tr[{\mathsf
T}_{g,g(f)}{\mathsf E}_{g,g(f)}(X)]
\end{equation} for all $g \in {\mathbf G}$, ${\mathsf T} \in {\mathbf
T}({\mathbf H}_\tau)^+_1$ and ${\mathsf E}(X) \in {\mathbf L}_r({\mathbf
H}_\tau)^+_{\leq 1}$.
\end{prop}

Let us compare the two effects ${\mathsf E}(X)$ and ${\mathsf
E}_{g,g(f)}(X)$. In the case that Borel sets $X$ have a well-defined
transformation with respect to $g$, $\gamma(g) : {\mathbf F} \mapsto {\mathbf
F}$ and that the two effects belong to the same observable, we assume that
$$
{\mathsf E}_{g,g(f)}(X) = {\mathsf E}(\gamma(g)X)\ .
$$
Let us next compare measurable properties of the two states ${\mathsf T}$ and
${\mathsf T}_{g,g(f)}$. Proposition 9 implies:
$$
tr[{\mathsf T}_{g,g(f)}{\mathsf E}(X)] = tr[{\mathsf T}{\mathsf
E}_{g^{-1},g^{-1}(f)}(X)]\ .
$$
Hence, values that can be registered on the state ${\mathsf T}_{g,g(f)}$ are
those on ${\mathsf T}$ transformed by $g^{-1}$.

At this stage, one can formulate assumptions about the dependence of ${\mathsf
T}_{g,f}$ and ${\mathsf E}_{g,f}(X)$ on $f$ and about the way ${\mathbf G}$
acts on ${\mathbf T}({\mathbf H}_\tau)^+_1$ and ${\mathbf L}_r({\mathbf
H}_\tau)^+_{\leq 1}$ (see, e.g. Ref.\ \cite{ludwig1}). Then, important
theorems about action of ${\mathbf G}$ on ${\mathbf H}_\tau$ can be
proved. However, it is simpler and for our purposes sufficient to assume
directly:
\begin{rl} Let ${\mathcal S}$ be an isolated system ($f=0$) of type $\tau$
within which no separation status changes happen (see Sec.\ 2.2.4) and let
$\bar{\mathbf G}^+$ denote the universal covering of proper Galilean
group. Then, for each element $g \in \bar{\mathbf G}^+$ there is a unique map
${\mathsf U}(g) : {\mathbf H}_\tau \mapsto {\mathbf H}_\tau$ so that
\begin{enumerate}
\item ${\mathsf P}[{\mathsf U}(g)\psi] = {\mathsf P}[\psi]_{g^{-1},0}$ for all
$\psi \in {\mathbf H}_\tau$,
\item $g \mapsto {\mathsf U}(g)$ is a unitary ray representation of
$\bar{\mathbf G}^+$ depending on $\tau$,
\item for all ${\mathsf T} \in {\mathbf T}({\mathbf H}_\tau)^+_1$, ${\mathsf
E}(X) \in {\mathbf L}_r({\mathbf H}_\tau)^+_{\leq 1}$ and $g \in \bar{\mathbf
G}^+ $, we have
\begin{eqnarray}\label{actss1} {\mathsf T}_{g^{-1},0} & = & {\mathsf
U}(g){\mathsf T}{\mathsf U}(g)^\dagger\ , \\ \label{actss2} {\mathsf
E}_{g^{-1},0}(X) & = & {\mathsf U}(g){\mathsf E}(X){\mathsf U}(g)^\dagger\ .
\end{eqnarray}
\end{enumerate}
\end{rl} The 'ray' representation means that ${\mathsf U}(g_2g_1) = \exp[
i\omega(g_1,g_2)]{\mathsf U}(g_2){\mathsf U}(g_1)$. The representation depends
of $\tau$. For example, the representation is irreducible for
particles. Irreducible representations are classified by three numbers, $\mu,
s$ and $V$, with the meaning of mass, spin and constant potential,
respectively (see Ref.\ \cite{levy}, p.\ 221). Clearly, the probability is
preserved,
$$
tr[({\mathsf U}(g){\mathsf T}{\mathsf U}(g)^\dagger)({\mathsf U}(g){\mathsf
E}(X){\mathsf U}(g)^\dagger)] = tr[{\mathsf T}{\mathsf E}(X)]
$$
and we also have
\begin{equation}\label{inversw} {\mathsf U}(g){\mathsf E}(X){\mathsf
U}(g)^\dagger) = {\mathsf E}(\gamma(g)^{-1}X)
\end{equation} (see Ref.\ \cite{DST} for generalisation of this relation).

Now, the question arises: what is the form of spacetime transformations for
systems that are not isolated? There are two different answers, one for
Euclidean and one for time-translation subgroup of ${\mathbf G}$. This is a
part of the asymmetry between time and space in quantum mechanics.

\subsection{Euclidean group} Let ${\mathbf G}_E$ be the Euclidean subgroup of
${\mathbf G}$. We have:
\begin{rl} Let ${\mathcal S}$ be a system of type $\tau$ not necessarily
isolated. Then the conclusion of Rule 5 is valid only for the universal
covering of proper Euclidean group, $\bar{\mathbf G}^+_E$, if $f=0$ is
replaced by arbitrary $f$ and 'unitary ray representation' is replaced by
'unitary representation'.
\end{rl}

The specific form of the representation ${\mathsf U}(g)$ depends on the type
of $\mathcal S$ and can be described in all detail only after model
assumptions on $\mathcal S$ are made (see Chap.\ 4). However, some general
relations can be written already now.

If we restrict ourselves to the subgroup of space translations then the
corresponding element of ${\mathbf G}_E$ can be denoted by $g(\vec{a})$ and
its representative on ${\mathbf H}_\tau$ by ${\mathsf U}(\vec{a})$. Similarly,
each rotation can be described by unit vector $\vec{n}$ along its axis and
angle $\theta$ of rotation in the counter-clockwise direction around the
axis. Let the corresponding elements of the rotation subgroup be ${\mathsf
O}(\vec{n},\theta)$ with representative ${\mathsf U}(\vec{n},\theta)$.

Given a one-parameter group of unitary operators ${\mathsf U}(\lambda)$, then
according to Stone's theorem, there exists a s.a.\ operator ${\mathsf G}$
satisfying
$$
{\mathsf U}(\lambda) = \exp(i{\mathsf G}\lambda)\ .
$$
It called of {\em generator} of group ${\mathsf U}(\lambda)$ and can be
calculated with the help of the formula
$$
i{\mathsf G} = {\mathsf U}(\lambda)^\dagger\frac{d{\mathsf
U}(\lambda)}{d\lambda}\ .
$$
If the parameter is not specified by any convention, it is defined only up to
a real multiplier and so is the generator.

Let us define a one-parameter subgroup by ${\mathsf U}(\lambda \vec{a})$. Its
generator can be written in the form $(1/i\hbar)(a_1{\mathsf P}_1+a_2{\mathsf
P}_2 + a_3{\mathsf P}_3)$, where ${\mathsf P}_k$ are three s.a.\
operators. Similarly, the generator of the subgroup ${\mathsf
U}(\vec{n},\lambda)$ has the form $(1/i\hbar) (n_1{\mathsf J}_1+n_2{\mathsf
J}_2 + n_3{\mathsf J}_3)$, where ${\mathsf J}_k$ are s.a.\ operators.
\begin{df} The three s.a.\ operators ${\mathsf P}_k$ are three components,
with respect of the coordinates $x_1,x_2,x_3$, of the total {\em momentum} and
similarly ${\mathsf J}_k$ are components of total {\em angular momentum} of
${\mathcal S}$.
\end{df} The commutators of ${\mathsf P}_k$ and ${\mathsf J}_k$ are determined
by the fact that they are generators of ${\mathbf G}_E$ and that $g \mapsto
{\mathsf U}(g)$ is a group representation:
\begin{equation}\label{commg} [{\mathsf P}_k,{\mathsf P}_l] = {\mathsf 0}\
,\quad [{\mathsf J}_k,{\mathsf J}_l] = i\hbar \epsilon_{klj}{\mathsf J}_j\
,\quad [{\mathsf J}_k,{\mathsf P}_l] = i\hbar \epsilon_{klj}{\mathsf P}_j\ .
\end{equation}

Most quantum systems admit {\em position} as one of their observables. All
massive particles and all systems composite of a fixed number of massive
particles, their position being their centre of mass, do. Given an inertial
frame $(\vec{x},t)$, one can measure the corresponding coordinates of the
quantum system. We assume:
\begin{rl} There are three s.a.\ operators ${\mathsf Q}_k$ such that the
average position of object ${\mathcal S}$ in state ${\mathsf T}$ is
$tr[{\mathsf T}{\mathsf Q}_k]$ and
$$
{\mathsf U}(\vec{a}){\mathsf Q}_k{\mathsf U}(\vec{a})^\dagger = {\mathsf Q}_k
- a_k {\mathsf 1}
$$
and
$$
{\mathsf U}(\vec{n},\theta){\mathsf Q}_k{\mathsf U}(\vec{n},\theta)^\dagger =
\sum_l O(\vec{n},-\theta)_{kl}{\mathsf Q}_l\ .
$$
\end{rl} Hence, the position operators are transformed by Euclidean group in
agreement with Eq.\ (\ref{actiong1}) and the corresponding operators ${\mathsf
U}(g)$ transform ${\mathsf Q}_k$ according to Eq.\ (\ref{inversw}). From these
transformation properties, the commutation relations of position ${\mathsf
Q}_k$ with total momentum $P_k$ and total angular momentum $J^k$ can be
derived,
\begin{equation}\label{commq} [{\mathsf Q}_k,{\mathsf P}_l] = i\hbar
\delta_{kl}{\mathsf 1}\ ,\quad [{\mathsf J}_k,{\mathsf Q}_l] = i\hbar
\epsilon_{klj}{\mathsf Q}_j\ .
\end{equation}

Commutation rules (\ref{commg}) and (\ref{commq}) determine the action of
operators ${\mathsf Q}_k$ and ${\mathsf P}_k$ on ${\mathbf H}_\tau$
uniquely. Consider the set ${\mathbf D}$ of rapidly decreasing $C^\infty$
functions $\phi(\vec{x},\xi)$, where $\xi$ represents a set of parameters
dependent on the system type $\tau$. For example, for a composite system, it
can represent the relative coordinates of the constituents with respect to the
centre of mass $\vec{x}$ and for a particle with spin, it can represent the
components of the spin in some direction, etc. The rapid decrease is required
with respect to $\vec{x}$ and those parameters $\xi$ that have unbounded
ranges of values. In fact, ${\mathbf D}$ or its generalisation to functions of
more than one argument is a common invariant domains of all s.a.\ operators in
quantum mechanics and the products of these operators can be defined with the
help of ${\mathbf D}$.

Let ${\mathbf H}_\tau$ be the completion of ${\mathbf D}$ with respect to the
inner product
$$
\langle\phi|\psi\rangle = \sum_\xi \int_{{\mathbb R}^3} d^3x
\phi^*(\vec{x},\xi) \psi(\vec{x},\xi)\ ,
$$
where the sum over $\xi$ represents sums over all discrete parameters and
integrals over all continuous ones. The elements of ${\mathbf H}_\tau$ are
called {\em wave functions}. To describe the operators, it is sufficient to
define the action of their components on $C^\infty$ functions
$\phi(\vec{x},\xi)$ because there is always only one s.a.\ extension. The
result is
\begin{equation}\label{posit} {\mathsf Q}_k\phi(\vec{x},\xi) =
x_k\phi(\vec{x},\xi)
\end{equation} and
\begin{equation}\label{moment} {\mathsf P}_k\phi(\vec{x},\xi) =
-i\hbar\frac{\partial}{\partial x_k}\phi(\vec{x},\xi)\ .
\end{equation} For the total angular momentum, the result is more complicated
and $\tau$-dependent. We postpone its description to Chap.\ 4.

The form of Hilbert space and the operator described above is called $Q$-{\em
representa\-tion}. All operators on ${\mathbf H}_\tau$ in $Q$-representation
can be written in the form of integral operators with kernels
$$
{\mathsf A} = A(\vec{x},\xi;\vec{x}',\xi')\ ,
$$
the kernel $A(\vec{x},\xi;\vec{x}',\xi')$ being a generalised function of its
arguments acting on functions $\phi(\vec{x},\xi)$ as follows
$$
({\mathsf A}\phi)(\vec{x},\xi) = \sum_{\xi'} \int_{{\mathbb R}^3} d^3x'
A(\vec{x},\xi;\vec{x}',\xi') \phi(\vec{x}',\xi')\ ,
$$
where the right-hand side represents the action of the generalised function on
a test function.

\subsection{Time translations} The next set of transformations are time
translations. In quantum mechanics, time is, unlike position, just a
parameter. This is another part of the asymmetry between time and space. For
general systems (not necessarily closed), we assume
\begin{rl} Let ${\mathcal S}$ be a system of type $\tau$ that does not take
part in any separation status change (see Sec.\ 2.2.4) and let the external
fields $f$ be given. Then, time translation from $t_1$ to $t_2$ is represented
by unitary operator ${\mathsf U}(f,t_2,t_1)$ on ${\mathbf H}_\tau$ satisfying
$$
{\mathsf U}(f,t_3,t_1) = {\mathsf U}(f,t_3,t_2){\mathsf U}(f,t_2,t_1)
$$
and we have
$$
{\mathsf P}[\psi]_{g(t_2 - t_1)^{-1},f} = {\mathsf U}(f,t_2,t_1){\mathsf
P}[\psi]{\mathsf U}(f,t_2,t_1)^\dagger\ ,
$$
where $g(t_2 - t_1)$ is the group element for $\lambda_1 = \cdots = \lambda_9
= 0$ and $\lambda_{10} = t_2 - t_1$.
\end{rl} A time translation does not define a unique map on ${\mathbf
T}({\mathbf H}_\tau)^+_1$ in this case. ${\mathsf U}(f,t_2,t_1)$ depends not
only on $t_2 - t_1$ but also on the position of the system with respect to the
external fields (even if the fields are stationary). We shall let out the
argument $f$ in ${\mathsf U}(f,t_2,t_1)$ in agreement with the current
practice.
\begin{df} The operator ${\mathsf H}(t)$ defined by
$$
{\mathsf H}(t) = i\hbar{\mathsf U}(t,t_0)^\dagger\frac{d{\mathsf
U}(t,t_0)}{dt}\ ,
$$
is the Hamiltonian of $\mathcal S$.
\end{df} Operators ${\mathsf H}(t)$ for different $t$ do not commute in
general.

The most important observable in non-relativistic quantum mechanics is
energy. The corresponding operator for system $\mathcal S$ is its
Hamiltonian. Any individually prepared quantum systems has a definite
Hamiltonian with the form
$$
{\mathsf H} = {\mathsf H}(\text{operators, external fields})\ ,
$$
where function ${\mathsf H}$ symbolises the construction of the operator from
the other s.a.\ operators of $\mathcal S$. The Hamiltonian of system $\mathcal
S$ is an objective property of $\mathcal S$. Also, the form of Hamiltonian is
a model assumption of quantum mechanics. The choice of a Hamiltonian is
usually the most important step in model construction.

Hamiltonian $\mathsf H$ of quantum system $\mathcal S$ determines the dynamics
of $\mathcal S$. Dynamics has to do with the time aspect of preparation and
registration. Any preparation procedure finishes at some time instant $t_p$
and any registration procedure starts at some time instant $t_r$. The times
$t_r$ and $t_p$ can serve as defining the time aspects because the whole
preparation or registration processes can themselves take some time. The
dynamics enables to calculate how the probabilities depend on the times $t_r$
and $t_p$. The dependence can be obtained in two ways. We can make either
state ${\mathsf T}(t)$ to a function of $t$ by shifting ${\mathsf T}(0)$
forwards or observable ${\mathsf E}(t)$ by shifting it backwards by ${\mathsf
U}(t,0)$. The probability $p_{\mathsf T}^{\mathsf E}(X,t)$ corresponding to
the time $t$ is then
$$
p_{\mathsf T}^{\mathsf E}(X,t) = tr[{\mathsf T}(t){\mathsf E}] = tr[{\mathsf
T}{\mathsf E}(t)]\ .
$$
The first method is called Schr\"{o}dinger picture, the second Heisenberg
picture.

Rule 8 implies:
\begin{prop} Let quantum system $\mathcal S$ have a Hamiltonian $\mathsf
H(t)$. Then, the dynamical evolution of $\mathcal S$ in Schr\"{o}dinger
picture obeys von Neumann-Liouville equation of motion
\begin{equation}\label{dynamS} i\hbar\frac{d{\mathsf T}(t)}{dt} = [{\mathsf
H}(t),{\mathsf T}(t)]
\end{equation} and in Heisenberg picture Heisenberg equation of motion
\begin{equation}\label{dynamH} i\hbar\frac{d{\mathsf E}(X,t)}{dt} = [{\mathsf
E}(X,t),{\mathsf H}(t)]\ .
\end{equation}
\end{prop}

\begin{prop} Let $\mathsf T$ be a vector state
$|\phi\rangle\langle\phi|$. Then
$$
{\mathsf T}(t) = |\phi(t)\rangle\langle\phi(t)|\ ,
$$
where $\phi(t)$ obeys Schr\"{o}dinger equation
\begin{equation}\label{schrod} i\hbar\frac{d\phi(t)}{dt} = \mathsf H(t)\phi(t)
\end{equation} and
$$
\phi(t) = {\mathsf U}(t,0) \phi(0)\ .
$$
\end{prop} It follows that Liouville-von-Neumann equation for states also
determines an evolution equation for vectors. In particular, vector states
remain vector states in evolution. More generally, we assume
\begin{rl} Let $\mathsf T$ be a gemenge (\ref{properm}) with components
${\mathsf T}_k$. Then, its time evolution is again a gemenge with components
${\mathsf T}_k(t)$
\begin{equation}\label{evolmix} {\mathsf T}(t) = \left(\sum_k\right)
w_k{\mathsf T}_k(t)\ ,
\end{equation} where ${\mathsf T}_k(t)$ is given by (\ref{dynamS}) and $w_k$
are time independent, for each $k$.
\end{rl} Of course, Eq.\ (\ref{evolmix}) is easily obtained from Eq.\
(\ref{properm}) by multiplying both sides by $U(t,0)$ from the left and by
$U(t,0)^\dagger$ from the right. The non-trivial assumption is that the state
remains a gemenge of the described kind. Hence, if an object $\mathcal S$ is
prepared in the gemenge state $\mathsf T$ described by Eq.\ (\ref{properm}) at
time 0 and then evolved, then the state of each individual object $\mathcal S$
is objectively always one of the components ${\mathsf T}_k(t)$ at time $t$.

Let us now restrict ourselves to isolated systems. In this case, the
Hamiltonian ${\mathsf H}(t) = {\mathsf H}$ is time independent s.a.\ operator,
the unitary time translation is
$$
{\mathsf U}(t_2,t_1) = \exp[-(i/\hbar) {\mathsf H} (t_2-t_1)]
$$
and
$$
{\mathsf U}(t_2,t_1) = {\mathsf U}(\lambda_{10})\ ,
$$
where ${\mathsf U}(\lambda_{10})$ is the representative of the time
translation in the unitary representation of the Galilean group on ${\mathbf
H}_\tau$ and $\lambda_{10} = t_2 - t_1$. Clearly, the Hamiltonian must then
commute with the total momentum and angular momentum,
$$
[{\mathsf H},{\mathsf P}_k] = {\mathsf 0}\ ,\quad [{\mathsf H},{\mathsf J}_k]
= {\mathsf 0}\ .
$$

\begin{df} Each unitary transformation ${\mathsf U} : {\mathbf H}_\tau \mapsto
{\mathbf H}_\tau$ that leaves the Hamiltonian ${\mathsf H}$ of ${\mathcal S}$
invariant,
$$
{\mathsf U}{\mathsf H}{\mathsf U}^\dagger = {\mathsf H}
$$
is called a {\em symmetry} of system ${\mathcal S}$.
\end{df} All symmetries of a system ${\mathcal S}$ form a unitary group which
is an objective property of ${\mathcal S}$. The generators of its
one-parameter subgroups are s.a.\ operators that commute with the Hamiltonian
and, as sharp observables, yield probability distributions that are
independent of time.

We have ignored the improper transformations such as space inversions and time
reversal. More about them can be found in Ref.\ \cite{wigner}.

\chapter{Composition of quantum systems} This is a key chapter of this
Survey. It emphasises the difference between composition of systems of the
same and of different types. It lists the problems to which the standard rules
about composition of identical systems lead and it shows a rigorous way of
removing these difficulties.

\section{Composition of systems of different types} Suppose that two objects
of different types are prepared. The objects can be particles or composite
systems. Then, one can consider this pair of objects as one quantum
object. This is called composition of objects. First, we describe the
mathematical apparatus.

\subsection{Tensor product of Hilbert spaces} The tensor product ${\mathbf H}
= {\mathbf H}_1\otimes {\mathbf H}_2$ of ${\mathbf H}_1$ and ${\mathbf H}_2$
is the Cauchy completion of the linear span of the set of products
$$
\phi\otimes\psi:\phi \in {\mathbf H}_1, \psi\in {\mathbf H}_2
$$
with respect to the inner product of ${\mathbf H}$, which is determined by
$$
\langle\phi\otimes\psi|\phi'\otimes\psi'\rangle = \langle\phi|\phi'\rangle
\langle\psi|\psi'\rangle\ .
$$
The tensor product $\phi\otimes\psi$ is linear in both arguments.  Moreover,
if $\{\phi_k\}$ and $\{\psi_k\}$ are bases of ${\mathbf H}_1$ and ${\mathbf
H}_2$, then $\{\phi_k\otimes\psi_l\}$ is a basis of ${\mathbf H}$. If the
bases are orthonormal then any $\Psi \in {\mathbf H}$ can then be expressed as
$$
\Psi = \sum_{kl} \langle\phi_k\otimes\psi_l|\Psi\rangle\phi_k\otimes\psi_l\ .
$$

If ${\mathsf A} \in {\mathbf L}({\mathbf H}_1)$ and ${\mathsf B} \in {\mathbf
L}({\mathbf H}_2)$, then their tensor product ${\mathsf A}\otimes {\mathsf B}$
on ${\mathbf H}_1\otimes{\mathbf H}_2$ is determined via the relation
$$
({\mathsf A}\otimes {\mathsf B})(\phi\otimes\psi) = {\mathsf A}\phi\otimes
{\mathsf B}\psi
$$
for all $\phi \in {\mathbf H}_1$ and $\psi \in {\mathbf H}_2$. It follows that
$$
tr[{\mathsf A} \otimes {\mathsf B}] = tr[{\mathsf A}]tr[{\mathsf B}]\ .
$$
The tensor product ${\mathsf T}_1\otimes {\mathsf T}_2$ of ${\mathsf T}_1 \in
{\mathbf T}({\mathbf H}_1)^+_1$ and ${\mathsf T}_2 \in {\mathbf T}({\mathbf
H}_2)^+_1$ is determined in the same way and it is a trace-class operator with
trace 1. However, ${\mathbf T}({\mathbf H}_1\otimes{\mathbf H}_2)^+_1$
contains more than just tensor products of elements from ${\mathbf T}({\mathbf
H}_1)^+_1$ and ${\mathbf T}({\mathbf H}_2)^+_1$.

The {\em partial trace} over the Hilbert space ${\mathbf H}_2$, say, is the
positive linear mapping
$$
\Pi_2:{\mathbf T}({\mathbf H}_1\otimes{\mathbf H}_2)^+_1 \mapsto {\mathbf
T}({\mathbf H}_1)^+_1
$$
defined via the relation
$$
tr[\Pi_2({\mathsf W}){\mathsf A}] = tr[{\mathsf W}({\mathsf A}\otimes {\mathsf
1}_2)]
$$
for all ${\mathsf A}\in {\mathbf L}({\mathbf H}_1)$, ${\mathsf W}\in {\mathbf
T}({\mathbf H}_1\otimes{\mathbf H}_2)$ and ${\mathsf 1}_2$ is the identity
operator on ${\mathbf H}_2$. If $\{\phi_k\}\subset {\mathbf H}_1$ and
$\{\psi_k\}\subset {\mathbf H}_2$ are orthonormal bases, then $\Pi_2({\mathsf
W})$ can be written as
$$
\Pi_2({\mathsf W}) = \sum_{ijk}\langle\phi_i\otimes \psi_k|{\mathsf
W}(\phi_j\otimes \psi_k)\rangle |\phi_i\rangle \langle \phi_j|\ .
$$
Here $|\phi_i\rangle \langle \phi_j|$ is the bounded linear operator on
${\mathbf H}_1$ given by
$$
|\phi_i\rangle \langle \phi_j|(\phi) = \langle \phi_j|\phi\rangle \phi_i
$$
for all $\phi\in {\mathbf H}_1$. The partial trace over ${\mathbf H}_1$ is
defined similarly.

If ${\mathsf W} = {\mathsf T}_1\otimes {\mathsf T}_2$, then ${\mathsf T}_1 =
\Pi_2({\mathsf W})$ and ${\mathsf T}_2 = \Pi_1({\mathsf W})$ but, in general,
\begin{equation}\label{entang} {\mathsf W} \neq \Pi_2({\mathsf W})\otimes
\Pi_1({\mathsf W})\ .
\end{equation} In particular, if ${\mathsf W} = {\mathsf P}[\Psi]$, then
$$
{\mathsf P}[\Psi] = \Pi_2({\mathsf P}[\Psi])\otimes \Pi_1({\mathsf P}[\Psi])
$$
if and only if
$$
\Psi = \phi\otimes \psi
$$
for some $\phi\in {\mathbf H}_1$ and $\psi\in {\mathbf H}_2$. In that case
also
$$
\Pi_2 ({\mathsf P}[\Psi]) = {\mathsf P}[\phi]\ ,\quad \Pi_1 ({\mathsf
P}[\Psi]) = {\mathsf P}[\psi]\ .
$$
Thus, tensor products of extremal elements of the sets ${\mathbf T}({\mathbf
H}_1)^+_1$ and ${\mathbf T}({\mathbf H}_2)^+_1$ do not exhaust the set of
extremal elements of ${\mathbf T}({\mathbf H}_1\otimes{\mathbf H}_2)^+_1$.

Tensor products of more Hilbert spaces and the corresponding notions and
relations can be obtained using the above rules.

Tensor product is also an operation for POV measures. Let ${\mathsf E}_1 :
{\mathbf F}_1 \mapsto {\mathbf L}_r({\mathbf H}_1)$ with dimension $n_1$ and
${\mathsf E}_2 : {\mathbf F}_2 \mapsto {\mathbf L}_r({\mathbf H}_2)$ with
dimension $n_2$ be two POV measures on Hilbert spaces ${\mathbf H}_1$ and
${\mathbf H}_2$ with value sets ${\mathbf \Omega}_1$ and ${\mathbf
\Omega}_2$. Then POV measures $({\mathsf E}_1\otimes {\mathsf E}_2) :
({\mathbf F}_1\times{\mathbf F}_2) \mapsto {\mathbf L}_r({\mathbf
H}_1\otimes{\mathbf H}_2)$ on the tensor product ${\mathbf H}_1\otimes{\mathbf
H}_2$ has dimension $n_1 + n_2$ and values set ${\mathbf
\Omega}_1\times{\mathbf \Omega}_2$ and is defined by
$$
({\mathsf E}_1\otimes {\mathsf E}_2)(X_1\times X_2) = {\mathsf
E}_1(X_1)\otimes {\mathsf E}_2(X_2)
$$
for all $X_1\subset {\mathbb R}^{n_1}$ and $X_2\subset {\mathbb
R}^{n_2}$. Tensor product of POV measures is associative but not commutative.

Let us now turn to physical interpretation.
\begin{rl} Let ${\mathcal S}_1$ and ${\mathcal S}_2$ be two quantum objects of
different types and their Hilbert spaces be ${\mathbf H}_1$ and ${\mathbf
H}_2$, respectively. Then, the object ${\mathcal S}$ composed of ${\mathcal
S}_1$ and ${\mathcal S}_2$ has the Hilbert space ${\mathbf H} = {\mathbf
H}_1\otimes{\mathbf H}_2$, its states are elements of ${\mathbf T}({\mathbf
H}_1\otimes{\mathbf H}_2)^+_1$ and its effects are elements of ${\mathbf
L}_r({\mathbf H}_1\otimes{\mathbf H}_2)^+_{\leq 1}$
\end{rl}

\begin{rl} Let ${\mathsf T}$ be a state of a composite object ${\mathcal S} +
{\mathcal S}'$. The necessary and sufficient condition for the gemenge
structure of the partial trace over ${\mathcal S}'$ to be
$$
tr_{{\mathcal S}'}[{\mathsf T}] = \left(\sum_k\right) w_k {\mathsf T}_k
$$
is that ${\mathsf T}$ itself has the gemenge structure
$$
{\mathsf T} = \left(\sum_k\right) w_k {\mathsf T}_k \otimes {\mathsf T}'_k\ ,
$$
where ${\mathsf T}'_k$ are some states of ${\mathcal S}'$.
\end{rl} One can say that gemenge structures are invariant with respect to
compositions. By registration of observables of ${\mathcal S}$ alone, only the
state operator ${\mathsf T}$ defined by Eq.\ (\ref{properm}) can be
determined. However, by registration of observables pertaining to a composite
object ${\mathcal S} + {\mathcal S}'$, some information about the gemenge
structure of ${\mathsf T}$ can be obtained from Rule 11. Suppose, e.g., that
${\mathcal S} + {\mathcal S}'$ is in a vector state ${\mathsf P}[\Psi]$. Then
partial trace $\Pi'({\mathsf P}[\Psi])$ cannot be a gemenge. This fact is at
the root of the objectification problem in quantum theory of measurement (cf.\
\cite{BLM}, our Secs.\ 6.1 and 6.3).

If the state of the composite object ${\mathcal S}_1 + {\mathcal S}_2$ is
${\mathsf W}\in {\mathbf T}({\mathbf H}_1\otimes {\mathbf H}_2)^+_1$, then the
two partial trace operations determine unique states, $\Pi_2({\mathsf W})$ of
${\mathcal S}_1$ and $\Pi_1({\mathsf W})$ of ${\mathcal S}_2$. These states
are called {\em reduced states}. The fact that the reduced state
$\Pi_2({\mathsf W})$ is uniquely determined by ${\mathsf W}$ means, in
particular, that objects ${\mathcal S}_1$ and ${\mathcal S}_2$ can be
identified as subsystems of ${\mathcal S}_1 + {\mathcal S}_2$.

Based on the mathematical properties of the tensor products, the description
of objects composed of arbitrary number of sub-systems can be obtained by
extension of the above methods. The only condition is that no two subsystems
belong to the same type.

\subsection{Entanglement}
\begin{df} If object ${\mathcal S}$ composite of two quantum objects
${\mathcal S}_1$ and ${\mathcal S}_2$ is in a state ${\mathsf W}$ that
satisfies condition (\ref{entang}) one says that ${\mathcal S}_1$ and
${\mathcal S}_2$ are entangled or that state ${\mathsf W}$ is entangled.
\end{df}

Entanglement is a physical phenomenon that has measurable consequences. One of
these consequences are correlations between outcomes of registrations that are
performed jointly on the two (or more) entangled systems.

As an example, consider two objects ${\mathcal S}_1$ and ${\mathcal S}_2$ and
two sharp observables ${\mathsf A}_i : {\mathbf H}_i \mapsto {\mathbf H}_i$,
$i=1,2$. Let $|a_i\rangle \in {\mathbf H}_i$ and $|b_i\rangle \in {\mathbf
H}_i$ be four eigenstates,
$$
{\mathsf A}_i|a_i\rangle = a_i|a_i\rangle\ ,\quad {\mathsf A}_i|b_i\rangle =
b_i|b_i\rangle\ ,
$$
and $b_i > a_i$, $i=1,2$. The state ${\mathsf P}[\Psi]$ of the composite
object ${\mathcal S}_1 + {\mathcal S}_2$, where
$$
|\Psi\rangle = \frac{1}{\sqrt{2}}(|a_1\rangle\otimes |b_2\rangle
+|b_1\rangle\otimes |a_2\rangle)
$$
is entangled. Indeed,
\begin{equation}\label{parti} \Pi_2({\mathsf P}[\Psi]) =
\frac{1}{2}(|a_1\rangle\otimes\langle a_1| + |b_1\rangle\otimes\langle b_1|)\
,\quad \Pi_1({\mathsf P}[\Psi]) = \frac{1}{2}(|a_2\rangle\otimes\langle a_2| +
|b_2\rangle\otimes\langle b_2|)
\end{equation} and ${\mathsf P}[\Psi]$ is not a tensor product of these two
states.

Let us calculate the correlations of two sharp observables, ${\mathsf
A}_1\otimes{\mathsf 1}$ and ${\mathsf 1}\otimes{\mathsf A}_2$ on ${\mathbf
H}_1\otimes{\mathbf H}_2$. The observables commute hence we can measure them
jointly (see Sec.\ 3.2). First, we need:
\begin{df} Let ${\mathsf A}$ be a s.a.\ operator with a dense invariant domain
and ${\mathsf T}$ be an arbitrary state. Then
\begin{equation}\label{variance} \Delta {\mathsf A} = tr[{\mathsf
T}(\sqrt{{\mathsf A}^2-\langle{{\mathsf A}}\rangle^2})]
\end{equation} is called {\em variance} of ${\mathsf A}$ in ${\mathsf T}$.
\end{df} Second,
\begin{df} Normalised correlation of any two observables ${\mathsf A}$ and
${\mathsf B}$ in a state ${\mathsf T}$ is defined by
\begin{equation}\label{norcorr} C({\mathsf A},{\mathsf B}, {\mathsf T}) =
\frac{tr[{\mathsf T}{\mathsf A}{\mathsf B}] - tr[{\mathsf T}{\mathsf
A}]tr[{\mathsf T}{\mathsf B}]}{\Delta {\mathsf A}\Delta {\mathsf B}}\ .
\end{equation}
\end{df} The normalised correlation satisfies
$$
-1 \leq C({\mathsf A},{\mathsf B}, {\mathsf T}) \leq 1
$$
for all ${\mathsf A}$, ${\mathsf B}$ and ${\mathsf T}$ because of Schwarz'
inequality. The proof is based on the facts that $Re\,tr[{\mathsf T}{\mathsf
A}{\mathsf B}]$ is a positive symmetric bilinear form on real linear space
${\mathbf L}_r({\mathbf H}_1\otimes{\mathbf H}_2)$, $Re\,tr[{\mathsf
T}{\mathsf A}{\mathsf B}] = tr[{\mathsf T}{\mathsf A}{\mathsf B}]$ if
${\mathsf A}$ and ${\mathsf B}$ commute and their correlation can only be
measured if they commute. If $C({\mathsf A},{\mathsf B}, {\mathsf T}) = 1$,
the operators are strongly correlated, if $C({\mathsf A},{\mathsf B}, {\mathsf
T}) = 0$, they are uncorrelated and if $C({\mathsf A},{\mathsf B}, {\mathsf
T}) = -1$, they are strongly anticorrelated (see \cite{BLM}, p. 50).

First, we calculate their first two moments of observables ${\mathsf
A}_1\otimes{\mathsf 1}$ and ${\mathsf 1}\otimes{\mathsf A}_2$ in state
${\mathsf P}[\Psi]$:
$$
\langle\Psi|{\mathsf A}_1\otimes{\mathsf 1}|\Psi\rangle = \frac{1}{2}(a_1 +
b_1)\ ,\quad\langle\Psi|{\mathsf 1}\otimes{\mathsf A}_2|\Psi\rangle =
\frac{1}{2}(a_2 + b_2)\
$$
and
$$
\langle\Psi|({\mathsf A}_1\otimes{\mathsf 1})^2|\Psi\rangle =
\frac{1}{2}(a_1^2 + b_1^2)\ ,\quad\langle\Psi|({\mathsf 1}\otimes{\mathsf
A}_2)^2|\Psi\rangle = \frac{1}{2}(a_2^2 + b_2^2)\ .
$$
The variances are
$$
\Delta({\mathsf A}_1\otimes{\mathsf 1}) = \left(\frac{b_1-a_1}{2}\right)^2\
,\quad \Delta({\mathsf 1}\otimes{\mathsf A}_2) =
\left(\frac{b_2-a_2}{2}\right)^2\ .
$$
The average of the product $({\mathsf A}_1\otimes{\mathsf 1})({\mathsf
1}\otimes{\mathsf A}_2) = ({\mathsf A}_1\otimes{\mathsf A}_2)$ is
$$
\langle\Phi|{\mathsf A}_1\otimes{\mathsf A}_2\Phi\rangle =
\frac{a_1b_2+b_1a_2}{2}
$$
so that, finally
$$
C({\mathsf A}_1\otimes{\mathsf 1},{\mathsf 1}\otimes{\mathsf A}_2,{\mathsf
P}[\Psi])= -1\ .
$$
The result is that the observables are strongly anticorrelated in state
${\mathsf P}[\Psi]$. What this means, can be seen from the probability
distributions for different possible outcomes by measuring the two
observables. The corresponding PV measures define the four projections
$$
{\mathsf P}_{aa} = {\mathsf E}_1(\{a_1\})\otimes{\mathsf E}_2(\{a_2\})\ ,\quad
{\mathsf P}_{ab} = {\mathsf E}_1(\{a_1\})\otimes{\mathsf E}_2(\{b_2\})\ ,
$$
$$
{\mathsf P}_{ba} = {\mathsf E}_1(\{b_1\})\otimes{\mathsf E}_2(\{a_2\})\ ,\quad
{\mathsf P}_{bb} = {\mathsf E}_1(\{b_1\})\otimes{\mathsf E}_2(\{b_2\})\ ,
$$
and we obtain
$$
\langle\Psi|{\mathsf P}_{aa}|\Psi\rangle = 0\ ,\quad \langle\Psi|{\mathsf
P}_{ab}|\Psi\rangle = \frac{1}{2}\ ,
$$
$$
\langle\Psi|{\mathsf P}_{ba}|\Psi\rangle = \frac{1}{2}\ ,\quad
\langle\Psi|{\mathsf P}_{bb}|\Psi\rangle = 0\ .
$$
It follows: if the registration of ${\mathsf A}_1\otimes {\mathsf 1}$ gives
$a_1$ ($b_1$) then the registration of ${\mathsf 1}\otimes {\mathsf A}_1$
gives $b_2$ ($a_2$)with certainty, and vice versa, the correlation being
symmetric with respect to ${\mathcal S}_1$ and ${\mathcal S}_2$.

Suppose now that ${\mathcal S}_1$ and ${\mathcal S}_2$ are far from each
other, ${\mathcal S}_1$ near point $\vec{x}_1$ and ${\mathcal S}_2$ near point
$\vec{x}_2$. Could one use the strong anticorrelation to send signals from
$\vec{x}_1$ to $\vec{x}_2$, say? There are two reasons why one cannot. First,
one has no choice of what value, $a_1$ or $b_1$ one obtains at $\vec{x}_1$. As
is easily seen from the state $\Pi_2({\mathsf P}[\Psi])$ of ${\mathcal S}_1$,
Eq.\ (\ref{parti}), the probability of each outcome is 1/2. One has,
therefore, no control about what signal will be sent. Second, suppose that the
state ${\mathsf P}[\Psi]$ is prepared many times and let, in the first chain
of experiments, the observer at $\vec{x}_1$ registers ${\mathsf A}_1$ every
time, in the second chain, he does nothing. Is there any difference between
the two cases that could be recognized at $\vec{x}_2$? In the first case, the
state of ${\mathcal S}_2$ is given by Eq.\ (\ref{parti}) including its gemenge
structure, whereas, in the second case, the state operator is the same but it
is not a gemenge because the composite system is in a vector state. However,
by registration of any observable pertinent to ${\mathcal S}_2$, the observer
at $\vec{x}_2$ cannot distinguish different gemenge structures of the same
state operator from each other.

The next interesting question is, how the influence of a registered value at
$\vec{x}_1$ on that at $\vec{x}_2$ is to be understood. Even in classical
mechanics, one can arrange strong correlations. For example, if a body with
zero angular momentum with respect to its centre of mass decays into two
bodies flying away from each other, the angular momentum of the first is
exactly the opposite to that of the second. This strong anticorrelation cannot
be used to send signals, either. It is moreover clear, that a measurement of
the first angular momentum giving the value $\vec{L}_1$ is not a cause of the
second angular momentum having the value $-\vec{L}_1$. Rather, the decay is
the common cause of the two values being opposite. The process of creating the
two opposite values at distant points is also completely local: the decay is a
local phenomenon and the movement of each of the debris is governed by a local
equation of motion. And, the values $\vec{L}_1$ and $-\vec{L}_1$ are
objective, that is, they exist on the debris independently of any measurement
and can be, in this way, transported from the decay point to the measurements
in a local way.

In quantum mechanics, such an explanation of the correlations is not
possible. A value of an observable is created only during its registration. It
does not exist in any form before the registration, except in the special case
of state which is an eigenstate. However, in our example, $\Psi$ is not an
eigenstate either of ${\mathsf A}_1\otimes {\mathsf 1}$ or of ${\mathsf
1}\otimes {\mathsf A}_2$. The registrations performed simultaneously at
$\vec{x}_1$ and at $\vec{x}_2$, which can be very far from each other, are
connected by a relation that is utterly non-local. How can the apparatuses
together with parts of the quantum system at two distant points $\vec{x}_1$
and at $\vec{x}_2$ 'know' what values they are to create in order that the
correlations result? Note that the rejection of objectivity of observables
leads more directly to non-locality than assumption of some sort of realism.

The nature of simultaneity with which the correlations take place can be
described in more details, see Ref.\ \cite{KS}.

The non-local correlation between the registration outcomes can be ascertained
only if both values are known and can be compared. It may therefore be more
precise if we say that the non-local correlations can only be seen if the
non-local observable ${\mathsf A}_1\otimes {\mathsf A}_2$ is registered. This
non-locality of quantum correlations in entangled states does not lead to any
internal contradictions in quantum mechanics and is compatible with other
successful theories (such as special relativity) as well as with existing
experimental data. Nonetheless it is very surprising and it has been very
thoroughly studied. In this way, various conditions (e.g., Bell inequality)
have been found that had to be satisfied by values of observables if they were
objective and locality were satisfied. Experiments show that such conditions
are violated, and, moreover, that their violation can even be exploited in
quantum communication techniques \cite{brukner}.

Finally, let us repeat here that non-objectivity of observables is accepted by
our interpretation of quantum mechanics in the full extent and that this does
not lead to any contradiction with Basic Ontological Hypothesis of Quantum
Mechanics and with our realist interpretation.

\section{Composition of identical systems} If a prepared object has more than
one subsystem of the same type (identical subsystems), then these subsystems
are indistinguishable. This idea can be mathematically expressed as invariance
with respect to permutations.

\subsection{Identical subsystems} Let ${\mathbf S}_N$ be the permutation group
of $N$ objects, that is, each element $g$ of ${\mathbf S}_N$ is a bijective
map $g : \{1,\cdots,N\} \mapsto \{1,\cdots,N\}$, the inverse element to $g$ is
the inverse map $g^{-1}$ and the group product of $g_1$ and $g_2$ is defined
by $(g_1 g_2)(k) = g_1(g_2(k))$, $k \in \{1,\cdots,N\}$.

Given a Hilbert space $\mathbf H$, let us denote by ${\mathbf H}^N$ the tensor
product of $N$ copies of $\mathbf H$,
$$
{\mathbf H}^N = {\mathbf H} \otimes {\mathbf H} \otimes \cdots \otimes
{\mathbf H}\ .
$$
On ${\mathbf H}^N$, the permutation group ${\mathbf S}_N$ acts as follows. Let
$\psi_k \in {\mathbf H}$, $k=1,\cdots,N$, then
$$
\psi_1 \otimes \cdots \otimes \psi_N \in {\mathbf H}^N
$$
and
\begin{equation}\label{perm} {\mathsf g} (\psi_1 \otimes \cdots \otimes
\psi_N) = \psi_{g(1)} \otimes \cdots \otimes \psi_{g(N)}\ .
\end{equation} $\mathsf g$ preserves the inner product of ${\mathbf H}^N$ and
is, therefore, bounded and continuous. Hence, it can be extended by linearity
and continuity to the whole of ${\mathbf H}^N$. The resulting operator on
${\mathbf H}^N$ is denoted by the same symbol $\mathsf g$ and is a unitary
operator by construction. The action (\ref{perm}) thus defines a unitary
representation of the group ${\mathbf S}_N$ on ${\mathbf H}^N$.

All vectors of ${\mathbf H}^N$ that transform according to a fixed unitary
representation ${\mathcal R}$ of ${\mathbf S}_N$ form a closed linear subspace
of ${\mathbf H}^N$ that will be denoted by ${\mathbf H}^N_{\mathcal R}$. The
representations being unitary, the subspaces ${\mathbf H}^N_{\mathcal R}$ are
orthogonal to each other. Let us denote by ${\mathsf P}^{(N)}_{\mathcal R}$
the orthogonal projection operator,
$$
{\mathsf P}^{(N)}_{\mathcal R} : {\mathbf H}^N \mapsto {\mathbf H}^N_{\mathcal
R}\ .
$$

The location order of a given state in a tensor product can be considered as
an information about the identity of the corresponding system. Such
information has no physical meaning and a change of the ordering is just a
kind of gauge transformation\footnote{A different and independent part
(ignored here) of the theory of identical particles is that states of two
identical systems can also be swapped in a physical process of continuous
evolution, and can so entail a non-trivial phase factor at the total state
(anyons, see, e.g.\ Ref.\ \cite{wilczek}).}. Motivated by this idea, we look
for one-dimensional unitary representations of ${\mathbf S}_N$ because only
these transform vectors by a phase factor multiplication. ${\mathbf S}_N$ has
exactly two one-dimensional unitary representations: the symmetric (trivial)
one, $g \mapsto {\mathbf 1}$, and the alternating one $g \mapsto
\eta(g){\mathbf 1}$ for each $g \in {\mathbf S}_N$, where $\eta(g) = 1$ for
even permutations $g$ and $\eta(g) = -1$ for odd permutations $g$. If
${\mathcal R}$ is the symmetric (alternating) representation we use symbol
${\mathbf H}^N_s$ (${\mathbf H}^N_a$) for ${\mathbf H}^N_{\mathcal R}$. Let
${\mathsf P}_s^{(N)}$ (${\mathsf P}_a^{(N)}$) be the orthogonal projection on
${\mathbf H}^N_s$ (${\mathbf H}^N_a$). Note that the usual operation of
symmetrisation or antisymmetrisation on a vector $\Psi \in{\mathbf H}^N$, such
as
$$
\psi \otimes \phi \mapsto (1/2)(\psi \otimes \phi \pm \phi \otimes \psi)
$$
in ${\mathbf H}^2$, is nothing but ${\mathsf P}^{(N)}_s\Psi$ or ${\mathsf
P}^{(N)}_a\Psi$, respectively.

Now, we are ready to formulate the basic assumption concerning identical
subsystems.  From relativistic quantum field theory \cite{Weinberg}, we take
over the following result.
\begin{rl} Let ${\mathcal S}^N$ be a composite quantum object consisting of
$N$ subsystems ${\mathcal S}$, each of type $\tau$ with Hilbert space
${\mathbf H}_\tau$. Then, with important exceptions to be specified later, the
Hilbert space of ${\mathcal S}^N$ is ${\mathbf H}_{\tau s}^N$ for subsystems
with integer spin and ${\mathbf H}^N_{\tau a}$ for those with half-integer
spin.
\end{rl}
\begin{df} Systems with integer spin are called {\em bosons} and those with
half-integer spin are called {\em fermions}. The symmetry properties of states
leads to Bose-Einstein statistics for bosons and Fermi-Dirac one for fermions.
\end{df} We can, therefore, introduce a useful notation, a common symbol
${\mathbf H}^N_{{\mathcal R}(\tau)}$ for the subspaces of the symmetric and
anti-symmetric representations and ${\mathsf P}^{(N)}_\tau$ for the
corresponding projections because the representation is determined by the
system type $\tau$.
\begin{prop} The states of any system ${\mathcal S}^N$ consisting of
subsystems of type $\tau$ for which Rule 12 holds are elements of ${\mathbf
T}({\mathbf H}^N_{{\mathcal R}(\tau)})^+_1$ and the effects of ${\mathcal
S}^N$ are elements of ${\mathbf L}_r({\mathbf H}^N_{{\mathcal
R}(\tau)})^+_{\leq 1}$.
\end{prop} Proposition 12 follows directly from the definition of a system
with a given Hilbert space. To see its significance, let us make a comparison
with the case of subsystems of different types. First, one has to realise the
obvious fact that observing anything on a subsystem is tantamount to observing
something on the whole system. The application of this rule to system
${\mathcal S}'$ composite of two subsystems ${\mathcal S}_1$ and ${\mathcal
S}_2$ of different types has the following mathematical expression. Any effect
${\mathsf E}_1$ of ${\mathcal S}_1$ determines uniquely the effect ${\mathsf
E}_1 \otimes {\mathsf 1}_2$ of ${\mathcal S}'$. Thus, even if ${\mathcal S}_1$
is a part of ${\mathcal S}'$, we can measure ${\mathsf E}_1$ on it. Similarly,
if state ${\mathsf P}[\phi_1]$ of ${\mathcal S}_1$ and state ${\mathsf
P}[\phi_2]$ of ${\mathcal S}_2$ are prepared independently, then the state of
the composed system is ${\mathsf P}[\phi_1] \otimes {\mathsf P}[\phi_2]$ and
the partial traces give back the originally prepared states of the
subsystems. Thus, composing such systems does not disturb their individuality
and rules valid for each of them separately.

Looking at the composition of identical systems we observe a very different
picture. For instance, consider system ${\mathcal S}^2$ of two bosons of type
$\tau$ with common Hilbert space ${\mathbf H}_\tau$. The Hilbert space of
${\mathcal S}^2$ is ${\mathbf H}^2_{{\mathcal R}(\tau)}$. Let $\{\psi_k\}$ be
an orthonormal basis of ${\mathbf H}_\tau$. Then, the set of vectors
$$
\psi_k \otimes \psi_k\ ,\quad \forall k
$$
and
\begin{equation}\label{bosons} \frac{1}{\sqrt{2}}(\psi_k \otimes \psi_l +
\psi_l \otimes \psi_k)\ ,\quad \forall k \neq l
\end{equation} is an orthonormal basis for ${\mathbf H}^2_{{\mathcal
R}(\tau)}$. Hence, if Preparation I prepares a boson in state ${\mathsf
P}[\phi_k]$ and independent Preparation II prepares another boson in ${\mathsf
P}[\phi_l]$, $k \neq l$, then neither boson seems to be in its individually
prepared state. Clearly, the notion of individual states as partial traces
does not make sense for particles of the same type. Moreover, the notion of
preparation itself becomes problematical.

Next, let ${\mathsf A}$ be any operator on ${\mathbf H}_\tau$. Then, neither
${\mathsf A} \otimes {\mathsf 1}$ nor ${\mathsf 1} \otimes {\mathsf A}$ are
operators on ${\mathbf H}^2_{{\mathcal R}(\tau)}$ because they map any element
of the above basis out of ${\mathbf H}^2_{{\mathcal R}(\tau)}$. Thus, they are
not observables of any of the bosons. As ${\mathbf H}_\tau$ is the Hilbert
space of each boson constituent of ${\mathcal S}^2$, the effects of each boson
would have to be operators on ${\mathbf H}_\tau$ according to Rule 2.

It seems therefore that unrestricted validity of Proposition 12 would
contradict Rules of Secs.\ 1.1.2 and 1.2.2 on prepared states and form of
registrable observables. Clearly all this can be generalised to any number of
both boson and fermion constituents. Let us study this problem in more detail.

\subsection{Cluster separability} The problem described in the previous
section is aggravated if the two systems are prepared in different
laboratories. It seems that experiments with one object might be disturbed by
another object of the same type, even if it were prepared independently, far
away from the first. One can avoid similar problems by adding some assumption
of locality to the axioms of quantum theory.

The relativistic theory starts with the requirement that space-time symmetries
of an isolated system (i.e., that is alone in space) be realised by unitary
representations of Poincar\'{e} group on the Hilbert space of states, see
Refs.\ \cite{Weinberg} and \cite{haag}. Then, the {\em cluster decomposition
principle}, a locality assumption, states that if multi-particle scattering
experiments are studied in distant laboratories, then the $S$-matrix element
for the overall process factorizes into those concerning only the experiments
in the single laboratories. This ensures a factorisation of the corresponding
transition probabilities, so that an experiment in one laboratory cannot
influence the results obtained in another one. Cluster decomposition principle
implies non-trivial local properties of the theory underlying the $S$-matrix,
in particular it plays a crucial part in making local field theory inevitable
(cf. Ref.\ \cite{Weinberg}, Chap. 4).

In the phenomenological theory of relativistic or non-relativistic many-body
systems, Hilbert space of an isolated system must also carry a unitary
representation of Poincar\'{e} or Galilei group. Then, the so-called cluster
separability is a locality assumption, see, e.g., Refs.\ \cite{KP} or
\cite{coester} and references therein. It is a condition on interaction terms
in the generators of the space-time symmetry group saying: if the system is
separated into disjoint subsystems (=clusters) by a sufficiently large
spacelike separation, then each subsystem behaves as an isolated system with a
suitable representation of space-time symmetries on its Hilbert space, see
Ref.\ \cite{KP}, Sec.\ 6.1. Let's call this principle {\em cluster
separability I}.

Another special case of locality assumption has been described by Peres, Ref.\
\cite{peres}, p. 128. Let us reformulate it as follows
\par \vspace{.5cm}\noindent {\bf Cluster Separability II} No quantum
experiment with a system in a local laboratory is affected by the mere
presence of an identical system in remote parts of the universe.
\par \vspace{.5cm} \noindent It is well known (see, e.g., Ref.\ \cite{peres},
p. 136) that this principle leads to restrictions on possible statistics
(fermions, bosons). What is less well known is that it also motivates
non-trivial locality conditions on observables.

The locality condition is formulated in Ref.\ \cite{peres}, p. 128:
\begin{quote} \mbox{...} a state $\mathsf w$ is called remote if $\|{\mathsf
A}{\mathsf w}\|$ is vanishingly small, for any operator ${\mathsf A}$ which
corresponds to a quantum test in a nearby location. ... We can now show that
the entanglement of a local quantum system with another system in a remote
state (as defined above) has no observable effect.
\end{quote} This is a condition on ${\mathsf A}$ inasmuch as there has to be
at least one remote state for ${\mathsf A}$.

However, Peres does not warn that the standard operators of quantum mechanics,
which are in fact generators of space-time symmetries, do not satisfy his
condition on ${\mathsf A}$ \cite{hajicek2}. Similarly, basic observables of
relativistic-field or many-body theories are generators of Poincar\'{e} or
Galilei groups and so they do not satisfy the locality condition, either. It
follows that cluster separability II is logically independent from the cluster
decomposition or of cluster separability I. Of course, this does not mean that
the basic observables are to be rejected. They are very useful if the
assumption of isolated system is a good approximation. However, this
assumption is definitely a bad one for quantum theory of measurement.

The present section reformulates and extends Peres' ideas. Let us explain
everything, working in $Q$-representation of Hilbert space ${\mathbf H}_\tau$
and of operators on it. Then, one can also write tensor products as ordinary
products and indicate the order of factors by indices at system coordinates.

Consider two experiments. In Experiment I, vector state\footnote{We let out
the variable $\xi$ because it will not play any role in the subsequent
discussion.} $\psi(\vec{x}_1)$ of particle ${\mathcal S}_1$ of type $\tau$ is
prepared in our laboratory. In Experiment II, state $\psi(\vec{x}_1)$ of
particle ${\mathcal S}_1$ is prepared as in Experiment I and vector state
$\phi(\vec{x}_2)$ of system ${\mathcal S}_2$ of the same type is prepared
simultaneously in a remote laboratory. Then, according to Proposition 12, the
state of the composite system ${\mathcal S}$ must be
\begin{equation}\label{symst} \Psi(\vec{x}_1,\vec{x}_2) =
\nu\bigl(\psi(\vec{x}_1)\phi(\vec{x}_2) +\epsilon(\tau)
\phi(\vec{x}_1)\psi(\vec{x}_2)\bigr)\ ,
\end{equation} where $\nu = [2(1 + \epsilon |c|^2)]^{-1/2}$ is a normalisation
factor, $c = \langle \psi |\phi \rangle$ while $\epsilon(\tau)$ is 1 for
bosons and -1 for fermions.

Now, our lab claims that the state of ${\mathcal S}_1$ is $\psi(\vec{x}_1)$
because this state has been prepared in the lab according to all rules of
experimental art. However, somebody who knows what has happened in both labs
claims that the preparation has been a combination of two sub-preparations and
that the state is (\ref{symst}). Hence, the notion of preparation and state
becomes ambiguous, at least if Rule 12 holds without any restrictions. This is
true even if states $\psi$ and $\phi$ are localized (the wave functions have
supports) within the respective laboratories.

Localisation of the states $\psi$ and $\phi$ removes at least the following
difficulty. Suppose that a fermion is prepared in a distant laboratory in a
state $\phi$. Then a fermion of the same type cannot be prepared in the same
state $\psi = \phi$ in our laboratory: the composite state of the two fermions
had then to be zero (Pauli exclusion principle). However, if the wave function
of the fermion prepared in our laboratory falls off rapidly outside our
laboratory and that in the remote laboratory does the same outside the remote
one, then the two wave functions would be different, $\psi(x) \neq \phi(x)$,
and their antisymmetric combination would not be zero even if $\psi$ was just
a Euclidean group transform of $\phi$. The requirement of the fall-off is very
plausible indeed. It would be technically impossible to prepare a state with
the same wave function $\psi$ in a laboratory located in Prague, and
simultaneously in a laboratory in Bern, say, even for bosons.

Let us return to the ambiguity of preparations. It would not be really harmful
if it had no measurable consequences. But it apparently has! Suppose that the
registration on ${\mathcal S}_1$ corresponding to an observable $\mathsf a$
with kernel $a(\vec{x}_1;\vec{x}'_1)$ of ${\mathcal S}_1$ is performed in our
laboratory after Experiment II. It is then equally possible that the
registration is made on ${\mathcal S}_1$ or ${\mathcal S}_2$ and both can make
a contribution to the outcome. Hence, the correct observable $\mathsf A$ that
is associated with such a registration is described by kernel
\begin{equation}\label{symop} A(\vec{x}_1,\vec{x}_2;\vec{x}'_1,\vec{x}'_2) =
a(\vec{x}_1;\vec{x}'_1)\delta(\vec{x}_2-\vec{x}'_2) +
a(\vec{x}_2;\vec{x}'_2)\delta(\vec{x}_1-\vec{x}'_1)\ ,
\end{equation} which is an operator on ${\mathbf H}^2_{{\mathcal R}(\tau)}$ so
that it satisfies Proposition 12.

The average of $\mathsf a$ in the state prepared in Experiment I is $\langle
\psi | {\mathsf a}\psi \rangle$. On the other hand, Experiment II leads to the
average:
$$
\langle \Psi | {\mathsf A}\Psi \rangle = \frac{\langle \psi | {\mathsf a}\psi
\rangle + \langle \phi | {\mathsf a}\phi \rangle + \epsilon(\tau) c \langle
\phi | {\mathsf a}\psi \rangle + \epsilon(\tau) c^*\langle \psi | {\mathsf
a}\phi \rangle}{1 + \epsilon(\tau) |c|^2}\ .
$$
The two averages are different and the difference need not be small. For
example, let the supports of the wave functions $\psi$ and $\phi$ be disjoint,
each inside an open set around the corresponding laboratory. Then, $c =
0$. Consider the position operator of ${\mathcal S}_1$. The kernel is
$a(\vec{x}_1;\vec{x}'_1) = \vec{x}_1\delta(\vec{x}_1-\vec{x}'_1)$, and the
difference between the two averages becomes
$$
\int d^3x_1\vec{x}_1 \phi^*(\vec{x}_1)\phi(\vec{x}_1)\ .
$$
This can be made arbitrary large by shifting the remote laboratory
sufficiently far away.

Thus, cluster separability is violated. However, the position does not satisfy
Peres' condition: such an 'observable' controls position of the system in the
whole infinite space. By that, it is utterly different from observables that
can be registered in a human laboratory. It can analogously be shown that all
standard operators of quantum mechanics are similar to position in this
property.

\subsection{$D$-local states and observables} This section is a more detailed
and accurate account of the conditions on states and observables than that
given by Peres. They will be assumed as at least approximately satisfied.

We shall consider general $N$-particle systems and only those operators
$\mathsf A$ that fulfil the following condition. The kernel
$A(\vec{x}_1,\cdots,\vec{x}_N;\vec{x}'_1,\cdots,\vec{x}'_N)$ of $\mathsf A$ is
a distribution on Schwartz space (its elements are rapidly decreasing
$C^\infty$ functions, see, e.g., Ref.\ \cite{RS}) ${\mathbf S}({\mathbb
R}^{3N})$ in variables $\vec{x}'_1,\cdots,\vec{x}'_N$ for any fixed value of
$\vec{x}_1,\cdots,\vec{x}_N$ such that
$$
g(\vec{x}_1,\cdots,\vec{x}_N) = \int d^{3N}x'
A(\vec{x}_1,\cdots,\vec{x}_N;\vec{x}'_1,\cdots,\vec{x}'_N)
f(\vec{x}'_1,\cdots,\vec{x}'_N) \in {\mathbf S}({\mathbb R}^{3N})
$$
for any $f(\vec{x}_1,\cdots,\vec{x}_N)$. This is usually satisfied, see Sec.\
1.3.1.
\begin{df} Let $A(\vec{x}_1,\cdots,\vec{x}_N;\vec{x}'_1,\cdots,\vec{x}'_N)$
and $B(\vec{x}_1,\cdots,\vec{x}_N;\vec{x}'_1,\cdots,\vec{x}'_N)$ be two
kernels. Then, for any $K = 0,1,\cdots,N$,
\begin{equation}\label{convol} \int d^3x'_{k_1}\cdots x'_{k_K}
A(\vec{x}_1,\cdots,\vec{x}_N;\vec{x}'_1,\cdots,\vec{x}'_N)
B(\vec{x}'_1,\cdots,\vec{x}'_N;\vec{x}''_1,\cdots,\vec{x}''_N)
\end{equation} is a distribution acting on test functions of ${\mathbf
S}({\mathbb R}^{6N-3K})$ that maps test function
$f(\vec{x}'_1,\cdots,\vec{x}'_N,\vec{x}''_1,\cdots,\vec{x}''_N)^{\Delta
\prime}$ on test function
$g(\vec{x}_1,\cdots,\vec{x}_N,\vec{x}'_1,\cdots,\vec{x}'_N)^{\Delta \prime}$,
where symbol $(\cdots)^{\Delta \prime}$ means that $K$ variables
$\vec{x}'_{k_1},\cdots, \vec{x}'_{k_K}$ are left out from $N$ primed variables
$\vec{x}'_1,\cdots,\vec{x}'_N$. Distribution (\ref{convol}) is called {\em
convolution} ${\mathsf A}*{\mathsf B}$ over variables $\vec{x}_{k_1},\cdots,
\vec{x}_{k_K}$.
\end{df}

Let our laboratory takes an open set $D \subset {\mathbb R}^{3}$ and similarly
$D'$ for the remote one. The change of ${\mathcal S}_1$ state due to some
actions in a remote laboratory would not be measurable in our laboratory if
the states and observables are $D$-local in the following sense:
\begin{df} Let $D \subset {\mathbb R}^{3}$ be open, let $\mathsf A$ be an
operator of ${\mathcal S}$ and let the following conditions hold:
\begin{enumerate}
\item $A(\vec{x}_1,\cdots,\vec{x}_N;\vec{x}'_1,\cdots,\vec{x}'_N)$ is the zero
distribution for any $\vec{x}_1,\cdots,\vec{x}_N \in {\mathbb R}^{3N}
\setminus D^N$.
\item
$$
\int d^{3N}x' A(\vec{x}_1,\cdots,\vec{x}_N;\vec{x}'_1,\cdots,\vec{x}'_N)
f(\vec{x}'_1,\cdots,\vec{x}'_N) = 0
$$
for any test function $f$ such that
$$
\text{supp}f(\vec{x}'_1,\cdots,\vec{x}'_N) \subset {\mathbb R}^{3N} \setminus
D^N\ .
$$
\end{enumerate} Then $\mathsf A$ is called $D$-local.
\end{df} Let $\mathsf A$ be $D$-local for some open set $D$, let $D'$ be open
and $D \subset D'$. Then, $\mathsf A$ is $D'$-local. Let $\mathsf A$ be
$D$-local and also $D'$-local for two open sets $D$ and $D'$. Then, $\mathsf
A$ is ($D \cap D'$)-local. Thus, all open sets $D$ such that $\mathsf A$ is
$D$-local form a filter in the Boolean lattice of open subsets of ${\mathbb
R}^3$.
\par \noindent {\bf Example 1} $N = 1$, $A(\vec{x};\vec{x}') =
x^1\delta({\vec{x} - \vec{x}'})$. $\mathsf A$ is ${\mathbb R}^3$-local and the
filter is ${\mathbb R}^3$.
\par \noindent {\bf Example 2} $N = 1$, $\psi(\vec{x})$ is a wave function
with support $D$. Then $|\psi(\vec{x})\rangle \langle \psi(\vec{x})|$ is
$D'$-local, where $D'$ is the interior of $D$, and the filter is the family of
all open sets containing $D'$.

\begin{df} A state with a $D$-local state operator ${\mathsf T}$ is called
$D$-local. An observable ${\mathsf E}(X)$, all effects ${\mathsf E}(X)$, $X
\in {\mathbf F}$, of which are $D$-local, is called $D$-local.
\end{df}

The key property is the following:
\begin{prop} Let system ${\mathcal S}_1$ of type $\tau_1$ composite of $N_1$
particles (not necessarily identical) be prepared in state ${\mathsf T}_1$
that is $D_1$-local. Let system ${\mathcal S}_2$ of type $\tau_2$ composite of
$N_2$ particles be prepared in state ${\mathsf T}_2$ that is $D_2$-local,
where $D_1 \cap D_2 = \emptyset$. Then, the state of ${\mathcal S}_1 +
{\mathcal S}_2$ according to Rule 12 is
$$
{\mathsf P}({\mathsf T}_1 \otimes {\mathsf T}_2){\mathsf P}\ ,
$$
where ${\mathsf P}$ is the symmetrisation over all identical bosonic and the
anti-symmetrisation over all identical fermionic particles contained in
${\mathcal S}_1 + {\mathcal S}_2$ followed by multiplication by a
normalisation factor. Furthermore:
\begin{enumerate}
\item Let $\tau_1 = \tau_2$ ($N_1 = N_2$) and let ${\mathsf E}(X)$ be a
$D$-local observable for ${\mathcal S}_1$ as well as for ${\mathcal
S}_2$. Then,
\begin{equation}\label{noninf1} tr[({\mathsf E}(X) \otimes {\mathsf 1} +
{\mathsf 1} \otimes {\mathsf E}(X)){\mathsf P}({\mathsf T}_1 \otimes {\mathsf
T}_2){\mathsf P}] = tr[{\mathsf E}(X) {\mathsf T}_1]\ .
\end{equation}
\item Let $\tau_1 \neq \tau_2$ and let ${\mathsf E}(X)$ be a $D$-local
observable for ${\mathcal S}_1$. Then
\begin{equation}\label{noninf2} tr[({\mathsf E}(X) \otimes {\mathsf
1}){\mathsf P}({\mathsf T}_1 \otimes {\mathsf T}_2){\mathsf P}] = tr[{\mathsf
E}(X) {\mathsf T}_1]\ .
\end{equation}
\end{enumerate}
\end{prop} Thus, the existence of identical systems somewhere else does not
influence the registrations of $D$-local observables in our laboratory.
\par \vspace{.5cm} \noindent {\bf Proof} First, we formulate an immediate
consequence of the definition of $D$-locality and of convolution. Let
${\mathsf A}$ be $D$-local and ${\mathsf B}$ $D'$-local, where $D \cap D' =
\emptyset$. Then convolution ${\mathsf A} * {\mathsf B}$ over any variables
$\vec{x}_{k_1},\cdots, \vec{x}_{k_K}$ for $K \neq 0$ is the zero distribution
on ${\mathbf S}({\mathbb R}^{6N-3K})$.

Let us first prove Part 1 of the proposition. Then, the kernel of ${\mathsf
E}(X) \otimes {\mathsf 1} + {\mathsf 1} \otimes {\mathsf E}(X)$ is
\begin{multline*}
E(X)(\vec{x}_1,\cdots,\vec{x}_N;\vec{x}'_1,\cdots,\vec{x}'_N) \delta(\vec{y}_1
- \vec{y}'_1) \cdots \delta(\vec{y}_N - \vec{y}'_N) \\ + \delta(\vec{x}_1 -
\vec{x}'_1) \cdots \delta(\vec{x}_N - \vec{x}'_N)
E(X)(\vec{y}_1,\cdots,\vec{y}_N;\vec{y}'_1,\cdots,\vec{y}'_N)
\end{multline*} and that of ${\mathsf T}_1 \otimes {\mathsf T}_2$ is
\begin{equation}\label{TT}
T_1(\vec{x}'_1,\cdots,\vec{x}'_N;\vec{x}''_1,\cdots,\vec{x}''_N)
T_2(\vec{y}'_1,\cdots,\vec{y}'_N;\vec{y}''_1,\cdots,\vec{y}''_N)\ .
\end{equation} Operator ${\mathsf P}$ from the left acts in three
steps. First, it makes permutations of all variables inside each group of
variables $\vec{x}'_1,\cdots,\vec{x}'_N,\vec{y}'_1,\cdots,\vec{y}'_N$ that
correspond to one type of particles in expression (\ref{TT}). Second, it sums
with the correct sign over all such permutations that result in different
terms. Finally, it multiplies the resulting sum by a normalisation
factor. Operator ${\mathsf P}$ from the right acts similarly on variables
$\vec{x}''_1,\cdots,\vec{x}''_N,\vec{y}''_1,\cdots,\vec{y}''_N$.

Thus, the trace on the left-hand side of Eq.\ (\ref{noninf1}) is a sum of
traces of all terms obtained in this way. Let one such term be
$$
\int d^3x'_1\dots d^3x'_N d^3y'_1\dots d^3y'_N
E(X)(x;x')\delta(y;y')T_1(\cdots;\cdots)T_2(\cdots;\cdots)\ ,
$$
where we have indicated variables $\vec{x}'_1,\cdots,\vec{x}'_N$ by $x'$,
etc., and '$\cdots$' in arguments of ${\mathsf T}_1$ or ${\mathsf T}_2$ stands
for $N$ variables from
$\vec{x}'_1,\cdots,\vec{x}'_N,\vec{y}'_1,\cdots,\vec{y}'_N$ or from
$\vec{x}''_1,\cdots,\vec{x}''_N,\vec{y}''_1,\cdots,\vec{y}''_N$ obtained by
the corresponding permutations.

If the permutations only change the order of $x$'s or of $y$'s, then all such
terms are equal and the value of the trace is
\begin{multline}\label{oneterm} tr\left[\int d^3x'_1\dots d^3x'_N d^3y'_1\dots
d^3y'_N E(X)(x;x')\delta(y,y')T_1(x';x'')T_2(y';y'')\right] \\ = tr[{\mathsf
E}(X){\mathsf T}_1]
\end{multline} If, however, the permutations mix $K$ $x$'s with $K$ $y$'s,
then the $6N$-variables convolution contains a $K$-variables convolution
${\mathsf E}(X)* {\mathsf T}_2$, which is zero. Hence, the contribution of all
such terms is just $tr[{\mathsf E}(X){\mathsf T}_1]$.

Next, consider the terms of the form
$$
\int d^3x'_1\dots d^3x'_N d^3y'_1\dots d^3y'_N
\delta(x,x')E(X)(y:y')T_1(\cdots;\cdots)T_2(\cdots;\cdots)\ ,
$$
Analogously, only those permutations can contribute to the convolution that
exchange all $x$'s with all $y$'s so that $E(X)(y:y')$ is convoluted with
$T_1(y';y'')$. This is possible because both systems are of the same type and
so there is exactly one group of $y$'s to each groups of $x$'s and these two
groups contain the same number of variables. Again, all such permutations give
the same result and the total trace is, therefore, $2tr[{\mathsf E}(X){\mathsf
T}_1]$.

The normalisation factor is
$$
\big(tr[{\mathsf P}({\mathsf T}_1\otimes {\mathsf T}_2){\mathsf P}]\big)^{-1}
$$
and it is clear that only those permutations contribute to the trace which do
not lead to convolutions between ${\mathsf T}_1$ and ${\mathsf T}_2$. There
are only two kinds of them: those mixing $x$'s among themselves and $y$'s
among themselves, or those exchanging all $x$'s with all $y$'s. In each of the
two cases, the trace is unity, hence the normalisation factor is 1/2, which
implies Eq.\ (\ref{noninf1}). The proof of Part 2 is similar but simpler,
Q.E.D.

For example, if $|\psi \rangle \langle \psi|$ is $D$-local, $|\phi \rangle
\langle \phi|$ is $D'$-local and the $D$-local kernel
$a_D(\vec{x}_1,\vec{x}'_1)$ is used instead of $a(\vec{x}_1;\vec{x}'_1)$ in
formula (\ref{symop}) defining operator ${\mathsf A}_D$ instead of ${\mathsf
A}$ we obtain
\begin{multline*} \int_D
d^3x_1\int_Dd^3x'_1\int_Dd^3x_2\int_Dd^3x'_2\Psi^*(\vec{x}_1,\vec{x}_2)
A_D(\vec{x}_1,\vec{x}'_1;\vec{x}_2,\vec{x}'_2)\Psi(\vec{x}'_1,\vec{x}'_2) \\ =
\int_{-\infty}^\infty d^3x_1\int_{-\infty}^\infty
d^3x'_1\psi^*(\vec{x}_1)a_D(\vec{x}_1;\vec{x}'_1)\psi(\vec{x}'_1)
\end{multline*} as if no ${\mathcal S}_2$ existed. Thus, in this case, cluster
separability holds.

However, the ambiguity of the state of ${\mathcal S}_1$ in Experiment II still
remains: the state prepared in our laboratory is $|\psi \rangle \langle \psi|$
and, at the time that ${\mathcal S}_2$ is prepared, ${\mathcal S}_1$ loses its
individual states if Rule 12 is valid without exceptions. This is clearly at
variance with cluster separability; we shall remove this difficulty in Sec.\
2.2.4.

Next, we shall prove that each observable ${\mathsf E}$ of system ${\mathcal
S}$ of type $\tau$ is associated, for a reasonable open set $D$, with a unique
$D$-local observable $\Lambda_D({\mathsf E})$, the so-called $D$-localisation
of ${\mathsf E}$, such that the condition
$$
tr[{\mathsf E}(X) {\mathsf T}] = tr[\Lambda_D({\mathsf E}(X)) {\mathsf T}]
$$
is satisfied for all $X \in {\mathbf F}$ and for all $D$-local states
${\mathsf T}$. Thus, any standard observable ${\mathsf E}$ such as position,
momentum, etc. can be registered in $D$ by registering $\Lambda_D({\mathsf
E})$ and the probability distribution would be the same as if we had
registered ${\mathsf E}$.

Let us denote by ${\mathbf H}_\tau(D)$ the Hilbert space obtained by
completion of the linear space of $C^\infty$-functions with support in $D^N$
with respect to the inner product of ${\mathbf H}_\tau$. ${\mathbf H}_\tau(D)$
is a closed linear subspace of ${\mathbf H}_\tau$. Let ${\mathsf P}_D$ be the
orthogonal projection from ${\mathbf H}_\tau$ onto ${\mathbf H}_\tau(D)$.
\begin{df} Let
$$
\Lambda_D : {\mathbf L}_r({\mathbf H}_\tau) \mapsto {\mathbf L}_r({\mathbf
H}_\tau(D))
$$
be defined by
$$
\Lambda_D({\mathsf A}) = {\mathsf P}_D{\mathsf A}{\mathsf P}_D\ .
$$
Mapping $\Lambda_D$ is called $D$-{\em localisation}.
\end{df} Clearly, $D$-localisation of any operator in ${\mathbf L}_r({\mathbf
H}_\tau)$ is Hermitean, bounded and positive, hence
$$
\Lambda_D : {\mathbf L}_r({\mathbf H}_\tau)^+_{\leq 1}\mapsto {\mathbf
L}_r({\mathbf H}_\tau)^+_{\leq 1}\ .
$$
Of course, the $D$-localisation is not a unitary map. For example, it does not
preserve operator norm, but it holds
$$
\|\Lambda_D({\mathsf A})\| \leq \|{\mathsf A}\| .
$$
The operators and their $D$-localisations are considered as acting on
${\mathbf H}_\tau$. $D$-local operators leave ${\mathbf H}_\tau(D)$ invariant
and define, therefore, also operators on Hilbert space ${\mathbf H}_\tau(D)$.

Second, we can use these facts for a construction of $D$-local POV measure on
${\mathbf H}_\tau(D)$ from any observable ${\mathsf E}$ on ${\mathbf H}_\tau$
by $D$-localising the effects ${\mathsf E}(X)$. The normalisation condition
becomes
$$
\Lambda_D({\mathsf E}({\mathbb R}^n)) = {\mathsf P}_D{\mathsf 1}{\mathsf P}_D
= {\mathsf 1}_D\ ,
$$
where $n$ is the dimension of ${\mathsf E}$ (see Sec.\ 1.2.1). Hence,
$D$-localisation of an observable on ${\mathbf H}_\tau$ need not be an
observable on ${\mathbf H}_\tau$ but it is always one on ${\mathbf
H}_\tau(D)$. Also, $D$-localisation of a projection will not be a projection
in general and so a $D$-localisation of a PV measure need not be a PV
measure. Let us call this construction $D$-localisation of POV measures. All
$D$-local POV measures commute with spectral projections of position PV
measures, ${\mathsf E}_k^{\vec{\mathsf Q}}(X)$, if $X \cap D =\emptyset$,
where ${\mathsf E}_k^{\vec{\mathsf Q}}(X)$ are sharp position observables that
are well defined for ${\mathcal S}$. For example, if ${\mathcal S}$ is
composite of $N$ particles of different type, then $k = 1,\cdots,N$ and
${\mathsf E}_k^{\vec{\mathsf Q}}(X)$ is a spectral projection of the position
of the $k$-th particle. Thus, the restriction to $D$-local observables may be
formally understood in terms of superselection rules (see Sec.\ 3.3).

If the map $\Lambda_D$ is involved in construction of $D$-local states, it
must be followed by a suitable normalisation. A $D$-local state on ${\mathbf
H}_\tau$ is a state on ${\mathbf H}_\tau(D)$ and vice versa.

\subsection{Separation status} With the help of $D$-local states and
observables, we could remove some contradictions due to unrestricted validity
of Rule 12. The ambiguity of states, however, remained. This is a great and
ubiquitous problem. Indeed, any microsystem $\mathcal S$ is always a subsystem
of a huge family of identical subsystems that exist somewhere in the world.

If we could write down the state of this huge family in accordance with
unrestricted Rule 12, then the results of any registrations in our laboratory
would be well defined by it. However, such a state is not and cannot be
known. Luckily, the results of registration of $D$-local observables do not
depend on the unknown parts of the state because of Proposition 13. This is
the reason why the validity of Rule 12 can and must be restricted.

Clearly, one could ignore the entanglement of a single microsystem ${\mathcal
S}$ with all microsystems of the same type, if ${\mathcal S}$ had a
non-trivial {\em separation status} in the following sense:
\begin{df} Let $D \subset {\mathbb R}^3$ be an open set and system ${\mathcal
S}$ be prepared in state ${\mathsf T}$. Let the following conditions be
satisfied:
\begin{itemize}
\item The probability to register value of observable ${\mathsf E}(X)$ in set
$X$ be $tr[{\mathsf T}'{\mathsf E}(X)]$ for any $D$-local observable ${\mathsf
E}(X)$ and for any state ${\mathsf T}'$ of ${\mathcal S}$.
\item There is at least one $D$-local observable ${\mathsf E}'(X)$ such that
$tr[{\mathsf T}{\mathsf E}'(X)] \neq 0$ for some $X$.
\end{itemize} Then, open set $D$ is called {\em separation status} of
${\mathcal S}$.
\end{df} The first condition means that the registration of ${\mathsf E}(X)$
on ${\mathsf T}'$ is not disturbed by any identical system in a state
different from ${\mathsf T}'$. This includes the following condition that must
be fulfilled by the apparatus ${\mathcal A}$ that registers observable
${\mathsf E}(X)$: The states of those subsystems of ${\mathcal A}$ that are
identical with ${\mathcal S}$ do not disturb the registration of ${\mathsf
E}(X)$ by ${\mathcal A}$ (but may disturb other registrations). A model of
${\mathcal A}$ satisfying the condition is given in Sec.\ 6.3. We can then
view ${\mathcal S}$ as a physical object with a {\em modified} set of
observables. The modification depends on open set $D$ that is determined by a
preparation of $\mathcal S$. This motivates the following
\begin{rl} Effects of $\mathcal S$ that can be registered must lie inside
subsets ${\mathbf L}_r({\mathbf H}_\tau(D))^+_{\leq 1} \subset {\mathbf
L}_r({\mathbf H}_\tau)^+_{\leq 1}$, where $D$ is the separation status of
$\mathcal S$.
\end{rl} Thus, strictly speaking, no quantum system is proper (see Sec.\
1.2.2).

For example, a microsystem that is alone in the Universe has separation status
$D = {\mathbb R}^3$. This is a form of the assumption that a system is
isolated. Observables of such a system are the standard ones. The same
microsystem in a open set $D$ which is surrounded by matter containing a lot
of microsystems of the same type such that supports of their states do not
intersect $D$ has separation status $D$ and its observables are the $D$-local
ones. The trivial case of separation status for a microsystem is $D =
\emptyset$. It has then no observables of its own.

The composition of systems of the same type is determined by Rule 12 and its
exceptions that are specified as follows.
\begin{rl} Let two systems, $\mathcal S$ and ${\mathcal S}'$, have separation
statuses $D \neq \emptyset$ and $D' \neq \emptyset$, respectively so that $D
\cap D' = \emptyset$. Then, the compositions of $\mathcal S$ and ${\mathcal
S}'$ follow rules valid for composition of systems of different type even if
$\mathcal S$ and ${\mathcal S}'$ have in common some subsystems of the same
type.
\end{rl} For instance, let the wave function of particle ${\mathcal S}$ with
separation status $D$ be $\psi(\vec{x})$ and let ${\mathcal S}'$ be a family
of $N$ particles identical with ${\mathcal S}$. Let ${\mathcal S}'$ have
separation status $D'$ such that $D'\cap D = \emptyset$. Its state
$\Psi(\vec{x}_1,\cdots,\vec{x}_N)$ is symmetric or anti-symmetric in its $N$
arguments according to the type. Then the wave function of composite system
${\mathcal S} + {\mathcal S}'$ of $N+1$ particles of the same type must be
written as
\begin{equation}\label{firstst} \psi(\vec{x})\Psi(\vec{x}_1,\cdots,\vec{x}_N)\
.
\end{equation} Observe that wave function (\ref{firstst}) is not
(anti-)symmetric in all $N+1$ arguments! This is at variance with unrestricted
validity of Rule 12. According to Rule 12, the wave function had to be
\begin{equation}\label{sectst} {\mathsf P}_{\mathbf S}{\mathsf
P}^{(N+1)}_\tau\big(\psi(\vec{x})\Psi(\vec{x}_1,\cdots,\vec{x}_N)\big)\ ,
\end{equation} where ${\mathsf P}_{\mathbf S}$ is the projection to the unit
sphere of ${\mathbf H}^{N+1}_{{\mathcal R}(\tau)}$.  Now, it also ought to be
clear why we do not employ Fock-space method to deal with identical systems:
it automatically (anti-)symmetrises over all systems of the same type.

Rule 14 is in complete agreement with the common practice in quantum
mechanics. For example, in the theory of the experiment described in Sec.\
0.1.2, a state is prepared as a state of an individual electron and its
entanglement with all other electrons, which exist, in fact, everywhere in
huge amounts, is serenely ignored. Due to Proposition 13, the results of such
method would not lead to any contradictions.

Let us give a mathematical example of separation status change. As for states,
let it consist of the transformation of state (\ref{firstst}) into state
(\ref{sectst}). This transformation has the form
$$
{\mathsf P}_{\mathbf S} \circ {\mathsf P}^{(N+1)}_{s,a} : {\mathbf H} \otimes
{\mathbf H}^N_{s,a} \mapsto {\mathbf H}^{N+1}_{s,a}\ ,
$$
which is a non-invertible and non-linear map between two different Hilbert
spaces. As for observables, in state (\ref{firstst}), the observables that are
registrable on $\mathcal S$ form the algebra ${\mathbf A}_D$ of $D$-local
operators. Indeed, for any ${\mathsf a}_D \in {\mathbf A}_D$, there is
observable ${\mathsf a}_D \otimes {\mathsf 1}$ on ${\mathcal S} + {\mathcal
S}'$. In state (\ref{sectst}), the algebra would be ${\mathbf A}_\emptyset =
\emptyset$: there are no observables whatsoever that would be registrable
individually on $\mathcal S$. Thus, the set of observables changes from
${\mathbf A}_D$ to ${\mathbf A}_\emptyset$.

In classical mechanics, the possible states of system ${\mathcal S}$ are all
positive normalised functions (distribution functions) on the phase space
${\mathbf P}$ and possible observables are all real function on ${\mathbf P}$
(at least, all such observables have definite averages on ${\mathcal S}$
independently of external circumstances). ${\mathbf P}$ is uniquely associated
with the system alone and forms the basis of its kinematic description. Hence,
transitions between different sets of observables similar to those described
above would be impossible in classical mechanics. They are only enabled in
quantum mechanics by the non-objective character of observables: not only
their values cannot be ascribed to microsystem ${\mathcal S}$ alone but some
of them are not even registrable in principle due to external conditions in
which ${\mathcal S}$ is.

We assume that the quantum kinematics of a microsystem is defined
mathematically by possible states represented by all positive normalised
(trace one) operators, and possible observables represented by some POV
measures, on the Hilbert space associated with the system. Then the
transitions of states and observables that go with changes of separation
status cannot be viewed as a part of a dynamical trajectory due to some new
version of the dynamics of ${\mathcal S}$, but as a change of its kinematic
description. Thus, although the change of separation status is similar to the
collapse of the wave function (the non-local character included\footnote{The
time instant, when such a non-local change takes place, can in some cases be
established by observations \cite{KS}.}), it is more radical and its physical
roots are better understood.

What has been said up to now shows that standard quantum mechanics is
incomplete in the following sense:
\begin{enumerate}
\item It accepts and knows only \underline{two} separation statuses:
\begin{enumerate}
\item that of isolated systems, $D = {\mathbb R}^3$, with the standard
operators as observables, and
\item that of a member of a system of identical particles, $D = \emptyset$,
with no observables of its own.
\end{enumerate}
\item It provides \underline{no} rules for changes of separation status.
\end{enumerate} Let us call these theories {\em fixed status quantum
mechanics} (FSQM). Our main idea is: Quantum mechanics must be supplemented by
a theory of general separation status and by new rules that govern processes
in which separation status changes. The new rules must not contradict the rest
of quantum mechanics and ought to agree with and explain observational facts.

\chapter{Further general ideas} This chapter contains further important
general ideas and closes the account of the general language of quantum
mechanics. It explains the relation between the ontological and epistemic
aspects of probability notion. It describes some crucial properties of quantum
observables such as joint measurability and contextuality. Finally, it
introduces the notion of superselection rules in order to make clear that our
approaches to classical properties and quantum measurement have noting to do
with the superselection-rules one.

\section{Probability and information} Quantum mechanics is a statistical
theory. Thus, a correct understanding of probability and information is an
important part in the conceptual framework of the theory. The discussion
whether probability describes objective properties that can be observed in
nature or subjective state of the knowledge of some human has raged since the
invention of probability calculus by Jacob Bernoulli and Pierre-Simon Laplace
\cite{Jaynes}. The origin of the problem might be that the dispute cannot be
decided: probability has both aspects, ontological and epistemic.

Probability is a function of a proposition $A$ and its value, $p(A)$, is a
measure of the degree of certainty that $A$ is true. As a function on a
Boolean lattice of propositions, it satisfies Cox' axioms \cite{Jaynes}. Then
it becomes a real additive measure on the lattice. Whether a proposition is
true or false must be decided by observation, at least in principle. Hence,
the probability always concerns objective events, at least indirectly.

As an example, consider the state ${\mathsf T}$ of a quantum object $\mathcal
S$ that has been prepared and a registration of an observable ${\mathsf E}$ on
the state. The probability $p_{\mathsf T}^{\mathsf E}(X)$ concerns the
proposition that the value of ${\mathsf E}$ individually registered on
${\mathsf T}$ lies in the set $X$. The value of the probability can be
verified by studying the events that happen during such registrations. Thus,
the probability concerns both the lack of knowledge of what objectively
happens and properties of real systems. It is in this sense that probability
has both epistemic and ontological aspects.

In quantum mechanics, $p_{\mathsf T}^{\mathsf E}(X)$ is not the probability
that ${\mathsf E}$ possesses a value within $X$ in the state ${\mathsf T}$ but
the probability that a registration outcome will give such a value. The only
exception is that ${\mathsf T}$ is a gemenge of ${\mathsf E}$-eigenstates. A
related question is whether the outcomes are predictable and we just have not
enough information to make the prediction, or objectively
unpredictable. Quantum mechanics does not deliver the prediction and it would
even be incompatible with any 'deeper' theory that did (see Sec.\ 3.2.2).

Let us describe the general framework that is necessary for any application of
probability theory. First, there is a system, denote it by $\mathcal S$. In
the example above, there are certain conditions that are imposed on the
quantum object $\mathcal S$. State ${\mathsf T}$ is prepared: this is a
definite objective condition on $\mathcal S$. Observable ${\mathsf E}$ is
registered: another objective condition on $\mathcal S$. In general, there
must always be an analogous set of conditions, let us denote it by $\mathcal
C$. To each system $\mathcal S$ subject to condition $\mathcal C$
possibilities in some range are open. These possibilities are described by a
set of propositions that form a Boolean lattice $\mathbf F$. A probability
distribution $p : {\mathbf F} \mapsto [0,1]$ is a real additive measure on
$\mathbf F$.

If ${\mathsf E}$ is a discrete observable, its value set ${\mathbf \Omega}$ is
at most countable, ${\mathbf \Omega} = \{\omega_1, \omega_2 \cdots \}$. Then
the single-element sets $\{\omega_k\}, k = 1, 2, \cdots $, are atoms of
$\mathbf F$ that generate $\mathbf F$ and the probability distribution
$p_{\mathsf T}^{\mathsf E}(X)$ can be calculated from $p_{\mathsf T}^{\mathsf
E}(\{\omega_k\})$ by means of Cox' axioms. The atoms are called outcomes. If
${\mathsf E}$ is a sharp continuous observable, then there are no atoms but
continuous observables can be considered as idealisations of more realistic
POV measures with well-defined atoms. (The reading scale of each registration
apparatus can be divided into separate finite intervals, the width of which
corresponds to the accuracy that can really be achieved, etc.\ see, e.g.,
Ref.\ \cite{BLM}, p. 35.)

If condition $\mathcal C$ is reproducible or it obtains spontaneously
sufficiently often, anybody can test the value of the probabilities. The
probability distribution $p(X)$ is therefore an objective property of
condition $\mathcal C$ on $\mathcal S$. It is so even in cases when the
outcomes can be predicted uniquely. This would depend on whether condition
$\mathcal C$ can be sharpened to other conditions $\mathcal C_k$ so that
\begin{enumerate}
\item whenever $\mathcal C_k$ is satisfied, so is $\mathcal C$,
\item each $\mathcal C_k$ is still recognisable and reproducible and
\item each outcome allowed by $\mathcal C$ is uniquely determined by one of
$\mathcal C_k$'s.
\end{enumerate} Even in such a case, condition $\mathcal C$ itself leaves the
system a definite amount of freedom that can be described in all detail by the
probability distribution $p(X)$ : it is an objective, verifiable property of
$\mathcal C$ alone. And, if we know only that condition $\mathcal C$ obtains
and that a probability distribution is its objective property, then this
probability distribution describes the state of our knowledge, independently
of whether the conditions $\mathcal C_k$ do exist and we just do not know
which of them obtains in each case or not. There is, therefore, no
contradiction between the objective and subjective aspects of probability.

Entropy is a certain functional of $p(X)$ that inherits both objective and
subjective aspects from probability. Discussions similar to those about
probability spoil the atmosphere about entropy. The existence of a subjective
aspect of entropy---the lack of information---seduces people to ask confused
questions such as: Can a change of our knowledge about a system change the
system energy?

The general definition of entropy as a measure of missing information has been
given by Shannon \cite{shannon} and its various applications to communication
theory are, e.g., described in the book by Pierce \cite{pierce}. Our version
is:
\begin{df} Let $p : {\mathbf F} \mapsto [0,1]$ be a probability distribution
and let $X_k$ be the atoms of ${\mathbf F}$. Then the entropy of $p(X)$ is
\begin{equation}\label{entropy} S = -\sum_k p(X_k)\ln(p(X_k))\ .
\end{equation}
\end{df}

Let us return to the quantum-mechanical example in order to explain which
information is concerned. After the choice of preparation and registration
devices, we do not know what will be the outcome of a registration but we just
know that any outcome $X_k$ of the registration has probability $p_{\mathsf
T}^{\mathsf E}(X_k)$. After an individual registration, one particular outcome
will be known with certainty. The amount of information gained by the
registration is the value of $S$ given by Eq.\ (\ref{entropy}). For more
detail, see Ref.\ \cite{peres}. Thus, the value of $S$ measures the lack of
information before the registration.

The entropy and the so-called Maximum Entropy Principle\footnote{There is also
a principle of statistical thermodynamics that carries the same name but ought
not to be confused with the mathematical MEP.} (MEP) have become important
notions of mathematical probability calculus, see, e.g., Ref.\
\cite{Jaynes}. The mathematical problem MEP solves can be generally
characterised as follows. Let system $\mathcal S$, condition $\mathcal C$ and
lattice $\mathbf F$ with atoms $X_k$ be given. Let there be more than one set
of $p(X_k)$'s that is compatible with $\mathcal C$. How the probabilities
$p(X_k)$ are to be assigned so that condition $\mathcal C$ is properly
accounted for without any additional bias? Such $p(X_k)$'s yield the maximum
of $S$ as given by (\ref{entropy}). MEP clearly follows from the meaning of
entropy as a measure of lack of information. We shall use this kind of MEP.

In quantum mechanics, probability distributions are generally associated with
both states and observables. This has to do with the non-objective nature of
observables. Still, can there be a general measure of information lack
associated with a given state alone? Clearly, such a question is meaningful
and the answer is the minimum of entropies, each associated with the state and
an observable. The minimum, $S({\mathsf T})$ is a well-defined function of
state ${\mathsf T}$ and is called von Neumann entropy. One can show (see,
e.g., Ref.\ \cite{peres}) that, up to a constant factor,
\begin{equation}\label{vNentropy} S(\mathsf T) = -tr[{\mathsf T}\ln({\mathsf
T})].
\end{equation} As each ${\mathsf T}$ must have a discrete spectrum with
positive eigenvalues $t_k$, we have
$$
S({\mathsf T}) = -\sum_k t_k \ln(t_k)\ .
$$

The lack of information $S({\mathsf T})$ is associated with the state $\mathsf
T$ of object $\mathcal S$ and consequently with the (classical) condition
$\mathcal C$ that defines the preparation. Hence, according to our criterion
of objectivity, von Neumann entropy is an objective property of object
$\mathcal S$ prepared in state $\mathsf T$.

MEP together with the von Neumann entropy can be applied to
thermodynamics. Consider the problem: what is the state $\mathsf T$ that
maximises von Neumann entropy under the condition that it has a given average
$tr[{\mathsf T}{\mathsf E}]$ of energy (this is an objective property in our
approach)? The answer is: the Gibbs' state and the value of Lagrange
multiplier that multiplies the average of energy in the variational principle
is $1/kT_B$, where $T$ is the temperature and $k_B$ Boltzmann constant (for
proof see, e.g.\ \cite{peres}).

Von Neumann entropy of composite objects satisfies a number of interesting
inequalities, see, e.g., Ref.\cite{wehrl}. One is the so-called sub-additivity
\cite{AL}.
\begin{prop} Let $\mathcal S$ be composed of objects ${\mathcal S}_1$ and
${\mathcal S}_2$ of different types. Let ${\mathsf T}$ be the state of
$\mathcal S$ and let
$$
{\mathsf T}_1 = \Pi_2({\mathsf T})\ ,\quad {\mathsf T}_2 = \Pi_1({\mathsf T})\
.
$$
Then
$$
S({\mathsf T}) \leq S({\mathsf T}_1) + S({\mathsf T}_2)
$$
and the equality sign is valid only if
$$
{\mathsf T} = {\mathsf T}_1 \otimes {\mathsf T}_2\ .
$$
\end{prop}

Thus, von Neumann entropy can be used to measure the amount of entanglement in
a state of a composite object.

\section{Joint measurability and contextuality} Joint measurability is related
to contextuality by the mathematics used to describe each. Joint measurability
is defined by a map from a Boolean lattice into the lattice of effects, and
contextuality is related to maps from the sublattice of projections into the
Boolean lattice $\{0,1\}$.

\subsection{Joint measurability} Let ${\mathsf A}$ and ${\mathsf B}$ be two
s.a.\ operators on Hilbert space ${\mathbf H}$ that are not necessarily
bounded. Then the product ${\mathsf A}{\mathsf B}$ as well as the operators
${\mathsf A}{\mathsf B}+{\mathsf B}{\mathsf A}$ and ${\mathsf A}{\mathsf
B}-{\mathsf B}{\mathsf A}$ are well defined on their common invariant domain
${\mathbf D}$ (see Sec.\ 1.3.1). The latter operator (or its extension) is
called commutator of ${\mathsf A}$ and ${\mathsf B}$ and is denoted by
$[{\mathsf A},{\mathsf B}]$.
\begin{prop} Let ${\mathsf A}$ and ${\mathsf B}$ be two s.a.\ operators with a
common invariant domain, let $i[{\mathsf A},{\mathsf B}]$ have a s.a.\
extension and let ${\mathsf T}$ be an arbitrary state. Then
\begin{equation}\label{uncert} \Delta {\mathsf A}\Delta {\mathsf B} \geq
\frac{|\langle[{\mathsf A},{\mathsf B}]\rangle|}{2}\ .
\end{equation}
\end{prop} Equation (\ref{uncert}) is called {\em uncertainty relation}.

The interpretation of the uncertainty relation is as follows. If we prepare
many copies of an object in state ${\mathsf T}$ and register either ${\mathsf
A}$ or ${\mathsf B}$ or $i[{\mathsf A},{\mathsf B}]$ on each, then the average
of $[{\mathsf A},{\mathsf B}]$ and variances of ${\mathsf A}$ as well as of
${\mathsf B}$ will satisfy Eq. (\ref{uncert}). It is not necessary to register
all observables jointly.

An important notion of quantum mechanics is that of joint
measurability. Often, it is also called simultaneous measurability but the
notion has nothing to do with time. It simply means that there is one
registration device that can measure both quantities.
\begin{df} Two elements ${\mathsf E}_1$ and ${\mathsf E}_2$ of ${\mathbf
L}_r({\mathbf H})^+_{\leq 1}$ are jointly measurable if there is a POV measure
${\mathsf E} : {\mathbf F} \mapsto {\mathbf L}_r({\mathbf H})^+_{\leq 1}$ such
that ${\mathsf E}_1 = {\mathsf E}(X_1)$ and ${\mathsf E}_2 = {\mathsf E}(X_2)$
for some $X_1$ and $X_2$ in ${\mathbf F}$.
\end{df} This is a mathematical property that, in fact, is only a necessary
condition for existence of the required registration device. Even if ${\mathsf
E}$ existed, the question whether it is an observable is non-trivial.
\begin{prop} Two effects ${\mathsf E}_1$ and ${\mathsf E}_2$ are jointly
measurable if and only if there are three elements ${\mathsf E}'_1$, ${\mathsf
E}'_2$ and ${\mathsf E}'_{12}$ in ${\mathbf L}_r({\mathbf H})^+_{\leq 1}$ such
that
$$
{\mathsf E}'_1 +{\mathsf E}'_2 + {\mathsf E}'_{12} \in {\mathbf L}_r({\mathbf
H})^+_{\leq 1}
$$
and
$$
{\mathsf E}_1 = {\mathsf E}'_1 + {\mathsf E}'_{12}\ ,\quad {\mathsf E}_2 =
{\mathsf E}'_2 + {\mathsf E}'_{12}\ .
$$
\end{prop} For proof, see \cite{ludwig1}, p. 89. Proposition 16 has two
corollaries: 1) two projections are jointly measurable if they commute and 2)
a sufficient condition for joint measurability of ${\mathsf E}_1$ and
${\mathsf E}_2$ is that
$$
{\mathsf E}_1 +{\mathsf E}_2 \in \ {\mathbf L}_r({\mathbf H})^+_{\leq 1}.
$$

If an observable is sharp then its effects commute, but the effects of a
general observable need not commute. Thus, even non commuting operators can be
jointly measurable. This is completely compatible with standard quantum
mechanics. Indeed, any POV measure that is an observable of an object
${\mathcal S}$ can be measured as a sharp observable of a composite system
containing ${\mathcal S}$ and another object called {\em ancilla} as
subsystems. For proof, see e.g. \cite{peres}, p.\ 285. An important example
will be given in Sec.\ 4.4.

Two observables ${\mathsf E}_1 : {\mathbf F}_1 \mapsto {\mathbf L}_r({\mathbf
H})$ and ${\mathsf E}_2 : {\mathbf F}_2 \mapsto {\mathbf L}_r({\mathbf H})$
with ${\mathbf F}_1$ comprising the Borel subsets of ${\mathbb R}^{n_1}$ and
${\mathbf F}_2$ those of ${\mathbb R}^{n_2}$ are called jointly measurable, if
each two effects ${\mathsf E}_1(X_1)$ and ${\mathsf E}_2(X_2)$ are jointly
measurable. In this case, there is an effect ${\mathsf E}_1(X_1)\wedge
{\mathsf E}_2(X_2)$ in ${\mathbf L}_r (\mathbf H)^+_{\leq 1}$ giving the
probability that observable ${\mathsf E}_1$ has value in $X_1$ and observable
${\mathsf E}_2$ has value in $X_2$. Then, there is a unique observable
$$
{\mathsf E} : {\mathbf F} \mapsto {\mathbf L}_r({\mathbf H})\ ,
$$
where ${\mathbf F}$ is the set of Borel subsets of ${\mathbb R}^{n_1+n_2}$ and
$$
{\mathsf E}(X_1\times X_2) = {\mathsf E}_1(X_1)\wedge {\mathsf E}_2(X_2)\ .
$$
We call ${\mathsf E}$ {\em compound} of ${\mathsf E}_1$ and ${\mathsf E}_2$

An important example are two sharp observables ${\mathsf A}$ and ${\mathsf
B}$. They are jointly measurable, if and only if all their effects (which are
projections) commute with each other, i.e., the operators commute. In this
case, the wedge is just the ordinary operator product,
$$
{\mathsf E}^{\mathsf A}(X_1)\wedge {\mathsf E}^{\mathsf B}(X_2) = {\mathsf
E}^{\mathsf A}(X_1) {\mathsf E}^{\mathsf B}(X_2)\ .
$$
The compound ${\mathsf E}^{{\mathsf A}\wedge{\mathsf B}}$ of ${\mathsf
E}^{\mathsf A}$ and ${\mathsf E}^{\mathsf B}$ is an observable that represents
registration of a pair of values, one of ${\mathsf A}$, the other of ${\mathsf
B}$.

If two sharp observables ${\mathsf A}$ and ${\mathsf B}$ do not commute then
they can be jointly measurable at best with the inaccuracy corresponding to
their uncertainty relation. This means that only some effects of ${\mathsf A}$
are jointly measurable with only some effects of ${\mathsf B}$. An example
will be given in Sec.\ 4.4.

\subsection{Contextuality} The investigations in the field of contextuality
were motivated by the following problem. In Newton mechanics, a general state
of any system $\mathcal S$ is described by probability distributions $\rho : Z
\mapsto [0,1]$, $Z \in {\mathbf \Gamma}$ on its phase space ${\mathbf
\Gamma}$. Any such distribution results from a fixed preparation and describes
the ensemble of individual systems $\mathcal S$ prepared in this way. Each
individual system of the ensemble is, however, always assumed, at least in
Newton mechanics, to be in some state given by a point $Z$ of ${\mathbf
\Gamma}$. We just do not know which point and the distribution $\rho(Z)$
describes the state of our incomplete knowledge. Thus, all observables, which
are just functions on ${\mathbf \Gamma}$, have determinate values for each
element of the ensemble.

Can anything analogous be also assumed for quantum mechanics? Could a state
operator of a quantum object $\mathcal S$ describe our knowledge of an
ensemble defined by the preparation and could the individual elements of the
ensemble each have determinate values of all observables, which are just not
known, or even, be it for whatever reason, could not be known? To solve this
problem, it is sufficient to restrict the observables to sharp observables,
because if the question had negative answer for a restricted class, it would
have negative answer for any class containing the restricted one.

It is technically advantageous to restrict the problem even further by
limiting oneself to (perpendicular) projection operators to subspaces of
${\mathbf H}_\tau$. Let us denote the set of all such projections by ${\mathbf
P}({\mathbf H}_\tau) \subset {\mathbf L}_r({\mathbf H}_\tau)$. The problem can
then be formulated as follows. Is there a dispersion-free probability
distribution $h : {\mathbf P}({\mathbf H}_\tau) \mapsto \{0,1\}$?
Dispersion-free means that its values are just zero and one so that such a
distribution determines values of all projections.

The dispersion-free distribution $h$ has to fulfil certain conditions, or else
it could not be interpreted as concerning properties. Let us first describe
the structure of ${\mathbf P}({\mathbf H}_\tau)$.

Each projection ${\mathsf a} \in {\mathbf P}({\mathbf H}_\tau)$ can be mapped
on linear subspace ${\mathsf a}({\mathbf H}_\tau)$ of ${\mathbf H}_\tau$ and
this bijective map allows to define the lattice relations on ${\mathbf
P}({\mathbf H}_\tau)$. First, ${\mathsf a} \leq {\mathsf b}$ if ${\mathsf
a}({\mathbf H}_\tau) \subset {\mathsf b}({\mathbf H}_\tau)$. This defines a
partial ordering on ${\mathbf P}({\mathbf H}_\tau)$. Second, ${\mathsf a} =
{\mathsf b}^\bot$ if ${\mathsf a}({\mathbf H}_\tau)$ contains all vectors
orthogonal to ${\mathsf b}({\mathbf H}_\tau)$. Projection ${\mathsf a}$ is
called the orthocomplement of ${\mathsf b}$. Third, ${\mathsf c} = {\mathsf a}
\wedge {\mathsf b}$ if ${\mathsf c}({\mathbf H}_\tau)$ is the set-theoretical
intersection of ${\mathsf a}({\mathbf H}_\tau)$ and ${\mathsf b}({\mathbf
H}_\tau)$. Observe that ${\mathsf a} \wedge {\mathsf b} = {\mathsf a} {\mathsf
b}$ (operator product of the projections) only if ${\mathsf a}$ and ${\mathsf
b}$ commute (are orthogonal). Finally, ${\mathsf c} = {\mathsf a} \vee
{\mathsf b}$ if ${\mathsf c}({\mathbf H}_\tau)$ is the linear hull of
${\mathsf a}({\mathbf H}_\tau)$ and ${\mathsf b}({\mathbf H}_\tau)$.

${\mathbf P}({\mathbf H}_\tau)$ with these relations forms the so-called
orthocomplemented lattice (for proof, see, e.g., Refs.\ \cite{BvN,bub}), but
not a Boolean lattice\footnote{In the so-called {\em quantum logic},
properties of ${\mathcal S}$ are described by elements of ${\mathbf
P}({\mathbf H}_\tau)$. They re\-present the mathematical counterpart of the
so-called YES-NO registrations \cite{Piron}. If we pretend that the values
obtained in the YES-NO experiments are properties of a well-defined single
quantum system, then we are forced to replace the Boolean lattice of ordinary
logic by the orthocomplemented latice of quantum logic. But this pretence is
against all logic because these values are not properties of one but of many
different systems each consisting of ${\mathcal S}$ plus some registration
apparatus.}. It has however Boolean sublattices, which represent sets of
jointly measurable sharp effects, and must, therefore contain only mutually
commuting projectors. On these sublattices, $h$ is, in fact, an assignment of
truth values 0, 1 and it has to satisfy the usual logical conditions
$$
h({\mathsf a} \vee {\mathsf b}) = h({\mathsf a}) \vee h({\mathsf b})\ ,\quad
h({\mathsf a} \wedge {\mathsf b}) = h({\mathsf a}) \wedge h({\mathsf b})\
,\quad h({\mathsf a^\bot}) = h({\mathsf a})^\bot\ .
$$
In these relations, we consider the set $\{0,1\}$ as a Boolean lattice of one
empty set and one arbitrary non-empty set $\mathbf A$ with $\bot$ the
set-theoretical complement in $\mathbf A$, $\vee$ the set-theoretical union
and $\wedge$ the set-theoretical intersection. Thus, on each Boolean
sublattice, $h$ must also be a Boolean lattice homomorphism.

The first result relevant to the question of existence of such maps is the
Gleason's theorem \cite{gleason}\footnote{Gleason's theorem is valid for
Hilbert spaces of dimensions greater than 2.}. It states that the set of all
probability distributions on ${\mathbf P}({\mathbf H}_\tau)$ is just ${\mathbf
T}({\mathbf H}_\tau)^+_1$. Thus, our dispersion-free distributions are to be
found in ${\mathbf T}({\mathbf H}_\tau)^+_1$. It is easy to show that there
are none there.

One can object that ${\mathbf P}({\mathbf H}_\tau)$ is an idealisation
containing an infinity of observables. This leads to the question whether
there is a finite subset of ${\mathbf P}({\mathbf H}_\tau)$ that does not
admit such distributions, either. This is the next relevant result,
Kochen-Specker no-go theorem \cite{kochen}, in which an example of such a
subset is given. There are more examples now provided by different physicists
(see, e.g., Ref.\ \cite{bub}). The reason why $h$ does not exist is that the
assignment of a truth value to projector ${\mathsf a}$, say, depends on to
which Boolean sublattice ${\mathsf a}$ belongs. This can be understood as
'context': in one case, ${\mathsf a}$ is registered jointly with elements of
one Boolean sublattice, in another case with those of another sublattice.

The definitive result in this field is Bub-Clifton-Goldstein theorem
\cite{bub} which lists all maximal sub-lattices of ${\mathbf P}({\mathbf
H}_\tau)$ that admit dispersion-free probability distributions. They need not
be Boolean but they are always only proper sublattices of ${\mathbf
P}({\mathbf H}_\tau)$. Hence, only a limited number of properties can be
assumed to be determinate before registration, and this limits possible
'non-collapse' interpretations and modifications of quantum mechanics. In
fact, these interpretations or modifications can be classified according to
the sublattices determined by the Bub-Clifton-Goldstein theorem \cite{bub}.

\section{Superselection rules} A preparation that deals only with systems of
one and the same type prepares so called {\em one-type systems}. Not every
preparation is such. For example, we can randomly mix electrons in some states
with protons in some states. Such systems, and some generalisations of them,
are called {\em mixed systems}\footnote{This has nothing to do with the term
'mixed state', which is sometimes used for non-vector states.}. In this
section, the mathematical description of mixed systems will be explained. We
shall mix just two one-type systems, ${\mathcal S}_1$ and ${\mathcal S}_2$,
but the generalisation to any number is straightforward.

Let ${\mathbf H}_1$ and ${\mathbf H}_2$ be the Hilbert spaces of ${\mathcal
S}_1$ and ${\mathcal S}_2$. Then the Hilbert space ${\mathbf H}$ of the system
${\mathcal S}$ mixing ${\mathcal S}_1$ and ${\mathcal S}_2$ is
$$
{\mathbf H} = {\mathbf H}_1 \oplus {\mathbf H}_2\ .
$$
The direct sum on the right-hand side is the space of pairs,
$(\psi_1,\psi_2)$, $\psi_1 \in {\mathbf H}_1$ and $\psi_2 \in {\mathbf H}_2$
with linear superposition defined by
$$
a(\psi_1,\psi_2) + b(\phi_1,\phi_2) = (a\psi_1 + b\phi_1,a\psi_2 + b\phi_2)
$$
and the inner product defined by
$$
\langle (\psi_1,\psi_2),(\phi_1,\phi_2)\rangle = \langle
\psi_1,\phi_1\rangle_1 + \langle \psi_2,\phi_2\rangle_2\ ,
$$
where $\langle \cdot,\cdot\rangle_i$ is the inner product of ${\mathbf
H}_i$. There are then embeddings $\iota_i : {\mathbf H}_i \mapsto {\mathbf H}$
defined by
$$
\iota_1(\psi_1) = (\psi_1,0)\ ,\quad \iota_2(\psi_2) = (0,\psi_2)
$$
for any $\psi_i \in {\mathbf H}_i$ and $i=1,2$.

Let ${\mathbf L}({\mathbf H})$ be the algebra of bounded linear operators on
${\mathbf H}$ and ${\mathbf L}({\mathbf H}_i)$ that of ${\mathbf H}_i$. If two
operators ${\mathsf A}_i \in {\mathbf L}({\mathbf H}_i)$, $i=1,2$, are given,
then an operator $({\mathsf A}_1 \oplus {\mathsf A}_2) \in {\mathbf
L}({\mathbf H})$, called direct sum of ${\mathsf A}_1$ and ${\mathsf A}_2$, is
defined by
$$
({\mathsf A}_1 \oplus {\mathsf A}_2) (\psi_1,\psi_2) = ({\mathsf
A}_1(\psi_1),{\mathsf A}_2(\psi_2))
$$
for any $(\psi_1,\psi_2) \in {\mathbf H}$. The special property of the
direct-sum operators is that the subspaces $\iota_i({\mathbf H}_i) \subset
{\mathbf H}$ are invariant with respect to them. Clearly, not all operators in
${\mathbf L}({\mathbf H})$ are of this form. Effects of the form ${\mathsf
A}_1 \oplus {\mathsf 0}_2$ (${\mathsf 0}_1 \oplus {\mathsf A}_2$) can be
interpreted as representing registrations done on ${\mathcal S}_1$ (${\mathcal
S}_2$) alone.

The embeddings $\iota_i$ define maps on projections ${\mathsf P}[\psi_i]$,
which can be denoted by the same symbol, viz.
$$
\iota_i({\mathsf P}[\psi_i]) = {\mathsf P}[\iota_i(\psi_i)]
$$
for all $\psi_i \in {\mathbf H}_i$. $\iota_i$ can be extended to the whole
spaces ${\mathbf T}({\mathbf H}_i)^+_1$, $\iota_i : {\mathbf T}({\mathbf
H}_i)^+_1 \mapsto {\mathbf T}({\mathbf H})^+_1$ as follows. Let ${\mathsf T}_i
\in {\mathbf T}({\mathbf H}_i)^+_1$, then
$$
\iota_1({\mathsf T}_1) = ({\mathsf T}_1,{\mathsf 0})\ ,\quad \iota_2({\mathsf
T}_2) = ({\mathsf 0},{\mathsf T}_2)\ .
$$
The convex combinations of states from ${\mathbf T}({\mathbf H}_1)^+_1$ and
${\mathbf T}({\mathbf H}_2)^+_1$ are states of mixed systems defined at the
beginning of this section. However, the states ${\mathbf T}({\mathbf H})^+_1$
are not exhausted by convex combinations of states from ${\mathbf T}({\mathbf
H}_1)^+_1$ and ${\mathbf T}({\mathbf H}_2)^+_1$.

The structural properties in which the systems ${\mathcal S}_1$ and ${\mathcal
S}_2$ differ from each other can now be viewed as non-trivial operators on
${\mathbf H}$. Let for example the masses $\mu_1$ and $\mu_2$ satisfy $\mu_1
\neq \mu_2$. Then the s.a.\ operator $\mu_1{\mathsf 1}_1 \oplus \mu_2{\mathsf
1}_2$ defines the mass operator ${\mathsf m}$ on ${\mathbf H}$ with two
eigenspaces $\iota_i({\mathbf H_i})$ and eigenvalues $\mu_i$,
$i=1,2$. Similarly for charges, spins, etc. Now, it is easy to see that all
operators of ${\mathbf L}({\mathbf H})$ that commute with ${\mathsf m}$ have
the form ${\mathsf A}_1 \oplus {\mathsf A}_2$ and all operators of this form
commute with ${\mathsf m}$. Clearly, s.a.\ operators of the form ${\mathsf
A}_1 \oplus {\mathsf A}_2$ are observables of system ${\mathcal S}$. The
nature of mixing systems suggests the basic rule
\begin{rl} All effects that can be registered on system ${\mathcal S}$ mixing
${\mathcal S}_1$ and ${\mathcal S}_2$ have the form ${\mathsf A}_1 \oplus
{\mathsf A}_2$ where ${\mathsf A}_1 \in {\mathbf L}_r({\mathbf H}_1)^+_{\leq
1}$ and ${\mathsf A}_2 \in {\mathbf L}_r({\mathbf H}_2)^+_{\leq 1}$.
\end{rl}

Hence, the space of all observables of ${\mathcal S}$ is not ${\mathbf
L}_r({\mathbf H})^+_{\leq 1}$ but the subset of all observables that commute
with ${\mathsf m}$.

In general, we define
\begin{df} A discrete sharp observable ${\mathsf Z}$ of a system ${\mathcal
S}$ with Hilbert space $\mathbf H$ is called {\em superselection observable}
if ${\mathsf Z}$ commutes with all observables of ${\mathcal S}$, sharp or not
sharp. The existence of such an observable is called {\em superselection
rule}. The eigenspaces of ${\mathsf Z}$ are called {\em superselection
sectors}. All superselection observables form the centre ${\mathbf Z}$ of the
algebra ${\mathbf L}({\mathbf H})$.
\end{df}

A restriction of the set of observables for system ${\mathcal S}$ suggests the
introduction of an equivalence relation on the set of states ${\mathbf
T}({\mathbf H}_\tau)^+_1$.
\begin{df} Two states ${\mathsf T}_1$ and ${\mathsf T}_2$ are equivalent with
respect to a set ${\mathbf O}$ of observables if these states assign the same
probability measures to each observable of ${\mathbf O}$, that is ,
$p_{{\mathsf T}_1}^{\mathsf E} = p_{{\mathsf T}_2}^{\mathsf E}$ for each
${\mathsf E} \in {\mathbf O}$. In that case, we write ${\mathsf T}_1
\cong_{\mathbf O} {\mathsf T}_2$.
\end{df} For example, linear superposition
$$
\Psi = a(\psi_1,0) + b(0,\psi_2)
$$
for any $\psi_i \in {\mathbf H}_i$, $i=1,2$, $|a|^2 + |b|^2 = 1$, and convex
combination
$$
{\mathsf T} = |a|^2\iota_1(|\psi_1\rangle \langle\psi_1|) +
|b|^2\iota_2(|\psi_2\rangle \langle\psi_2|)
$$
are equivalent with respect to the set of superselection observables ${\mathbf
O}$,
$$
|\Psi\rangle \langle\Psi |\cong_{\mathbf O} {\mathsf T}\ .
$$

For a given set of observables ${\mathbf O}$, $\cong_{\mathbf O}$ is, indeed,
an equivalence relation on the set of states ${\mathbf T}({\mathbf
H})^+_1$. Let us denote the set of equivalence classes of states in ${\mathbf
T}({\mathbf H})^+_1$ with respect to the set of observables $\mathbf O$ by
${\mathbf T}_{\mathbf O}({\mathbf H})^+_1$. No two states of the same class
can be distinguished by any registrations. No two different states from
${\mathbf T}({\mathbf H})^+_1$ are equivalent with respect to all observables
unless the set of effects ${\mathbf O}$ is restricted.
\begin{prop} Given a system ${\mathcal S}$ with the sets $\mathbf O$ and
${\mathbf T}_{\mathbf O}({\mathbf H})^+_1$ of observables and state classes,
respectively. Then the following statements are equivalent:
\begin{description}
\item[A]${\mathsf Z}$ is a superselection observable.
\item[B]Each state is equivalent to a unique convex combination of eigenstates
of ${\mathsf Z}$.
\end{description}
\end{prop} See Ref.\ \cite{BLM}, p. 18.

\begin{rl} The states of system ${\mathcal S}$ mixing ${\mathcal S}_1$ and
${\mathcal S}_2$ are the equivalence classes with respect to the set of
observables $\mathbf O$ of the form ${\mathsf A}_1 \oplus {\mathsf A}_2$. The
unique convex combination of eigenstates of ${\mathsf Z}$ to which each
element of ${\mathbf T}({\mathbf H})^+_1$ is equivalent describes the gemenge
structure of the class.
\end{rl} This motivates attempts to utilise superselection rules for solution
of the problem of classical properties and of quantum measurement
\cite{Hepp,Bell3,Bona,Sewell,Primas}. The formalism of this section follows
Ref.\ \cite{BLM} because it gives some meaning to the linear superpositions of
states from different superselection sectors, which is used by some of the
references mentioned.

\part{The models} In this part, the basic notions and rules of Part 1 will be
employed to construct models of observed systems or at least of some aspects
thereof. The relation between Part 1 and Part 2 within any theory is more or
less loose. The general language is used to construct models but it does not
determine which models are to be constructed. Indeed, it could be used also to
work with arbitrary or even fancy models that need not have much in common
with practical experience. The scientific practice is to encounter various
systems in nature and to construct models of these observed systems in order
to understand and predict their properties. We can also improve earlier
constructed models in order that they include newly observed properties, etc.

The division of a physical theory into a general part and a model part may
have some relevance to Popper's philosophy (cf.\ \cite{popper}). The first
part is a kind of language that does not have any specific, numerical
consequences. Only the models give specific, numerical values that can be
compared with experimental results and observations. It seems therefore that
the general part is practically not falsifiable but a specific model surely
is. Thus, a failure of a theory can always be interpreted as a failure of some
models. Two different reactions to such a falsification are possible. Either
one constructs a completely new theory starting with the general part, or one
tries to improve the models. For example, the difficulties of Newton mechanics
in the domain of quantum phenomena can either lead to the invention of quantum
mechanics or to the improvement of the corresponding Newton-mechanical
models. Indeed, the Bohm-de Broglie pilot-wave theory can be viewed as an
example of the latter approach.

The menagerie of successful models is a very important part of any physical
theory. Textbooks of quantum mechanics dedicate most of their space to an
account of it. We shall restrict ourself only to those models that are
immediately important to our main aim: to solve the problem of classical
properties and quantum measurement.

\chapter{Models of microscopic systems} In this chapter, we limit ourselves
just to few examples of microsystems, first to show the crucial steps of model
construction within our conception of quantum mechanics and second to
introduce some models that will be important later.

The construction of quantum models can be divided into two steps. First,
assumptions are made about states and observables of particles. Second, the
states and observables of a composite quantum system can be constructed as
described in Secs.\ 2.1 and 2.2 in agreement with its structural properties
such as composition and a specific Hamiltonian.

The operators that will be defined in the present section will have some
relations to similar quantities that describe classical systems of an
analogous composition and they will have names that do not seem to be
justified by anything but this classical analogy. This is true but it will be
interpreted in this paper as follows. The definition of the operators is a
purely quantum-mechanical business and the role of the corresponding
observables as well as their registration method belong to important new
assumptions of the theory. The only concession we can do is that the classical
mechanics can give us a hint on how quantum operators may be constructed, but
the definitive decision is only possible if the model works, that is, predicts
properly what can be observed.

One often tries to find some way from classical systems to analogous quantum
systems, a so-called 'quantisation' method, or at least to formulate some kind
of 'correspondence principle' concerning relations between classical and
quantum mechanics. In our opinion, the natural way is just the opposite and
must lead from quantum mechanics to the classical one because classical
systems are just some macroscopic quantum systems. This ought to explain any
correspondence.

Model assumptions are assumptions about specific systems that are to be
modelled. They will, therefore, be called 'Assumptions' to distinguish them
from general 'Rules'.

\section{Basic PV measures} Let us start with the following structural
properties: a particle with mass $\mu$, spin $s$ and charge $e$ (this
determines the type $\tau$). To construct the Hilbert space ${\mathbf
H}_{\mu,s,e}$ of such model, we shall work in $Q$-representation. Consider the
set of rapidly decreasing $C^\infty$ functions $\phi(\vec{x},m)$ (wave
functions), where $m \in \{-s,\cdots, s\}$. Then, ${\mathbf H}_{\mu,s,e}$ is
the completion of the set of the wave functions with respect to the inner
product
$$
\langle\phi|\psi\rangle = \sum_m \int_{{\mathbb R}^3} d^3x
\phi^*(\vec{x},m)\psi(\vec{x},m)\ .
$$
The states of the particle that can be prepared are elements of ${\mathbf
T}({\mathbf H}_{\mu,s,e})^+_1$.

Next, we define certain operators on ${\mathbf H}_{\mu,s,e}$: {\em position}
$\vec{\mathsf q} = ({\mathsf q}^1,{\mathsf q}^2,{\mathsf q}^3)$, (linear) {\em
momentum} $\vec{\mathsf p} = ({\mathsf p}^1,{\mathsf p}^2,{\mathsf p}^3)$ and
{\em spin} $\vec{\mathsf s} = ({\mathsf s}^1,{\mathsf s}^2,{\mathsf s}^3)$,
where the symbol $\vec{\mathsf q}$ denotes the compound observable (see Sec.\
3.2.1) ${\mathsf q}^1 \wedge {\mathsf q}^2 \wedge {\mathsf q}^3$ and similarly
for other vectorial operators.

The position and momentum operators have already been defined in Sec.\ 1.3,
Eqs.\ (\ref{posit}) and (\ref{moment}). The three components of spin are
$(2s+1)\times(2s+1)$ Hermitean matrices $ s^k_{mn}$ that act on wave functions
(vector-valued functions) as follows
\begin{equation}\label{spin} {\mathsf s}^k\phi(\vec{x},m) = \sum_{n=-s}^s
s^k_{m\,n}\phi(\vec{x},n)\ .
\end{equation} In non relativistic quantum mechanics, one mostly works only
with $s = 0$ or $s = 1/2$. The matrices for $s = 1/2$ are
\begin{equation}\label{spinm} {\mathsf s}^1 = \hbar/2 \begin{pmatrix}0 & 1 \\
1 & 0
\end{pmatrix}\ ,\quad {\mathsf s}^2 = \hbar/2 \begin{pmatrix}0 & -i \\ i & 0
\end{pmatrix}\ ,\quad {\mathsf s}^3 = \hbar/2 \begin{pmatrix}1 & 0 \\ 0 & -1
\end{pmatrix}\ .
\end{equation} All components of $\vec{\mathsf q}$, $\vec{\mathsf p}$ and
$\vec{\mathsf s}$ can be extended to self-adjoint (s.a.) operators on
${\mathbf H}_{\mu,s,e}$.

The common invariant domain of operators $\vec{\mathsf q}$, $\vec{\mathsf p}$
and $\vec{\mathsf s}$ is the set of rapidly decreasing $C^\infty$ functions
$\phi(\vec{x},m)$. Hence one can construct other operators from them by
products and linear combinations, e.g., orbital {\em angular momentum}
$$
{\mathsf L}^1 = {\mathsf q}^2{\mathsf p}^3 - {\mathsf q}^3{\mathsf p}^2 \
,\quad {\mathsf L}^2 = {\mathsf q}^3{\mathsf p}^1 - {\mathsf q}^1{\mathsf p}^3
\ ,\quad {\mathsf L}^3 = {\mathsf q}^1{\mathsf p}^2 - {\mathsf q}^2{\mathsf
p}^1\ .
$$
The operator has a unique s.a.\ extension. One of the assumptions of quantum
mechanics is that one can construct all observables for a particle from its
basic operators. This is the reason why they are called {\em basic}. The
construction is simple, if the resulting observable is a polynomial. More
complicated functions can be constructed using spectral theorem. For example,
any function $V(\vec{\mathsf q})$ can be defined by
$$
V(\vec{\mathsf q}) = \int_{{\mathbb
R}^3}V(\iota^1,\iota^2,\iota^3)dE^{\vec{\mathsf q}}\ .
$$

Let us construct the PV measures ${\mathsf E}^{\vec{\mathsf{q}}}$, ${\mathsf
E}^{\vec{\mathsf{p}}}$ and ${\mathsf E}^{\mathsf{s}^3}$. The dimension of the
first two is 3 and $\mathbf F$ is the set of all Borel subsets of ${\mathbb
R}^3$. We define
\begin{equation}\label{povq} {\mathsf E}^{\vec{\mathsf{q}}}(X)\psi(\vec{x},m)
= \chi[X](\vec{x})\psi(\vec{x},m)\ ,
\end{equation} where $\chi[X]$ is the characteristic function of $X \in
\mathbf{F}$, that is, $\chi[X](\vec{x}) = 1$ for $\vec{x} \in X$ and
$\chi[X](\vec{x}) = 0$ otherwise. Clearly, the value set of ${\mathsf
E}^{\vec{\mathsf q}}$ is the whole of ${\mathbb R}^3$.

For ${\mathsf E}^{\vec{\mathsf{p}}}$, we define
\begin{multline}\label{povp} {\mathsf E}^{\vec{\mathsf{p}}}(X)\psi(\vec{x},m)
= (2\pi\hbar)^{-3}\int d^3p\,
\exp\left(\frac{i}{\hbar}\vec{p}\cdot\vec{x}\right)\chi[X](\vec{p}) \\ \int
d^3x'\,exp\left(-\frac{i}{\hbar}\vec{p}\cdot\vec{x}'\right)\psi(\vec{x}',m)\ ,
\end{multline} where $\chi[X](\vec{p}) = 1$ for $\vec{p} \in X$ and
$\chi[X](\vec{p}) = 0$ otherwise. The value set of ${\mathsf
E}^{\vec{\mathsf{p}}}$ is the whole of ${\mathbb R}^3$.

The three components of spin cannot be combined to a single PV measure because
they do not commute. Let us describe the third component $\mathsf{s}^3$. PV
measure ${\mathsf E}^{\mathsf{s}^3}$ is one-dimensional and $\mathbf F$ is the
set of Borel subsets of ${\mathbb R}$. We define
\begin{equation}\label{povs} {\mathsf E}^{\mathsf{s}^3}(X)\psi(\vec{x},m) =
\chi[X](m)\psi(\vec{x},m)\ ,
\end{equation} where $\chi[X](m) = 1$ if $m \in X\cap\{-s,\cdots,s\}$ and
$\chi[X](m) = 0$ otherwise. ${\mathsf E}^{\mathsf{s}^3}$ is a discrete PV
measures with value space $\{-s,\cdots,s\}$.

Let us summarize:
\begin{assump} Any particle with mass $\mu$, spin $s$ and charge $e$ is a
quantum system with Hilbert space ${\mathbf H}_{\mu,s,e}$. Its basic operators
are position, momentum and spin, defined by Eqs.\ (\ref{posit}),
(\ref{moment}), as well as (\ref{spin}) and (\ref{spinm}) for particles with
spin 1/2.
\end{assump}

According to Sec.\ 2.2.2, these basic PV measures are not observables. To
obtain observables from them, separation status $D$ in which we are going to
prepare all particles is to be determined and $D$-localisation of the PV
measures have to be constructed. We shall speak about measurement of basic
observables understanding the so constructed POV measures.

Suppose that a free particle is prepared in $D$. Registration of the (smeared)
position of the particle can e.g.\ be carried out by a small Geiger counter
that is placed so that its sensitive volume $V$ is at a fixed position in
$D$. If it counts at time $t$, then we know that the registered position of
the quantum system were in $V$ at $t$. The momentum of the particle can be
inferred from its energy because its Hamiltonian is a quadratic function of
the momentum (see the next subsection), and from the direction of motion that
is given, e.g., by apertures in screens. This functional relation between
momentum and Hamiltonian is of course only approximate for $D$-localised
quantities. The spin can be registered via the Stern-Gerlach apparatus because
even neutral particles have a non-zero gyromagnetic ratio.

Finally, the basic operators of a composite system can be obtained from the
basic operators of its subsystems. A prominent example in any textbook of
quantum mechanics is the hydrogen atom. A possible model of it is composed of
an electron and a proton. The Hilbert space of hydrogen is ${\mathbf
H}_e\otimes {\mathbf H}_p$, where ${\mathbf H}_e$ and ${\mathbf H}_p$ are
those of the electron and the proton. These are just ${\mathbf H}_{\mu,s,e}$
with suitable numerical values of $\mu,s$ and $e$. The basic operators are
$\vec{{\mathsf q}_e}\otimes {\mathsf 1}_p$, $\vec{{\mathsf p}_e}\otimes
{\mathsf 1}_p$, $\vec{{\mathsf s}_e}\otimes {\mathsf 1}_p$, ${\mathsf
1}_e\otimes \vec{{\mathsf q}_p}$, ${\mathsf 1}_e\otimes \vec{{\mathsf p}_p}$,
${\mathsf 1}_e\otimes \vec{{\mathsf s}_p}$.

Further operators can be constructed by algebraic operations from the basic
ones, and the corresponding observables then by $D$-localisation.

\section{Hamiltonian}
\begin{assump} Hamiltonian of any particle with mass $\mu$, charge $e$ and
gyromagnetic ratio $g$ moving in classical electrostatic potential
$V(\vec{x})$ and classical stationary magnetic field $\vec{B}(\vec{x})$ is
$$
{\mathsf H} = \frac{{\mathsf p}^2}{2\mu} + eV(\vec{\mathsf q}) - g\vec{\mathsf
s}\cdot\vec{B}(\vec{\mathsf q})\ ,
$$
where the square means the squared norm of a 3-vector and '$\cdot$' is the
inner product of 3-vectors.

A textbook model of hydrogen atom consists of a spin-zero proton of mass
$\mu_p$ and charge $e$ and a spin-zero electron of mass $\mu_e$ and charge
$-e$. The Hamiltonian of the model is
$$
{\mathsf H} = \frac{{\mathsf p}_e^2}{2\mu_e} + \frac{{\mathsf p}_p^2}{2\mu_p}
- \frac{e^2}{|\vec{{\mathsf q}_e} - \vec{{\mathsf q}_p}|}\ .
$$
\end{assump}

For any quantum system ${\mathcal S}$ with Hamiltonian ${\mathsf H}$, the
spectrum of the s.a.\ operator ${\mathsf H}$ is called energy spectrum of
${\mathcal S}$. The energy spectrum is thus a structural property. Systems can
be recognised via their spectra. If light comes from far away in the space and
the (possibly Doppler shifted) Balmer series is found in its spectrum,
hydrogen must be somewhere there.

Kinetic energy of a free system can e.g.\ be measured by a proportional
counter. In fact, what is measured in this way is a $D$-localisation of the
total kinetic energy, which is a s.a.\ operator that forms a part of the
Hamiltonian. For a particle, it is
$$
\frac{{\mathsf p}^2}{2\mu}
$$
and the spectrum is $[0,\infty)$. For the hydrogen atom, we had first to
separate the motion of its centre of mass from its internal degrees of
freedom. An utterly different business is a measurement of a spectrum of a
bound system. Spectra of bound systems are discrete and can be measured if
large number of the same systems emits or absorbs electromagnetic
radiation. This is an indirect measurement via scattering of photons in which
the final 'kinetic' energy of photons is measured by a spectrometer.

\section{Galilean group} A free particle is defined by $V(\vec{\mathsf q}) =
\text{const}$ and $\vec{B}(\vec{\mathsf q}) = 0$. It is an isolated quantum
system ${\mathcal S}$ and Galilean group has, therefore, a unitary ray
representation on ${\mathbf H}_{\mu,s.e}$, see Sec.\ 1.3. We can describe it
by writing up its generators. From Sec.\ 1.3, we know already the form of
space translations, space rotations and time translations even for systems
that are not isolated. The additional assumption is
\begin{assump} For a particle (not necessarily free) with mass $\mu$ and spin
$s$, the total angular momentum is ${\mathsf J}^k = {\mathsf L}^k + {\mathsf
s}^k$. For a free particle, the boosts are generated by ${\mathsf b}^k = -
\mu{\mathsf q}^k$.
\end{assump} All operators are written in the Schr\"{o}dinger picture. The
only group representatives where multipliers appear is multiplication of
boosts, ${\mathsf U}(\vec{v})$, with space shifts ${\mathsf U}(\vec{a})$. This
leads to 'anomalous' commutator,
$$
[{\mathsf b}^k,{\mathsf p}_l] = -i\hbar\mu\delta_{kl}.
$$

It is easy to work out how the corresponding one-parameter groups act on wave
function $\psi(\vec{x},m)$:
$$
{\mathsf U}(\vec{a})\psi(\vec{x},m) = \psi(\vec{x} - \vec{a},m)
$$
and
$$
{\mathsf U}(\vec{v})\psi(\vec{x},m) =
\exp\left(\frac{i}{\hbar}\mu\vec{v}\cdot\vec{x}\right)\psi(\vec{x},m)\ .
$$
Thus, the boost just shifts the momentum by $\mu\vec{v}$. For example, the
average of momentum in the shifted state equals to that in the original one
plus $\mu\vec{v}$. The operators ${\mathsf b}^k$ and ${\mathsf p}_k$ generate
a subgroup of ${\mathbf G}$ called Heisenberg group.

\section{Joint measurement of position \\ and momentum} Existence of an
observable that represents a joint measurement of position and momentum will
play some role in Chap.\ 5. To this aim, we generalise the construction of
such an observable for a simplified model that was first proposed in Ref.\
\cite{AK}. The model system ${\mathcal S}$ is a free one-dimensional spin-zero
particle with position ${\mathsf q}$ and momentum ${\mathsf p}$. The Hilbert
space is $L^2({\mathbb R})$ and the operators are defined by equations
analogous to (\ref{posit}) and (\ref{moment}).

Operators ${\mathsf q}$ and ${\mathsf p}$ have an invariant common domain and
their commutator is easily calculated to be
$$
[{\mathsf q},{\mathsf p}] = i\hbar\ .
$$
Hence, the joint measurement may be a problem.

The general construction of a non-trivial POV measure for system ${\mathcal
S}$ introduces another system, ancilla, that forms a composite system with
${\mathcal S}$. Let our ancilla ${\mathcal A}$ be a similar particle with
position ${\mathsf Q}$ and momentum ${\mathsf P}$. We work in
$Q$-representation so that the Hilbert space of the composite system
${\mathcal S} + {\mathcal A}$ is $L^2({\mathbb R})\otimes L^2({\mathbb R})$
which can be identified with $L^2({\mathbb R}^2)$. Then, we have wave
functions $\Psi(q,Q)$ and integral operators with kernels of the form
${O}(q,Q;q',Q')$.

The dynamical variables ${\mathsf A} = {\mathsf q}-{\mathsf Q}$ and ${\mathsf
B} = {\mathsf p}+{\mathsf P}$ of the composite system ${\mathcal S} +
{\mathcal A}$ commute and can therefore be measured jointly. The value space
of PV observable ${\mathsf E}^{{\mathsf A}\wedge {\mathsf B}}$ is ${\mathbb
R}^2$ with coordinates $a$ and $b$ (see end of Sec.\ 3.2.1).

The next step is to smear ${\mathsf E}^{{\mathsf A}\wedge {\mathsf B}}$ to
obtain a realistic POV meausre ${\mathsf E}_k$. Let us divide the $ab$ plane
into disjoint rectangular cells $X_k= [a_{k-}, a_{k+}]\times [b_{k-}, b_{k+}]$
covering the entire plane. Each cell is centred at $(a_k,b_k)$, $a_k = (a_{k+}
+ a_{k-})/2$, $b_k = (b_{k+} + b_{k-})/2$ and $S_k = (a_{k+} - a_{k-})(b_{k+}
- b_{k-})$ is its area. Then,
\begin{equation}\label{cell} {\mathsf E}_k = {\mathsf E}^{{\mathsf A}\wedge
{\mathsf B}}(X_k) = {\mathsf E}^{\mathsf A}([a_{k-},a_{k+}]){\mathsf
E}^{\mathsf B}([b_{k-}, b_{k+}])\ .
\end{equation} The cells can be arbitrarily small.

The probability to obtain the outcome $k$ in state $\mathsf T$ of the
composite system is
\begin{equation}\label{149} tr[{\mathsf E}_k{\mathsf T}] = \int
dq\,dQ\,dq'\,dQ'\, {E}_k(q,Q;q',Q'){T}(q',Q';q,Q)\ .
\end{equation} We assume that the composite system ${\mathcal S} + {\mathcal
A}$ is prepared in a factorised state
\begin{equation}\label{factors} {\mathsf T} ={\mathsf T}_{\mathcal
S}\otimes{\mathsf T}_{\mathcal A}
\end{equation} and express the probability (\ref{149}) in terms of the state
${\mathsf T}_{\mathcal S}$. The action of the projections ${\mathsf
E}^{\mathsf A}([a_{k-}, a_{k+}])$ and ${\mathsf E}^{\mathsf B}([b_{k-},
b_{k+}])$ on vector states of the form $\Psi(q,Q)$ is
$$
{\mathsf E}^{\mathsf A}([a_{k-},a_{k+}])\Psi(q,Q) =
\chi[a_{k-},a_{k+}](q-Q)\Psi(q,Q)\ ,
$$
where
\begin{eqnarray*} \chi[a_{k-},a_{k+}](x) &=& 1\quad \forall\, x \in
[a_{k-},a_{k+}]\ ,\\ \chi[a_{k-},a_{k+}](x) &=& 0\quad \forall\, x\,\text{not}
\in [a_{k-},a_{k+}]
\end{eqnarray*} and
\begin{multline*} {\mathsf E}^{\mathsf B}([b_{k-},b_{k+}])\Psi(q,Q) =
(2\pi\hbar)^{-2}\int dp\,dP\,dq'\,dQ' \\ \exp\frac{i}{\hbar}[p(q-q') +
P(Q-Q')]\chi[b_{k-},b_{k+}](p + P)\Psi(q',Q')\ ,
\end{multline*} where
\begin{eqnarray*} \chi[b_{k-},b_{k+}](x) &=& 1\quad \forall\, x \in
[b_{k-},b_{k+}]\ ,\\ \chi[b_{k-},b_{k+}](x) &=& 0\quad \forall\, x\
\text{not}\in [b_{k-},b_{k+}]\ .
\end{eqnarray*}

The trace (\ref{149}) can be calculated in several steps as follows. First,
\begin{multline*} tr[{\mathsf E}_k{\mathsf T}] = tr[{\mathsf E}^{\mathsf
A}([a_{k-},a_{k+}]){\mathsf E}^{\mathsf B}([b_{k-},b_{k+}]) {\mathsf T}] =
tr[{\mathsf E}^{\mathsf B}([b_{k-},b_{k+}]) {\mathsf T}{\mathsf E}^{\mathsf
A}([a_{k-},a_{k+}])] \\ = (2\pi\hbar)^{-2}\int
dq\,dQ\,dp\,dP\,dq'\,dQ'\,\exp\frac{i}{\hbar}[p(q-q') + P(Q-Q')] \\
\chi[b_{k-},b_{k+}](p + P) \chi[a_{k-},a_{k+}](q - Q) {T}(q',Q';q,Q)\ .
\end{multline*} Second, introduce new integration variables $q$, $a$, $p$,
$b$, $q'$, $a'$,
\begin{multline*} tr[{\mathsf E}_k{\mathsf T}] = (2\pi\hbar)^{-2}\int
dq\,da\,dp\,db\,dq'\,da' \\ \exp\frac{i}{\hbar}[p(q-q') + (b - p)(q - a-q' +
a')] \chi[b_{k-},b_{k+}](b) \chi[a_{k-},a_{k+}](a) \ {T}(q',q' - a';q,q - a)\
.
\end{multline*} Third, if the cells are sufficiently small, the integrands do
not change appreciably inside the integration intervals of $a$ and $b$ so that
they can be approximated as follows:
$$
\int da\,db\,\chi[a_{k-},a_{k+}](a)\chi[b_{k-},b_{k+}](b)f(a,b) \approx
S_kf(a_k,b_k)\ .
$$
In this way, we obtain
\begin{multline*} tr[{\mathsf E}_k{\mathsf T}] \approx
\frac{S_k}{(2\pi\hbar)^2}\int dq\,dp\,dq'\,da' \\
\exp\frac{i}{\hbar}[(p(a_k-a') + b_k(q - q' - a_k + a')] \ T(q',q' - a';q,q -
a_k)\ .
\end{multline*} But, fourth, the factor containing the integral over $p$ is a
$\delta$-function,
$$
\int dp\,\exp\frac{i}{\hbar}p(a_k-a') = 2\pi\hbar\delta(a' -a_k)\ .
$$
Thus, we obtain,
$$
tr[{\mathsf E}_k{\mathsf T}] \approx \frac{S_k}{(2\pi\hbar)^2}\int
dq\,dq'\,\exp\frac{i}{\hbar}b_k(q - q') \ T(q',q' - a_k;q,q - a_k)\ .
$$
Fifth, we use Eq.\ (\ref{factors}),
$$
{\mathsf T}(q',Q';q,Q) = {\mathsf T}_{\mathcal S}(q',q) {\mathsf T}_{\mathcal
A}(Q',Q)\ ,
$$
with the result
\begin{prop} The probability of the outcome $k$ to be found in the factorised
state (\ref{factors}) is given approximately by
\begin{equation}\label{factorT} tr[{\mathsf E}_k{\mathsf T}] \approx
tr\left[{\mathsf T}_{\mathcal S}\frac{S_k}{(2\pi\hbar)^2}{\mathsf T}_{\mathcal
A}[a_k,b_k]\right]\ ,
\end{equation} where
$$
{\mathsf T}_{\mathcal A}[a_k,b_k] = \exp\frac{i}{\hbar}b_k(q - q') \ T(q',q' -
a_k;q,q - a_k)
$$
is the state ${\mathsf T}_{\mathcal A}$ first shifted by $a_k$ and then
boosted by $-b_k$. The approximation improves if the cells are smaller.
\end{prop}

If the cells are sufficiently small, we have $q \approx Q + a_k$ and $p
\approx b_k - P$. In this way, the coordinate of $\mathcal S$ is shifted by
$a_k$ while the inverted momentum is shifted by $-b_k$ with respect to those
of $\mathcal A$.

Eq.\ (\ref{factorT}) shows that there is an 'effective' POV measure ${\mathsf
E}_{{\mathcal S}k}$ for system $\mathcal S$ defined by
$$
{\mathsf E}_{{\mathcal S}k} = \frac{S_k}{(2\pi\hbar)^2}{\mathsf T}_{\mathcal
A}[a_k,b_k]
$$
that yields the probability of the outcome $k$ of the above registration
considered as a registration performed on $\mathcal S$.

The state ${\mathsf T}_{\mathcal A}$ is completely arbitrary. To construct a
useful quantity, one usually chooses a vector state in the form of a Gaussian
wave packet (see, e.g., Ref.\ \cite{peres}, p. 418),
$$
\Psi_\sigma = (\pi\sigma^2)^{-1/4}\exp\left(-\frac{Q^2}{2\sigma^2}\right)\ .
$$
Easy calculation yields
$$
\langle Q\rangle = \langle P\rangle = 0\ ,\quad \Delta Q =
\frac{\sigma}{\sqrt{2}}\ ,\quad \Delta P = \frac{\hbar}{\sigma\sqrt{2}}
$$
so that $\Psi_\sigma$ is a state of minimum uncertainty. If we shift
$\Psi_\sigma$ by $a_k$ and then boost it by $-b_k$, we obtain
$$
\Psi_\sigma[a_k,b_k] =
(\pi\sigma^2)^{-1/4}\exp\left(-\frac{(Q-a_k)^2}{2\sigma^2}
-\frac{i}{\hbar}b_kQ\right)\ ,
$$
which is the Gaussian packet concentrated at $(a_k,-b_k)$,
\begin{equation}\label{avvar1} \langle Q\rangle = a_k\ ,\quad \langle P\rangle
= -b_k\ ,
\end{equation} and
\begin{equation}\label{avvar2} \quad \Delta Q = \frac{\sigma}{\sqrt{2}}\
,\quad \Delta P = \frac{\hbar}{\sigma\sqrt{2}}\ .
\end{equation} As the Gaussian wave packet is uniquely determined by its
averages and variances, we can interpret the observable ${\mathsf E}_k$ as
giving the probability that the corresponding registration applied to state
${\mathsf T}_{\mathcal S}$ detects the state of $\mathcal S$ with the averages
and variances given by Eqs.\ (\ref{avvar1}) and (\ref{avvar2}).

\chapter{Quantum models of classical properties} Quantum mechanics is a
genuinely statistical theory. A well-defined repeatable preparation procedure
defines an ensemble. Values of any observable obtained by repeated
registrations on individual elements of the ensemble satisfy statistical laws
that are given by quantum mechanics. Moreover, in general, values of
observables are not objective, i.e., the individual elements of the ensemble
cannot be assumed to possess determinate values of their observables before or
independently of registrations and the results of individual registrations
cannot be predicted.

On the other hand, basic physical theories that describe classical world,
i.e. mechanics, electrodynamics, theory of gravitation, are objective and
deterministic in character. This comprises three assumptions. First, the
states are determined by values of observables and can be absolutely sharp
(they contain no random elements) and, second, the sharp states are objective
in the sense that any system always really is in one of them. Fuzzy states are
also possible. They have the form of probability distributions of sharp states
and are due to incomplete knowledge. Third, dynamical laws determine a unique
evolution of all states.

More generally (i.e., even for thermodynamics), there is a criterion of
objectivity, the so-called Modified Principle of Macroscopic
Realism\footnote{The original Principle of Macroscopic Realism has been
formulated by Leggett \cite{leggett}. We have replaced Leggett's
'macroscopically distinct (quantum) states' by 'distinct classical
states'. Indeed, if the 'macroscopically distinct (quantum) states' include
pure quantum states, point 1 of Leggett's macroscopic realism violates the
principle of superposition. Then, one has to assume that some as yet unknown
phenomena exist at the macroscopic level which are not compatible with
standard quantum mechanics (see, e.g., \cite{leggett} and the references
therein). However, no such phenomena have been observed.} (MPMR) \cite{PHJT}:
\begin{enumerate}
\item A macroscopic system which has available to it two or more {\em distinct
classical states} is at any given time in a definite one of those states.
\item It is possible in principle to determine which of these states the
system is in without any effect on the state itself or on the subsequent
system dynamics.
\end{enumerate}

To obtain classical theories as some limiting cases of quantum mechanics or to
construct quantum models of classical properties is a difficult problem. As
yet, only partial and incomplete solutions to this problem exist. Let us now
briefly review the most popular approaches to the problem. The shortcomings of
the approaches are well known \cite{d'Espagnat, bell2,Wallace}. We mention
them only fleetingly. First, the quantum decoherence theory \cite{Zurek, Zeh}
works only if certain observables concerning both the environment and the
quantum system cannot be measured (see the analysis in
\cite{d'Espagnat,bub}). The deep reason is that one works with values of
observables. Second, the theories based on coarse-grained operators
\cite{peres,poulin,Kofler}: the problem is the same as with the
decoherence. For example, the Leggett-Garg inequality \cite{Kofler} is a
condition for the validity of the principle of macroscopic realism that works
with values of observables. Third, the Coleman-Hepp theory
\cite{Hepp,Bell3,Bona} and its modifications \cite{Sewell,Primas}: they are
based on some particular theorems that hold for infinite systems but do not
hold even approximately for finite ones (see the analysis in \cite{Bell3}).

If we turn from theory to experiment, we may notice that any well-founded
scientific observation of classical properties always has a statistical
form. A measurement or observation is only viewed as well understood, if it is
given as an average with a variance. This fact does not by itself contradict
the deterministic character of the corresponding theory. The usual excuse is
that the observation methods are beset with inaccuracy but that improvement of
techniques can lead to better and better results approaching the 'objective'
values arbitrarily closely.

This popular excuse is clearly incompatible with the assumption that the
classical world is only an aspect of a deeper quantum world and that each
classical model is nothing but a kind of incomplete description of the
underlying quantum system. If we assume such universality of quantum theory,
then the statistical character of classical observational results must not
only be due to inaccuracy of observational methods but also to genuine
uncertainty of quantum origin. This point of view is originally due to Exner
\cite{Exner}, p.~669, and Born \cite{Born} and will be adopted here as a
starting point of our theory of classical properties.

\section{Modified correspondence principle} Born-Exner assumption has quite
radical consequences that are only seldom realised. First, the exactly sharp
states and trajectories of classical theories are not objective. They do not
exist in reality but are only idealisations. What does really exist are fuzzy
states and trajectories. The objectivity of fuzzy states in classical world is
a difficult point to accept and understand. Let us explain it in more detail.

In quantum mechanics, the basis of objectivity of dynamical properties is the
objectivity of the conditions that define preparation procedures. In other
words, if a property is uniquely determined by a preparation, then it is an
objective property. If we look closely, the only hindrance to try the same
idea in classical theories is just the custom always to speak about initial
data instead of preparations. An initial datum is nothing but a state, more or
less sharp one. The question on how an exactly sharp state can come into being
is ignored. This in turn seems justified by the hypothesis that sharp states
are objective, that is, they just exist by themselves.

Thus, one has to accept that preparation procedures play the same basic role
in the classical as in quantum physics. Then, the nature and form of necessary
preparation procedures must be specified and the corresponding states
described.

Our starting point was that all physical systems are quantum systems. Let us
now make it more precise. There is one level of description (approximative
model of some aspects of a physical system), which determines a classical
system ${\mathcal S}_c$ and its classical properties for which quantum theory
is not needed, namely the classical description. In addition, the same system
can also be understood as a quantum system ${\mathcal S}_q$ such that the
classical properties of ${\mathcal S}_c$ are some objective quantum properties
of ${\mathcal S}_q$. This follows from the fact that all classical properties
can be assumed to be objective without any danger of
contradictions. ${\mathcal S}_q$ is a quantum model of ${\mathcal S}_c$.

The construction of quantum model ${\mathcal S}_q$ consists of the following
points. 1) The composition of ${\mathcal S}_q$ must be defined. 2) The
observables that can be measured on ${\mathcal S}_q$ are to be determined. On
the one hand, this is a non-trivial problem: there are relatively strong
restrictions on what observables of macroscopic systems can be measured
(Secs.\ 2.2.2 and 6.2). On the other, as any observable is measurable only by
a classical apparatus, the existence of such apparatuses is tacitly assumed
from the very beginning. Quantum model ${\mathcal S}_q$ will thus always
depend on some classical elements. This does not mean that classicality has
been smuggled in because, in our approach, classical properties are specific
quantum ones. 3) A Hamiltonian operator of the system must be set up. Finally,
the known classical properties of ${\mathcal S}_c$ must be listed and each
derived as an objective property of ${\mathcal S}_q$ from the three sets of
assumptions above. This is a self-consistent framework for a non-trivial
problem.

It follows that there must be some at least approximate relation between
classical observables of ${\mathcal S}_c$ and quantum observables of
${\mathcal S}_q$ as well as between the classical states of ${\mathcal S}_c$
and quantum states of ${\mathcal S}_q$. The following model assumption on such
a relation might be viewed as our version of Correspondence Principle, let us
call it Modified Correspondence Principle (MCP).

\begin{assump} First, the state of classical system ${\mathcal S}_c$ in a
given classical theory is always described by a set of $n$ numbers
$\{a_1,\cdots,a_n\}$ (state coordinates) that represent values of some
classical observables. Let us assume that there are preparation procedures
that make the fuzziness of {\em real} classical states negligible. This can be
described by considering the values $\{a_1,\cdots,a_n\}$ as {\em averages}
with {\em variances} $\{\Delta a_1,\cdots,\Delta a_n\}$ that are relatively
small,
\begin{equation}\label{qss} \frac{\text{Max}_k\{\Delta
a_k\}}{\text{Max}_k\{a_k\}} \ll 1\ .
\end{equation} Such states are the best we can achieve on the way to sharp
classical states.

Second, we assume that there is a subset $\{{\mathsf a}_1,\cdots,{\mathsf
a}_n\}$ of sharp observables of quantum system ${\mathcal S}_q$ and a state
${\mathsf T}$ such that
\begin{equation}\label{MCP} tr[{\mathsf T} {\mathbf a}_k] = a_k, \quad
\sqrt{tr[{\mathsf T} {\mathsf a}_k^2)-(tr[{\mathsf T} {\mathbf a}_k])^2} =
\Delta a_k.
\end{equation} All such states can form a large subset of ${\mathbf
T}({\mathbf H}_{{\mathcal S}_q})^+_1$. Each of them can be considered as a
quantum counterpart of the original classical state.

Third, there are fuzzy classical states with a probability distribution
$\rho(a_1,\cdots,a_n)$ that either need not have negligible variances or need
not be determined by its averages and variances, or both. However, we assume
that every classical state of this kind results from a preparation procedure
such that the resulting state ${\mathbf T}$ that can be interpreted as a
gemenge of some states with small variances and that $\rho(a_1,\cdots,a_n)$
can be well approximated by the coefficients of such a gemenge.
\end{assump}

Let us give an example. The classical state of a mass point in mechanics can
be described by three coordinates $q^k = a_k $ and three momenta $p_k =
a_{3+k}$, and the operator algebra of its quantum model contains the operators
${\mathsf q}^k = {\mathsf a}_k$ and ${\mathsf p}_k = {\mathsf a}_{3+k}$.

Of course, we can define a sensible classical state only for macroscopic
systems so that their classical states have a huge number of quantum
counterparts and in this way contain much less information than their quantum
states. Each of the many quantum states satisfying the above equations can be
viewed as representing one and the same classical state.

Clearly, MCP does not need the assumption that all observables from
$\{{\mathsf a}_1,\cdots,{\mathsf a}_n\}$ commute with each other. Then, even
if they themselves are not jointly measurable, their fuzzy values can be (see
Sec.\ 4.4).

Moreover, it does not follow that each classical property is an average of a
quantum operator. That would be false. We assume only that the classical state
coordinates ${a_1,...,a_n}$ can be chosen in such a way.

It is important to realise that MCP is just compatible with but does not imply
MPMR. It suggests the way of how quantum models of classical properties can be
constructed. In each such model, validity of MPMR ought to be achieved
independently.

We emphasize that our approach to classical properties is very different from
the well-known WKB approximation method. In particular, pure states such as
coherent ones that have averages with small variances do not satisfy MPMR. Not
only are pure states readily linearly superposed but also any quantum
registration that were to find the parameters of a coherent state (a
generalized measurement: positive operator valued measure) would strongly
change the state (for a general argument, see Ref.\ \cite{BLM}, p. 32).

An interesting subset of classical properties of macroscopic system are the
thermodynamic ones. They are important for us because quantum models of these
properties are available. Moreover, quantum models of this kind of classical
properties are compatible with Copenhagen interpretation
\cite{ludwig2}. Existing models based on statistical physics need one
non-trivial assumption: the states of sufficiently small macroscopic systems
that we observe around us are approximately states of maximum entropy. As it
has been discussed in Sec.\ 3.1, entropy is an objective property of quantum
systems because it is defined by their preparation. Thus, the validity of
thermodynamics depends on the preparation conditions, or the origin, of
observed macroscopic systems. The averages that result from the models based
on the maximum-entropy assumption agree with observations and have small
variances spontaneously. In this way, they comply with MCP. We describe a
simplified model in the next section.

The physical foundations of thermodynamics are not yet well-understood but
there are many ideas around. They might follow partially from logic (Bayesian
approach, \cite{Jaynes}) and partially from quantum mechanics (thermodynamic
limit, \cite{thirring}, Vol.\ 4). Some very interesting models of how maximum
entropy quantum states come into being are based on entanglement \cite{GMM,
short1,short2,goldstein}).

In any case, statistical methods can be generalised as follows
\cite{hajicek}. We assume that the maximum entropy principle underlies all
classical properties:
\begin{assump} An overwhelming part of macroscopic systems occur in quantum
states that approximately maximize entropy under the condition of given
averages of some quantum observables.
\end{assump}

Let us briefly compare our approach with the theories mentioned at the
beginning of this section. It starts with the idea that all classical
properties of a macroscopic system ${\mathcal S}$ in a quantum state $\mathsf
T$ are certain {\em objective} quantum properties of ${\mathcal S}$ in
$\mathsf T$. Then, first, objective properties are quantum properties of all
quantum systems and there is no question about how they emerge in quantum
mechanics. This can avoid e.g. the artificial construction in the Coleman-Hepp
approach. The new point is that they are considered as, and proved to be,
objective \cite{PHJT}. Hence, second, they could in principle serve as
classical properties because they can satisfy MPMR. This avoids the problems
of both the quantum-decoherence and the coarse-grained theory that assume
values of quantum observables to be objective. Such an assumption, as analysed
in \cite{bub}, works only for restricted classes of observables, all other
measurements being forbidden without an independent justification. Third, we
conjecture that certain macroscopic quantum systems possess objective
properties, either structural or dynamical, that can fully model their
classical behaviour. Hence, classical states and properties defined in the
present paper are available only for some quantum systems and the relation
between classical and quantum states is not one-to-one but one-to-many. This
is different from other approaches such as Wigner-Weyl-Moyal scheme,
quantum-mechanics-on-phase-space theory or coherent state approach. Finally,
our construction of classical properties uses the way analogous to that of
statistical physics. Thus, our MCP is a basis of modelling classical
properties of quantum systems.

\section{A model of thermodynamic properties} Here, we shall construct the
length of a body as an average value with a small variance. No original
calculation is to be expected, but simple and well known ideas are carefully
interpreted. This entails that, first, the quantum structure of the system
must be defined, second, the structural properties such as the energy spectrum
calculated, third, some assumptions on the state of the system done and
fourth, some dynamical properties derived that satisfy our definition of
classical property.

\subsection{Composition, Hamiltonian and spectrum} We shall consider a linear
chain of $N$ identical particles of mass $\mu$ distributed along the $x$-axis
with the quantum Hamiltonian
\begin{equation}\label{Ham} {\mathsf H} = \frac{1}{2\mu}\sum_{n=1}^N {\mathsf
p}_n^2 + \frac{\kappa^2}{2}\sum_{n=2}^N ({\mathsf x}_n - {\mathsf x}_{n-1} -
\xi)^2,
\end{equation} involving only nearest-neighbour elastic forces. Here operator
${\mathsf x}_n$ is the position, operator ${\mathsf p}_n$ the momentum of the
$n$-th particle, $\kappa$ the oscillator strength and $\xi$ the equilibrium
interparticle distance. The parameters $\mu$, $\kappa$ and $\xi$ are
structural properties (the last two defining the potential function).

This kind of chain seems to be different from most that are studied in
literature: the positions of the chain particles are dynamical variables so
that the chain can move as a whole. However, the chain can still be solved by
methods that are described in \cite{Kittel,Rutherford}.

First, we find the variables ${\mathsf u}_n$ and ${\mathsf q}_n$ that
diagonalize the Hamiltonian and define normal modes. The transformation is
\begin{equation}\label{xu} {\mathsf x}_n = \sum_{m=0}^{N-1}Y^m_n{\mathsf u}_m
+ \left(n - \frac{N+1}{2}\right)\xi,
\end{equation} and
\begin{equation}\label{pu} {\mathsf p}_n = \sum_{m=0}^{N-1}Y^m_n{\mathsf q}_m,
\end{equation} where the mode index $m$ runs through $0,1,\cdots,N-1$ and
$Y^m_n$ is an orthogonal matrix; for even $m$,
\begin{equation}\label{evenm} Y^m_n = A(m)\cos\left[\frac{\pi
m}{N}\left(n-\frac{N+1}{2}\right)\right],
\end{equation} while for odd $m$,
\begin{equation}\label{oddm} Y^m_n = A(m)\sin\left[\frac{\pi
m}{N}\left(n-\frac{N+1}{2}\right)\right]
\end{equation} and the normalization factors are given by
\begin{equation}\label{factor} A(0) = \frac{1}{\sqrt{N}},\quad A(m) =
\sqrt{\frac{2}{N}},\quad m>0.
\end{equation}

To show that ${\mathsf u}_n$ and ${\mathsf q}_n$ do represent normal modes, we
substitute Eqs.\ (\ref{xu}) and (\ref{pu}) into (\ref{Ham}) and obtain, after
some calculation,
$$
{\mathsf H} = \frac{1}{2\mu}\sum_{m=0}^{N-1}{\mathsf q}_m^2 +
\frac{\mu}{2}\sum_{m=0}^{N- 1}\omega_m^2{\mathsf u}_m^2,
$$
which is indeed diagonal. The mode frequencies are
\begin{equation}\label{spectr} \omega_m =
\frac{2\kappa}{\sqrt{\mu}}\sin\frac{m}{N}\frac{\pi}{2}.
\end{equation}

Consider the terms with $m=0$. We have $\omega_0=0$, and
$Y^0_n=1/\sqrt{N}$. Hence,
$$
{\mathsf u}_0 = \sum_{n=1}^N\frac{1}{\sqrt{N}}{\mathsf x}_n, \quad {\mathsf
q}_0 = \sum_{n=1}^N\frac{1}{\sqrt{N}}{\mathsf p}_n,
$$
so that
$$
{\mathsf u}_0 = \sqrt{N}{\mathsf X},\quad {\mathsf q}_0 =
\frac{1}{\sqrt{N}}{\mathsf P},
$$
where ${\mathsf X}$ is the centre-of-mass coordinate of the chain and
${\mathsf P}$ is its total momentum. The 'zero' terms in the Hamiltonian then
reduce to
$$
\frac{1}{2M}{\mathsf P}^2
$$
with $M = N\mu$ being the total mass. Thus, the 'zero mode' describes a
straight, uniform motion of the chain as a whole. The other modes are harmonic
oscillators called 'phonons' with eigenfrequencies $\omega_m$, $m =
1,2,\dots,N-1$. The energy spectrum of our system is built from the mode
frequencies by the formula
\begin{equation}\label{phonons} E = \sum_{m=1}^{N-1}\nu_m \hbar\omega_m,
\end{equation} where $\{\nu_m\}$ is an $(N-1)$-tuple of non-negative
integers---phonon occupation numbers.

At this stage, a new and important assumption must be done. We imagine that
all states ${\mathsf T}$ of the modes $m = 1,\cdots,N-1$ are prepared that
have the same average internal energy $E$,
$$
\text{Tr}\left[{\mathsf T}\left(H-\frac{P^2}{2M}\right)\right] = E.
$$
We further assume that it is done in a perfectly random way, i.e., all other
conditions or bias are to be excluded. Hence, the resulting state must
maximize the entropy. In this way, the maximum of entropy does not represent
an additional condition but rather the absence of any, see Sec.\ 3.1. The
resulting state ${\mathsf T}_E$ is the Gibbs state of the internal degrees of
freedom. The conditions that define the preparation of Gibbs state are
objective and need not be connected with human laboratory activity.

The internal energy has itself a very small relative variance in the Gibbs
state; this need not be assumed from the start. Thus, it is a classical
property. All other classical internal properties will turn out to be
functions of the classical internal energy. Hence, for the internal degrees of
freedom, $E$ forms itself a complete set of state coordinates introduced in
Assumption 4.

The mathematics associated with the maximum entropy principle is variational
calculus. The condition of fixed average energy is included with the help of
Lagrange multiplier denoted by $\lambda$. It becomes a function $\lambda(E)$
for the resulting state. As it is well known, $\lambda(E)$ has to do with
temperature.

The phonons of one species are excitation levels of a harmonic oscillator, so
we have
$$
{\mathsf u}_m = \sqrt{\frac{\hbar}{2\mu\omega_m}}({\mathsf a}_m + {\mathsf
a}^\dagger_m),
$$
where ${\mathsf a}_m$ is the annihilation operator for the $m$-th species. The
diagonal matrix elements between the energy eigenstates $\mid\nu_m\rangle$
that we shall need then are
\begin{equation}\label{averu} \langle\nu_m|{\mathsf u}_m|\nu_m\rangle =
0,\quad \langle\nu_m| {\mathsf u}^2_m|\nu_m\rangle =
\frac{\hbar}{2\mu\omega_m}(2\nu_m + 1).
\end{equation}

For our system, the phonons of each species form statistically independent
subsystems, hence the average of an operator concerning only one species in
the Gibbs state ${\mathsf T}_{\bar{E}}$ of the total system equals the average
in the Gibbs state for the one species. Such a Gibbs state operator for the
$m$-th species has the form
$$
{\mathsf T}_m = \sum_{\nu_m=0}^{\infty}|\nu_m\rangle
p_{\nu_m}^{(m)}\langle\nu_m|,
$$
where
$$
p_{\nu_m}^{(m)} = Z^{-1}_{m}\exp\left(-\hbar\lambda\omega_m\nu_m\right)
$$
and $Z_m$ is the partition function for the $m$-th species
\begin{equation}\label{partf} Z_{m}(\lambda) = \sum_{\nu_{m}=0}^\infty
e^{-\lambda\hbar\omega_m\nu_m} = \frac{1}{1-e^{-\lambda\hbar\omega_m}}.
\end{equation}

\subsection{The length of the body} The classical property that will be
defined and calculated in our quantum model is the average length of the
body. Let us define the length operator by
\begin{equation}\label{length} {\mathsf L} = {\mathsf x}_N - {\mathsf x}_1.
\end{equation} It can be expressed in terms of modes ${\mathsf u}_m$ using
Eq.~(\ref{xu}),
$$
{\mathsf L} = (N-1)\xi + \sum_{m=0}^{N-1}(Y^m_N-Y^m_1){\mathsf u}_m.
$$
The differences on the right-hand side are non-zero only for odd values of
$m$, and equal then to $-2Y^m_1$. We easily find, using Eqs.~(\ref{oddm}) and
(\ref{factor}):
\begin{equation}\label{L} {\mathsf L} = (N-1)\xi - \sqrt{\frac{8}{N}}\
\sum_{m=1}^{[N/2]}(-
1)^m\cos\left(\frac{2m-1}{N}\frac{\pi}{2}\right)\,{\mathsf u}_{2m-1}.
\end{equation}

The average length is obtained using Eq.\ (\ref{averu}),
\begin{equation}\label{averL} \langle {\mathsf L}\rangle_E = (N-1)\xi.
\end{equation} It is a function of objective properties $N$, $\xi$ and $E$.

Eq.\ (\ref{L}) is an important result. It shows that contributions to the
length are more or less evenly distributed over all odd modes. Such a
distribution leads to a very small variance of ${\mathsf L}$ in Gibbs
states. A lengthy calculation \cite{PHJT} gives for large $N$
\begin{equation} \frac{\Delta {\mathsf L}}{\langle {\mathsf L}\rangle_E}
\approx \frac{2\sqrt{3}}{\pi\kappa\xi\sqrt{\lambda}}\frac{1}{\sqrt{N}}.
\end{equation}

Thus, the small relative variance for large $N$ need not be assumed from the
start. The only assumptions are values of some structural properties and that
an average value of energy is fixed. In the sense explained in Section 5.1,
the length is then a classical property of our model body. We have obtained
even more information, viz.\ the internal-energy dependence of the length (in
this model, the dependence is trivial). This is an objective relation that can
be in principle tested by measurements.

Similar results can be obtained for further thermodynamic properties such as
elasticity coefficient, specific heat etc. They all are well known to have
small variances in Gibbs states. The reason is that the contributions to these
quantities are evenly distributed over the normal modes and the modes are
mechanically and statistically independent. Further important classical
properties are the mechanical ones: centre of mass and total momentum. In
fact, these quantities can be chosen as the remaining state coordinates for
the whole chain. The contributions to them are evenly distributed over all
atoms, not modes: the bulk motion is mechanically and statistically
independent of all other modes and so its variances will not be small in Gibbs
states. Still, generalized statistical methods can be applied to it. This is
done in the next section.

The last remark is that the thermodynamic equilibrium can settle down starting
from an arbitrary state only if some weak but non-zero interaction exists
between the phonons. We assume that this can be arranged so that the influence
of the interaction on our result is negligible.

\section{Maximum entropy assumption \\ in classical mechanics} Let us start
with the warning that the subject of this section has nothing to do with what
is usually called 'statistical mechanics'.

If one is going to model classical mechanics then what are the properties that
one would like to reproduce? The most conspicuous property from the point of
view of quantum mechanics appears to be the sharpness of mechanical
trajectories in the phase space because quantum mechanics denies the existence
of such trajectories. This leads most researchers to aim at quantum states the
phase-space picture of which is as sharp as possible. That are states with
minimum uncertainty allowed by quantum mechanics. For one degree of freedom,
described by coordinate ${\mathsf q}$ and momentum ${\mathsf p}$, the
uncertainty is given by the quantity
\begin{equation}\label{uncertainty} \nu = \frac{2\Delta {\mathsf q}\Delta
{\mathsf p}}{\hbar},
\end{equation} where $\Delta {\mathsf a}$ denotes the variance of quantity
${\mathsf a}$, as defined by Eq.\ (\ref{variance}).

The states with $\nu = 1$ are, however, very special states. First, they must
be pure states such as Gaussian wave packets or coherent states. Such states
are very difficult to prepare unlike the usual states of macroscopic systems
described by classical mechanics. They are also prone to strong distortion by
measurements. Moreover, as pure states, they can be linearly superposed. This
is another peculiarity that is never observed for states of systems of
classical mechanics. Hence, trying to get a trajectory as sharp as possible
entails the loss of other desirable properties.

Moreover, the systems to which classical mechanics is applicable, are
macroscopic. The overwhelming number of their degrees of freedom decouple from
their mechanical ones, of which there are always only very few. The purely
mechanical degrees of freedom have to do, e.g., with the motion of the centre
of mass of a small but macroscopic body. As explained at the beginning of this
chapter, the sharpness of phase-space trajectories is just an idealisation, a
limit in which things become mathematically simpler. We can use it in
calculations which, however, must also take into account the necessary
non-zero variances of real observations. Indeed, such observations are
generally afflicted with uncertainties $\nu \gg 1$. Hence, if we want to
compare the predictions of quantum models with observations of classical
mechanics, we are forced to compare states that are fuzzy in both theories.

Our main idea is to consider states with given averages and variances of
coordinates and momenta, $Q^k$, $\Delta Q^k$, $P_k$, $\Delta P_k$, and leave
everything else as fuzzy as possible. Thus, the averages of coordinates, of
their squares, of momenta, and of their squares are fixed. To calculate the
corresponding probability distributions in classical, and the state operators
in quantum mechanics, we shall, therefore, apply the maximum entropy
principle. The resulting states are called {\em maximum-entropy packets}, ME
packets. The averages of coordinates and momenta take over the role of
coordinate and momenta in classical mechanics. In any case these averages
represent measurable aspects of these variables. The dynamical evolution of
variances is an important indicator of the applicability of the model one is
working with. It determines the time intervals within which reasonable
predictions are possible.

Consider a three-body system that is to model the Sun, Earth and Jupiter in
Newton mechanics. It turns out that generic trajectories starting as near to
each other as, say, the dimension of the irregularities of the Earth surface
will diverge from each other by dimensions of the Earth-Sun distance after the
time of only about $10^7$ years. This seems to contradict the $4\times
10^{12}$ years of relatively stable Earth motion around the Sun that is born
out by observations. The only way out is the existence of a few special
trajectories that are much stabler than the generic ones and the fact that
bodies following an unstable trajectory have long ago fallen into the Sun or
have been ejected from the solar system. This spontaneous evolution can be
considered as a preparation procedure of solar system.

As explained in Sec.\ 1.1.2, a prepared state can be viewed as objective. A
simple example is a gun in a position that is fixed in a reproducible way and
that shoots bullets using cartridges of a given provenance. The state of each
individual shot is defined by the conditions and is the same for all shots
even if observations may have different results for different shots. In the
theoretical description of a state, we can make the limit of $\Delta Q^k
\rightarrow 0,\Delta P_k \rightarrow 0$. This is considered as a non-existing,
but practically useful idealization.

A finer analysis is possible only as long as new preparations are specified
that determine a subset of individual shots. For example, if the conditions on
the gun are that the gun is held by a human hand, then one cause of the
relatively large $\Delta Q^k$ and $\Delta P_k$ is that the gun is always in a
slightly different position because the hand and eye are not completely
sure. The cause is 'classical' in its nature and one can imagine that each
shot will be repeated several times more under the condition of a mechanical
fixation. Then, there will still be variances but they may be much
smaller. Thus, the original preparation can be considered as composed of
sub-preparations in the way of gemenge.

In general, if a part of the variances in a state come about as random or
controlled changes of classical conditions in the preparation procedure, then
the preparation is clearly composed of sub-preparations and the state a
gemenge. Some variances will still remain, but they will be 'sufficiently'
small for the corresponding states to be viewed as defined by the
corresponding averages and negligible variances.

To limit ourselves just to given averages and variances of coordinates and
momenta is a great simplification that enables us to obtain interesting
results easily.

\section{Classical ME packets} Let us first consider a system ${\mathcal S}$
with one degree of freedom (the generalization to any number is easy). Let the
coordinate be $q$ and the momentum $p$. A state is a distribution function
$\rho(q,p)$ on the phase space spanned by $q$ and $p$. The function
$\rho(q,p)$ is dimensionless and normalized by
$$
\int\frac{dq\,dp}{v}\,\rho = 1\ ,
$$
where $v$ is an auxiliary phase-space volume to make the integration
dimensionless. The entropy of $\rho(q,p)$ can be defined by
$$
S := -\int\frac{dq\,dp}{v}\,\rho \ln\rho\ .
$$
The value of entropy will depend on $v$ but most other results will
not. Classical mechanics does not offer any idea of how to fix $v$. We shall
get its value from quantum mechanics.

\subsection{Definition and properties}
\begin{df} ME packet is the distribution function $\rho$ that maximizes the
entropy subject to the conditions:
\begin{equation}\label{21.4} \langle q\rangle = Q\ ,\quad \langle q^2\rangle =
\Delta Q^2 + Q^2\ ,
\end{equation} and
\begin{equation}\label{21.5} \langle p\rangle = P\ ,\quad \langle p^2\rangle =
\Delta P^2 + P^2\ ,
\end{equation} where $Q$, $P$, $\Delta Q$ and $\Delta P$ are given values.
\end{df} We have used the abbreviation
$$
\langle x\rangle = \int\frac{dq\,dp}{v}\,x\rho\ .
$$

The explicit form of $\rho$ can be found using the partition-function method
\cite{hajicek}. The resulting partition function for classical ME packets is
\begin{equation}\label{22.1} Z=
\frac{\pi}{v}\frac{1}{\sqrt{\lambda_3\lambda_4}}
\exp\left(\frac{\lambda_1^2}{4\lambda_3} +
\frac{\lambda_2^2}{4\lambda_4}\right)\ .
\end{equation} The four Lagrange multipliers $\lambda_1$, $\lambda_3$,
$\lambda_2$ and $\lambda_4$ correspond to the four conditions (\ref{21.4}) and
(\ref{21.5}).

The generalization to any number of dimensions is trivial.
\begin{prop} The distribution function of the ME packet for a system with
given averages and variances $Q_1,\cdots,Q_n$, $\Delta Q_1,\cdots,\Delta Q_n$
of coordinates and $P_1,\cdots,P_n$, $\Delta P_1,\cdots,\Delta P_n$ of
momenta, is
\begin{equation}\label{23.1} \rho =
\left(\frac{v}{2\pi}\right)^n\prod_{k=1}^n\left(\frac{1}{\Delta Q_k\Delta
P_k}\exp\left[-\frac{(q_k-Q_k)^2}{2\Delta Q_k^2} -\frac{(p_k-P_k)^2}{2\Delta
P_k^2}\right]\right)\ .
\end{equation}
\end{prop} For proof, see Ref.\ \cite{hajicek}. We observe that all averages
obtained from $\rho$ are independent of $v$ and that the result is a Gaussian
distribution in agreement with Jaynes' conjecture that the maximum entropy
principle gives the Gaussian distribution if the only conditions are fixed
values of the first two moments.

As $\Delta Q$ and $\Delta P$ approach zero, $\rho$ becomes a delta-function
and the state becomes sharp. For some quantities this limit is sensible for
others it is not. In particular, the entropy, which can easily be calculated,
$$
S = 1 + \ln\frac{2\pi\Delta Q\Delta P}{v}\ ,
$$
diverges to $-\infty$. This is due to a general difficulty in giving a
definition of entropy for a continuous system that would be satisfactory in
every respect. What one could do is to divide the phase space into cells of
volume $v$ so that $\Delta Q\Delta P$ could not be chosen smaller than
$v$. Then, the limit $\Delta Q\Delta P \rightarrow v$ of entropy would make
more sense.

\subsection{Classical equations of motion} Let us assume that the Hamiltonian
of ${\mathcal S}$ has the form
\begin{equation}\label{35.1} H = \frac{p^2}{2\mu} + V(q)\ ,
\end{equation} where $\mu$ is the mass and $V(q)$ the potential function. The
equations of motion are
$$
\dot{q} = \{q,H\}\ ,\quad\dot{p} = \{p,H\}\ .
$$
Inserting (\ref{35.1}) for $H$, we obtain
\begin{equation}\label{36.6} \dot{q} = \frac{p}{\mu}\ ,\quad\dot{p} =
-\frac{dV}{dq}\ .
\end{equation} The general solution to these equations can be written in the
form
\begin{equation}\label{36.65} q(t) = q(t;q,p)\ ,\quad p(t) = p(t;q,p)\ ,
\end{equation} where
\begin{equation}\label{36.3} q(0;q,p) = q\ ,\quad p(0;q,p) = p\ ,
\end{equation} $q$ and $p$ being arbitrary initial values. We obtain the
equations of motion for the averages and variances if the initial state is an
ME packet:
\begin{equation}\label{36.7} Q(t) = \langle q(t; q,p)\rangle\ ,\quad \Delta
Q(t) = \sqrt{\langle (q(t;q,p)- Q(t))^2\rangle}
\end{equation} and
\begin{equation}\label{36.8} P(t) = \langle p(t; q,p)\rangle\ ,\quad \Delta
P(t) = \sqrt{\langle (p(t;q,p)- P(t))^2}\rangle\ .
\end{equation} In general, $Q(t)$ and $P(t)$ will depend not only on $Q$ and
$P$, but also on $\Delta Q$ and $\Delta P$.

Let us consider the special case of at most quadratic potential:
\begin{equation}\label{36.1} V(q) = V_0 + V_1 q + \frac{1}{2} V_2 q^2\ ,
\end{equation} where $V_k$ are constants with suitable dimensions. If $V_1 =
V_2 =0$, we have a free particle, if $V_2 = 0$, it is a particle in a
homogeneous force field and if $V_2 \neq 0$, it is an harmonic or
anti-harmonic oscillator.

In this case, the general solution has the form
\begin{eqnarray}\label{37.1} q(t) &=& f_0(t) + q f_1(t) + p f_2(t)\ , \\
\label{37.2} p(t) &=& g_0(t) + q g_1(t) + p g_2(t)\ ,
\end{eqnarray} where $f_0(0) = f_2(0) = g_0(0) = g_1(0) = 0$ and $f_1(0) =
g_2(0) = 1$. If $V_2 \neq 0$, the functions are
\begin{equation}\label{37.4} f_0(t) = -\frac{V_1}{V_2}(1-\cos\omega t)\ ,\quad
f_1(t) = \cos \omega t\ ,\quad f_2(t) = \frac{1}{\xi}\sin\omega t\ ,
\end{equation}
\begin{equation}\label{37.5} g_0(t) = -\xi\frac{V_1}{V_2}\sin\omega t\ ,\quad
g_1(t) = -\xi\sin \omega t\ ,\quad g_2(t) = \cos\omega t\ ,
\end{equation} where
$$
\xi = \sqrt{\mu V_2}\ ,\quad \omega = \sqrt{\frac{V_2}{\mu}}\ .
$$
Only for $V_2 > 0$, the functions remain bounded. If $V_2 = 0$, we obtain
\begin{equation}\label{37.9a} f_0(t) = -\frac{V_1}{2\mu}t^2\ ,\quad f_1(t) =
1\ ,\quad f_2(t) = \frac{t}{\mu}\ ,
\end{equation}
\begin{equation}\label{37.9b} g_0(t) = -V_1t\ ,\quad g_1(t) = 0\ ,\quad g_2(t)
= 1\ .
\end{equation}

The equations for averages and variances resulting from Eqs.\ (\ref{36.65}),
(\ref{21.4}) and (\ref{21.5}) are \cite{hajicek}
\begin{equation}\label{38.3} Q(t) = f_0(t) + Q f_1(t) + P f_2(t)\ ,
\end{equation}
\begin{equation}\label{39.1} \Delta Q(t) = \sqrt{f_1^2(t)\Delta Q^2 +
f_2^2(t)\Delta P^2}\ ,
\end{equation}
\begin{eqnarray}\label{39.2} P(t) &=& g_0(t) + Q g_1(t) + P g_2(t)\ ,\\
\label{39.3} \Delta P(t) &=& \sqrt{f_g^2(t)\Delta Q^2 + g_2^2(t)\Delta P^2}\ .
\end{eqnarray} We observe: if functions $f_1(t)$, $f_2(t)$, $g_1(t)$ and
$g_2(t)$ remain bounded, the variances also remain bounded and the predictions
are possible in arbitrary long intervals of time. Otherwise, there will always
be only limited time intervals in which the theory can make predictions.

In the case of general potential, the functions (\ref{36.65}) can be expanded
in products of powers of $q$, $p$ and $t$, and the averages of these products
can be calculated by partition function method. In Ref.\ \cite{hajicek},
equations of motion are calculated up to fourth order.

\section{Quantum ME packets} Let us now turn to quantum mechanics and try to
solve an analogous problem.

\subsection{Definition and properties}
\begin{df} Let the quantum model ${\mathcal S}_q$ of system ${\mathcal S}$ has
the basic observables ${\mathsf q}$ and ${\mathsf p}$. State ${\mathsf T}$
that maximizes von Neumann entropy (see Sec.\ 3.1)
\begin{equation}\label{VNE} S = -tr({\mathsf T}\ln{\mathsf T})
\end{equation} under the conditions
\begin{equation}\label{12.1} tr[{\mathsf T}{\mathsf q}] = Q\ ,\quad
tr[{\mathsf T} {\mathsf q}^2] = Q^2 + \Delta Q^2\ ,
\end{equation}
\begin{equation}\label{12.2} tr[{\mathsf T}{\mathsf p}] = P\ ,\quad
tr[{\mathsf T} {\mathsf p}^2] = P^2 + \Delta P^2\ ,
\end{equation} where $Q$, $P$, $\Delta Q$ and $\Delta P$ are given numbers, is
called quantum ME packet.
\end{df}

To calculate the state operator of a ME packet, one can use the partition
function method. However, it must be modified because the involved quantities
do not commute, see Ref.\ \cite{hajicek}. The partition function for the
quantum ME packets is
\begin{equation}\label{17.3} Z =
\frac{\exp\left(\frac{\lambda_1^2}{4\lambda_3} +
\frac{\lambda_2^2}{4\lambda_4}\right)}{2\sinh(\hbar\sqrt{\lambda_3\lambda_4})}\
.
\end{equation}

The resulting state operator, generalised to $n$ degrees of freedom is
described by the following
\begin{prop} The state operator of the ME packet of a system with given
averages and variances $Q_1,\cdots,Q_n$, $\Delta Q_1,\cdots,\Delta Q_n$ of
coordinates and $P_1,\cdots,P_n$, $\Delta P_1,\cdots, \Delta P_n$ of momenta,
is
\begin{equation}\label{32.1} {\mathsf T}=
\prod_{k=1}^n\left[\frac{2}{\nu_k^2-1}\exp\left(-\frac{1}{\hbar}\ln\frac{\nu_k+1}{\nu_k-1}{\mathsf
K}_k\right)\right]\ ,
\end{equation} where
\begin{equation}\label{20.3} {\mathsf K}_k = \frac{1}{2}\frac{\Delta
P_k}{\Delta Q_k}({\mathsf q}_k-Q_k)^2 + \frac{1}{2}\frac{\Delta Q_k}{\Delta
P_k}({\mathsf p}_k-P_k)^2
\end{equation} and
\begin{equation}\label{20.3b} \nu_k = \frac{2\Delta P_k\Delta Q_k}{\hbar}\ .
\end{equation}
\end{prop} For the proof, see Ref.\ \cite{hajicek}. The calculation of the
state operator is simplified by the fact that ${\mathsf K}_k$ is Hamiltonian
operator of a harmonic oscillator. Of course, system ${\mathcal S}_q$ for
which the calculation is done, is completely general and its Hamiltonian
operator has nothing to do with ${\mathsf K}_k$.

Strictly speaking, the state operator (\ref{32.1}) is not a Gaussian
distribution. Thus, it seems to be either a counterexample to, or a
generalization of, Jaynes' hypothesis.

Everything can be decomposed into well-known eigenstates $|m\rangle$ of
${\mathsf K}$. We have ($n=1$)
\begin{equation}\label{20.2a} {\mathsf K} = \sum_{k=0}^\infty
R_m|m\rangle\langle m|\ .
\end{equation} We easily obtain for $R_m$ that
\begin{equation}\label{20.2} R_m = 2\frac{(\nu-1)^m}{(\nu+1)^{m+1}} .
\end{equation} Hence,
$$
\lim_{\nu\rightarrow 1}R_m = \delta_{m0}\ ,
$$
and the state ${\mathsf T}$ becomes $|0\rangle\langle 0|$. In general, states
$|m\rangle$ depend on $\nu$. The state vector $|0\rangle$ expressed as a
function of $Q$, $P$, $\Delta Q$ and $\nu$ is given by
\begin{equation}\label{20.1} \psi(q) = \left(\frac{1}{\pi} \frac{\nu}{2\Delta
Q^2}\right)^{1/4} \exp\left[-\frac{\nu}{4\Delta Q^2}(q-Q)^2 +
\frac{iPq}{\hbar}\right]\ .
\end{equation} This is a Gaussian wave packet that corresponds to other values
of variances than the original ME packet but has the minimal uncertainty. For
$\nu\rightarrow 1$, it remains regular and the projection $|0\rangle\langle
0|$ becomes the state operator of the original ME packet. Hence, Gaussian wave
packets are special cases of quantum ME packets.

\subsection{Quantum equations of motion} Let the Hamiltonian of ${\mathcal
S}_q$ be ${\mathsf H}$ and the unitary evolution group be ${\mathsf
U}(t)$. The dynamics in the Schr\"{o}dinger picture leads to the time
dependence of ${\mathsf T}$:
$$
{\mathsf T}(t) = {\mathsf U}(t){\mathsf T} {\mathsf U}(t)^\dagger\ .
$$
Substituting for ${\mathsf T}$ from Eq.\ (\ref{32.1}) and using a well-known
property of exponential functions, we obtain
\begin{equation}\label{32.2} {\mathsf T}(t) =
\frac{2}{\nu^2-1}\exp\left(-\frac{1}{\hbar}\ln\frac{\nu+1}{\nu-1}{\mathsf
U}(t){\mathsf K}{\mathsf U}(t)^\dagger\right)\ .
\end{equation}

In the Heisenberg picture, ${\mathsf T}$ remains constant, while ${\mathsf q}$
and ${\mathsf p}$ are time dependent and satisfy the equations
\begin{equation}\label{33.1} i\hbar\frac{d{\mathsf q}}{dt} = [{\mathsf
q},{\mathsf H}]\ ,\quad i\hbar\frac{d{\mathsf p}}{dt} = [{\mathsf p},{\mathsf
H}]\ .
\end{equation} They are solved by
$$
{\mathsf q}(t) = {\mathsf U}(t)^\dagger{\mathsf q}{\mathsf U}(t)\ ,\quad
{\mathsf p}(t) = {\mathsf U}(t)^\dagger{\mathsf p}{\mathsf U}(t)\ ,
$$
where ${\mathsf q}$ and ${\mathsf p}$ are the initial operators, ${\mathsf
q}={\mathsf q}(0)$ and ${\mathsf p}={\mathsf p}(0)$. The resulting operators
can be written in the form of operator functions analogous to classical
expressions (\ref{36.65}) so that Eqs.\ (\ref{36.7}) and (\ref{36.8}) can
again be used.

The example with potential function (\ref{36.1}) is solvable in quantum
theory, too, and we can use it for comparison with the classical dynamics as
well as for a better understanding of the ME packet dynamics. Eqs.\
(\ref{33.1}) have then the solutions given by (\ref{37.1}) and (\ref{37.2})
with functions $f_n(t)$ and $g_n(t)$ given by (\ref{37.4}) and (\ref{37.5}) or
(\ref{37.9a}) and (\ref{37.9b}). The calculation of the averages and variances
is analogous to the classical one and we obtain Eqs.\ (\ref{38.3}) and Eq.\
(\ref{39.1}) again. Similarly for ${\mathsf p}$, the results are given by Eqs.\
(\ref{39.2}) and (\ref{39.3}).

We have shown that the averages and variances of quantum ME packets have
exactly the same time evolution as those of classical ME packets in the
special case of at-most-quadratic potentials. From formulae (\ref{39.1}) and
(\ref{39.3}) we can also see an interesting fact. On the one hand, both
variances must increase near $t=0$. On the other, the entropy must stay
constant because the evolution of the quantum state is unitary. As the
relation between entropy and $\nu$ is fixed for ME packets, the ME packet form
is not preserved by the evolution (the entropy ceases to be maximal). This is
similar for Gaussian-packet form or for coherent-state form.

For general potentials, there will be two types of corrections to the dynamics
of the averages: terms containing the variances and terms containing
$\hbar$. To see these corrections, equations of motion have been expanded in
powers of ${\mathsf q}$, ${\mathsf p}$ and $t$ in Ref.\ \cite{hajicek}. The
conclusion has been that the quantum equations begin to differ from the
classical ones only in the higher order terms in $V$ or in the higher time
derivatives and the correction is of the second order in $1/\nu$. This seems
to be very satisfactory: our quantum model reproduces the classical dynamic
very well. Moreover, Eq.\ (\ref{20.1}) shows that Gaussian wave packets are
special cases of ME packets with $\nu=1$. Thus, they approximate classical
trajectories less accurately than ME packets with large $\nu$.

\section{Classical limit} Let us now look to see if our equations give some
support to the statement that $\nu\gg 1$ is the classical regime.

The quantum partition function (\ref{17.3}) differs from its classical
counterpart (\ref{22.1}) by the denominator
$\sinh(\hbar\sqrt{\lambda_3\lambda_4})$. If
\begin{equation}\label{ll} \hbar\sqrt{\lambda_3\lambda_4} \ll 1\ ,
\end{equation} we can write
$$
\sinh(\hbar\sqrt{\lambda_3\lambda_4}) = \hbar\sqrt{\lambda_3\lambda_4}[1 +
O((\hbar\sqrt{\lambda_3\lambda_4})^2)]
$$
The leading term in the partition function then is
$$
Z = \frac{\pi}{h}\frac{1}{\sqrt{\lambda_3\lambda_4}}
\exp\left(\frac{\lambda_1^2}{4\lambda_3} +
\frac{\lambda_2^2}{4\lambda_4}\right)\ ,
$$
where $h = 2\pi\hbar$. Comparing this with formula (\ref{22.1}) shows that the
two expressions are identical, if we set
$$
v = h\ .
$$
We can say that quantum mechanics gives us the value of $v$. Next, we have to
express condition (\ref{ll}) in terms of the averages and variances. The usual
relations between the Lagrange multipliers on the one hand and the averages
and variances on the other (see Ref.\ \cite{hajicek}) imply
$$
\hbar\sqrt{\lambda_3\lambda_4} = \frac{1}{2}\ln\frac{\nu+1}{\nu-1}\ .
$$
Hence, condition (\ref{ll}) is equivalent to
\begin{equation}\label{gg} \nu \gg 1\ .
\end{equation}

The result can be formulated as follows. Classical mechanics allows not only
sharp, but also fuzzy trajectories and the comparison of some classical and
quantum fuzzy trajectories shows a very good match. The fuzzy states chosen
here are the so-called ME packets. Their fuzziness is described by the
quantity $\nu = 2\Delta Q\Delta P/\hbar$. The entropy of an ME packet depends
only on $\nu$ and is an increasing function of it. The time evolution of
classical and quantum ME packets with the same initial values of averages and
variances defines the averages as time functions. The larger $\nu$ is, the
better the quantum and the classical evolutions of average values have been
shown to agree. Thus, the classical regime is neither $\Delta Q = \Delta P =
0$ (absolutely sharp trajectory) nor $\nu=1$ (minimum quantum
uncertainty). This is the most important result of Ref.\ \cite{hajicek}. The
time functions coincide for the two theories in the limit $\nu \rightarrow
\infty$. Hence, in our approach, this is the classical limit. It is very
different from the usual assumption that the classical limit must yield the
variances as small as possible. Of course, $\nu$ can be very large and still
compatible with classically negligible variances.

One also often requires that commutators of observables vanish in classical
limit. This is however only motivated by the assumption that all basic quantum
properties are single values of observables. Within our interpretation, this
assumption is rejected and if classical observables are related to quantum
operators then only by being average values of the operators in prepared
states. Then, first, all such averages are defined by a preparation and do
exist simultaneously, independently of whether the operators commute or
not. For example, $Q$ and $P$ are such simultaneously existing variables for
ME packets. Second, a joint measurement of fuzzy values of non-commuting
observables is possible. In Sec.\ 4.4, we have shown Eq. (\ref{factorT}),
which is a general formula valid for an arbitrary state ${\mathsf T}_{\mathcal
A}$ of the ancilla. Then, we have chosen ${\mathsf T}_{\mathcal A}$ to be a
Gaussian wave packet, which has $\nu = 1$. However, ${\mathsf T}_{\mathcal A}$
can also be chosen as a quantum ME packet with $Q = 0$, $P = 0$ and arbitrary
$\Delta Q$ and $\Delta P$ allowing arbitrary large $\nu$. The resulting
observable represents the detection of a shifted and boosted ME packet.

Let us compare the present paper notion of classical limit with a modern
textbook version such as Chap.\ 14 of \cite{ballentine}. Both approaches
define the classical limit of a quantum state as a classical ensemble
described by a fuzzy distribution function and calculate time evolutions of
averages in the states. However, in the textbook, any quantum system, even not
macroscopic, and any state, even pure, are allowed (pure states are preferred
as they have smaller uncertainties). Hence, our notion is much narrower: we
consider only macroscopic quantum systems and only some of their maximum
entropy states. This has obvious physical reasons explained in Secs.\ 5.1 and
5.3.

\chapter{Quantum models of preparation and registration of microsystems}
Discussions about the nature of quantum measurement were started already by
founding fathers of quantum theory, persisted throughout and seem even to
amplify at the present time.

An old approach to the problem of quantum measurement is Bohr's (its newer,
rigorously reformulated version is Ref.\ \cite{ludwig1}). This approach denies
that measuring apparatuses, and all classical systems in general, are quantum
systems in the sense that all their properties can be derived from, or are
compatible with, quantum mechanics. They must be described by other theories,
called pretheories. Of course some classical properties of macroscopic systems
can be obtained by quantum statistics. Ref.\ \cite{ludwig2} shows that such
occasional applications of quantum mechanics to classical systems are
compatible with the form of denying the universality of quantum mechanics
specified there.

Modern approaches assume the universality of quantum mechanics together with
various further ideas. A very brief account of the three most important
examples is:
\begin{enumerate}
\item Quantum decoherence theory \cite{Zeh,Zurek}. The idea is that system
${\mathcal S} + {\mathcal A}$ composite of a quantum object and an apparatus
cannot be isolated from environment ${\mathcal E}$. Then the unitary evolution
of ${\mathcal S} + {\mathcal A} + {\mathcal E}$ leads to a non-unitary
evolution of ${\mathcal S} + {\mathcal A}$ that can erase all correlations and
interferences from ${\mathcal S} + {\mathcal A}$ that hinder the
objectification (see the Introduction). However, then the correlations and
interferences are just transferred to ${\mathcal S} + {\mathcal A} + {\mathcal
E}$ and the method can only work, if some bizarre assumptions are made about
world, for instance one accepts some variant of the many world interpretation.
\item Superselection sectors approach \cite{Hepp,Primas}. Here, classical
properties are described by superselection observables of ${\mathcal A}$ which
commute with each other and with all other observables of it (see Sec.\
3.3). Then, the state of ${\mathcal A}$ after the measurement is equivalent to
that with a suitable gemenge structure. The problems are that, first, the
Hamiltonian of the measurement coupling cannot be an observable or else the
measurement model does not work. Second, the only known physical mechanism for
producing strict superselection rules occurs in systems having infinitely many
degrees of freedom. However, measuring apparatuses do not satisfy this
conditions and the superselection rules can therefore be only approximative.
\item Modal interpretation \cite{bub}. One assumes that there is a subset of
projections that have determinate values in the state (of ${\mathcal S} +
{\mathcal A}$) after the registration. The set of projections must just be
such that it does not violate contextuality (see Sec.\ 3.2.2), even if it can
violate the existence of correlations and interferences. Thus, one must
postulate that the measurements of the correlations and interferences on the
state are forbidden for some unknown reason.
\end{enumerate} In general, the problem is far from being satisfactorily
solved by any of the modern theories. Analysis of Refs.\
\cite{d'Espagnat,bub,BLM}, as well as of our previous papers
\cite{PHJT,hajicek}, give an account of their shortcomings. We adopt the
definition of the problem and the proof that it is far from being solved from
Ref.\ \cite{BLM}, Chap. IV, where the above short points are dealt with in
more detail and the corresponding references are given.

Our approach to the problem is very different from anything that has ever been
published in the field. It is based on a physical idea rather than on some
spectacular mathematics. It seems to promise a complete solution to the
problem. To explain the idea, we make some simplifying assumptions. First, we
restrict ourselves to measurements performed on microsystems, i.e.\ particles
and systems composite of few particles. There are other systems on which
recently a lot of interesting experiments has been done, such as Bose-Einstein
condensates, strong laser beams or currents in superconducting rings. Such
quantum states of 'large' systems, mezoscopic or even macroscopic, will be
ignored here. Second, we use a simplified model of measurement.

\section{Beltrametti-Cassinelli-Lahti model} In this section, we are going to
recapitulate the well-known ideas on measurement that will be needed later. A
summary is \cite{BLM}, p. 25: \begin{quote} \dots the object system ${\mathcal
S}$, prepared in a state ${\mathsf T}$ is brought into a suitable contact---a
{\em measurement coupling}---with another, independently prepared system, the
{\em measurement apparatus} from which the {\em result} related to the
measured observable ${\mathsf O}$ is {\em determined} by {\em reading} the
value of the {\em pointer observable}.  \end{quote} In Ref.\ \cite{BLM}, these
ideas are developed in detail with the help of models. One of them, which we
call Beltrametti-Cassinelli-Lahti (BCL) model (p.\ 38), is as follows. Let a
discrete observable ${\mathsf O}$ of system ${\mathcal S}$ of type $\tau$ with
Hilbert space ${\mathbf H}_\tau$ be measured. Let $o_k$ be eigenvalues and
$\{\phi_{kj}\}$ be the complete orthonormal set of eigenvectors of ${\mathsf
O}$,
$$
{\mathsf O}\phi_{kj} = o_k \phi_{kj}\ .
$$
We assume that $ k = 1,\cdots,N$ so that there is only a finite number of
different eigenvalues $o_k$. This is justified by the fact that no real
registration apparatus can distinguish all elements of an infinite set from
each other. It can therefore measure only a function of an observable that
maps its spectrum onto a finite set of real numbers. Our observable ${\mathsf
O}$ is such a function. The projection ${\mathsf E}^{\mathsf O}_k$ on the
eigenspace of $o_k$ is then ${\mathsf E}^{\mathsf O}_k = \sum_j
|\phi_{kj}\rangle\langle\phi_{kj}|$.

Let the registration apparatus\footnote{In our language, a measurement
consists of preparation and registration so that what Ref.\ \cite{BLM} often
calls 'measurement' is our 'registration'.} be a quantum system ${\mathcal A}$
with Hilbert space ${\mathbf H}_{\mathcal A}$ and an observable ${\mathsf
A}$. Let ${\mathsf A}$ be a non-degenerate, discrete observable with the same
eigenvalues $o_k$ and with orthonormal set of eigenvectors $\psi_k$,
$$
{\mathsf A}\psi_k = o_k \psi_k
$$
with possible further eigenstates (such as $\psi$ of Prop.\ 21) and
eigenvalues. The projection on an eigenspace is ${\mathsf E}^{\mathsf A}_k =
|\psi_k\rangle\langle\psi_k|$. ${\mathsf A}$ is the so-called {\em pointer
observable}.

Let the measurement start with the preparation of ${\mathcal S}$ in state
${\mathsf T}$ and the independent preparation of ${\mathcal A}$ in state
${\mathsf T}_{\mathcal A}$. The initial state of the composed system
${\mathcal S} + {\mathcal A}$ is thus ${\mathsf T}\otimes {\mathsf
T}_{\mathcal A}$.

Let ${\mathcal S}$ and ${\mathcal A}$ then interact for a finite time by the
so-called {\em measurement coupling} and let the resulting state be given by
${\mathsf U}({\mathsf T}\otimes {\mathsf T}_{\mathcal A}){\mathsf U}^\dagger$,
where ${\mathsf U}$ is a unitary transformation on ${\mathbf H}_\tau \otimes
{\mathbf H}_{\mathcal A}$.

The final state of the apparatus is $\Pi_{\mathcal S}\bigl[{\mathsf
U}({\mathsf T}\otimes {\mathsf T}_{\mathcal A}){\mathsf U}^\dagger\bigr]$,
where $\Pi_{\mathcal S}$ is the partial trace over states of ${\mathcal
S}$. The first requirement on the model is that this state gives the same
probability measure for the pointer observable as the initial state ${\mathsf
T}$ predicted for the observable ${\mathsf O}$:
$$
tr[{\mathsf T}{\mathsf E}^{\mathsf O}_k] = tr\bigl[\Pi_{\mathcal S}[{\mathsf
U}({\mathsf T}\otimes {\mathsf T}_{\mathcal A}){\mathsf U}^\dagger]{\mathsf
E}^{\mathsf A}_k\bigr]\ .
$$
This is called {\em probability reproducibility condition}. Now, there is a
theorem \cite{belt}:
\begin{prop} Let a measurement fulfil all assumptions and conditions listed
above. Then, for any initial vector state $\psi$ of ${\mathcal A}$, there is a
set $\{\varphi_{kl}\}$ of unit vectors in ${\mathbf H}_\tau$ satisfying the
orthogonality conditions
\begin{equation}\label{orth} \langle \varphi_{kl}|\varphi_{kj}\rangle =
\delta_{lj}
\end{equation} such that ${\mathsf U}$ is a unitary extension of the map
\begin{equation}\label{unitar} \phi_{kl}\otimes \psi \mapsto
\varphi_{kl}\otimes \psi_k\ .
\end{equation}
\end{prop}

The second requirement on the model is that it has to lead to a definite
result. More precisely, the apparatus must be in one of the states
$|\psi_k\rangle\langle\psi_k|$ after each individual registration. This is
called {\em objectification requirement}\footnote{Ref.\ \cite{BLM} calls this
{\em pointer objectification} and introduces a further requirement concerning
the state of ${\mathcal S}$ after the registration, {\em value
objectification} (pp. 30, 49). As our approach denies that objectivity of
observable values is necessary for realist interpretation, the value
objectification is not an important requirement for us, but it will follow
from the pointer objectification via strong correlations between observable
and pointer values valid in some models.}. Ref.\ \cite{BLM} introduces a more
general concept of measurement that leaves open whether the objectification
requirement is satisfied or not. Such a procedure is called {\em
premeasurement}. A measurement is then a premeasurement that satisfies
objectification requirement.

Suppose that the initial state of ${\mathcal S}$ is an eigenstate of ${\mathsf
O}$, ${\mathsf T} =|\phi_{kl}\rangle\langle\phi_{kl}|$, with the eigenvalue
$o_k$. Then, Eq.\ (\ref{unitar}) implies that the final state of apparatus
${\mathcal A}$ is $|\psi_k\rangle\langle\psi_k|$, and the premeasurement does
lead to a definite result. However, suppose next that the initial state is an
arbitrary vector state, ${\mathsf T} =|\phi\rangle\langle\phi|$. Decomposing
$\phi$ into the eigenstates,
$$
\phi = \sum_{kl} c_{kl}\phi_{kl}\ ,
$$
we obtain from Eq.\ (\ref{unitar})
\begin{equation}\label{finalSA} \Phi_{\text{end}} = {\mathsf U} (\phi \otimes
\psi) = \sum_k \sqrt{p_k}\varphi^1_k\otimes \psi_k\ ,
\end{equation} where
\begin{equation}\label{Phik} \varphi^1_k = \frac{\sum_l
c_{kl}\varphi_{kl}}{\sqrt{\langle \sum_l c_{kl}\varphi_{kl}|\sum_j
c_{kj}\varphi_{kj}\rangle}}
\end{equation} and
$$
p_k = \left\langle \sum_l c_{kl}\varphi_{kl}\Biggm|\sum_j
c_{kj}\varphi_{kj}\right\rangle
$$
is the probability that a registration of ${\mathsf O}$ performed on vector
state $\phi$ gives the value $o_k$. The final state of apparatus ${\mathcal
A}$ then is
\begin{equation}\label{finalA} tr_{\mathcal S}[{\mathsf U}({\mathsf T}\otimes
{\mathsf T}_{\mathcal A}){\mathsf U}^\dagger] = \sum_{kl}
\sqrt{p_k}\sqrt{p_l}\langle\varphi^1_k|\varphi^1_l\rangle
|\psi_k\rangle\langle\psi_l|\ .
\end{equation}

Because of the orthonormality of $|\psi_k\rangle$'s, the probability that the
apparatus shows the value $o_k$ if ${\mathsf A}$ is registered on it in this
final state is $p_k$, which is what the probability reproducibility
requires. However, if the objectification requirement is to be satisfied, the
final state of the apparatus must be \begin{equation}\label{gemengA}
tr_{\mathcal S}[{\mathsf U}({\mathsf T}\otimes {\mathsf T}_{\mathcal
A}){\mathsf U}^\dagger] = \left(\sum_j\right) p_j|\psi_j\rangle\langle\psi_j|
\end{equation} (it must be a gemenge, see Chapter 1, Sec.\ 1.2). Thus, in
general we have a model of a premeasurement, but not of a measurement.

\subsection{Repeatable premeasurement and von Neumann model} In order to
define what a repeatable premeasurement is, we need the notion of state
transformer. To this aim, let us first calculate the final state of system
${\mathcal S}$ after a BCL premeasurement is finished:
$$
\Pi_{\mathcal A}[{\mathsf U}(|\phi\rangle\langle\phi| \otimes|
\psi\rangle\langle\psi|){\mathsf U}^\dagger] = \sum_k
p_k|\varphi^1_k\rangle\langle\varphi^1_k|\ .
$$
The part of the sum on the right-hand side corresponding to the result of
premeasurement lying in the set $X$ is
\begin{equation}\label{sttrans} {\mathcal I}(X)(|\phi\rangle\langle\phi|) =
\sum_{o_k\in X} p_k|\varphi^1_k\rangle\langle\varphi^1_k|\ .
\end{equation} The right-hand side is not a state, because it is not
normalised. Its trace is the probability that the result lies in $X$,
$$
p^{\mathsf O}_{\mathsf T}(X) = tr[{\mathcal I}(X)({\mathsf T})]
$$
if the initial state of ${\mathcal S}$ is ${\mathsf T}$. The quantity
$(1/p^{\mathsf O}_{\mathsf T}(X)){\mathcal I}(X)$ is an
operation-valued\footnote{Operation ${\mathcal J}$ is a map, ${\mathcal J} :
{\mathbf T}({\mathbf H}_\tau)^+_1 \mapsto {\mathbf T}({\mathbf H}_\tau)^+_1$,
see Ref.\ \cite{BLM}.} measure and is called {\em state transformer}. For more
details, see Ref.\ \cite{BLM}.

\begin{df} A premeasurement is called {\em repeatable} if its state
transformer satisfies the equation
\begin{equation}\label{repeat} tr[{\mathcal I}(Y)({\mathcal I}(X)({\mathsf
T}))] = tr[{\mathcal I}(Y \cap X)({\mathsf T})]
\end{equation} for all subsets of possible values $X$ and $Y$ and all possible
states ${\mathsf T}$ of ${\mathcal S}$.
\end{df} That is, the repetition of the premeasurement on ${\mathcal S}$ does
not lead to any new result from the probabilistic point of view. To see
whether the state transformer (\ref{sttrans}) satisfies Eq. (\ref{repeat}),
let us rewrite in the form:
$$
{\mathcal I}(X)(|\phi\rangle\langle\phi|) = \sum_{o_k\in X}
p_k|\varphi^1_k\rangle\langle\varphi^1_k| = \sum_{o_k\in X}{\mathsf
K}_k|\phi\rangle\langle\phi|{\mathsf K}^\dagger_k\ ,
$$
where
$$
{\mathsf K}_k = \sum_l|\varphi_{kl}\rangle\langle\phi_{kl}|\ .
$$
One can show that this relation is general,
$$
{\mathcal I}(X)({\mathsf T}) = \sum_{o_k\in X}{\mathsf K}_k{\mathsf T}{\mathsf
K}^\dagger_k\ ,
$$
for proof, see Ref.\ \cite{BLM}. We have then
$$
{\mathcal I}(Y)({\mathcal I}(X)({\mathsf T})) = \sum_{o_l\in X}{\mathsf
K}_l\left(\sum_{o_k\in X}{\mathsf K}_k{\mathsf T}{\mathsf
K}^\dagger_k\right){\mathsf K}^\dagger_l = \sum_{o_l\in X}\sum_{o_k\in
X}({\mathsf K}_l{\mathsf K}_k){\mathsf T}({\mathsf K}_l{\mathsf K}_k)^\dagger\
.
$$
Eq.\ (\ref{repeat}) would be satisfied if
\begin{equation}\label{K} {\mathsf K}_l{\mathsf K}_k = {\mathsf K}_k
\delta_{kl}\ ,
\end{equation} which is in general not the case.

Let us therefore restrict ourselves to measurement couplings satisfying
\begin{equation}\label{vNeum} \phi_{kl} = \varphi_{kl}\ .
\end{equation} This model is called von Neumann premeasurement because it was
first described in Ref.\ \cite{JvN}\footnote{In fact, von Neumann
premeasurement is slightly more general in the sense that it is a
premeasurement of a function $f({\mathsf O})$, where $f$ need not be
bijective, cf.\ Ref.\ \cite{BLM}.}.

For von Neumann premeasurement, the operator ${\mathsf K}_k$ is the projection
${\mathsf E}^{\mathsf O}_k$ on the eigenspace of $o_k$,
$$
{\mathsf K}_k = \sum_l|\phi_{kl}\rangle\langle\phi_{kl}|
$$
and Eq.\ (\ref{K}) is satisfied. Thus, von Neumann premeasurement is a special
case of repeatable premeasurement.

The vector states $\varphi^1_k$ given by Eq.\ (\ref{Phik}) are orthonormal for
von Neumann premeasurements. Thus, the final state of the apparatus given by
Eq.\ (\ref{finalA}) reduces to (\ref{gemengA}). This is, however, not
sufficient for the objectification requirement to be satisfied. The right-hand
side of Eq.\ (\ref{gemengA}) must be the gemenge structure of the state. If
so, then Rule 11 implies that the final state of the composite system
${\mathcal S} + {\mathcal A}$ must be
$$
\sum_j p_j{\mathsf T}_j' \otimes |\psi_j\rangle\langle\psi_j|\ ,
$$
where ${\mathsf T}_j'$ are some states of ${\mathcal S}$. But
$\Phi_{\text{end}}$ of (\ref{finalSA}) is a vector state. State operators of
vector states admit only trivial decomposition. Hence, the objectification
requirement is not satisfied for von Neumann premeasurements, and it is
therefore not a measurement.

An analogous difficulty holds for more general models of premeasurement
described in Ref.\ \cite{BLM} and the book contains more general no-go
theorems. This is called {\em problem of objectification}. The conclusion of
Ref.\ \cite{BLM} is that the von Neumann model of premeasurement, together
with its generalisations, does not work as a model of measurement. Von Neumann
himself postulated that measurements define another, non-unitary and
indeterministic kind of evolution in which ${\mathcal S}$ randomly jumps into
one of the eigenstates of the measured observable (Ref.\ \cite{JvN}, pp. 217,
351). This was called {\em collapse of the wave function} by Bohm (Ref.\
\cite{bohm}, p. 120).

\section{Comparison with real experiment. \\ Importance of detectors} The
theoretical models of the previous section ought to describe and explain at
least some aspects of real experiments. This section will try to go into all
experimental details that can be relevant to our theoretical understanding.

First, we briefly collect what we shall need about detectors. Microsystem
${\mathcal S}$ to be detected interacts with the sensitive matter of the
detector so that some part of energy of ${\mathcal S}$ is transferred to the
detector. Mostly, ${\mathcal S}$ interacts with some subsystems of the
sensitive matter exciting each of them because the excitation energy is much
smaller than the energy of ${\mathcal S}$. The resulting subsystem signals are
collected, or amplified and collected so that they can be distinguished from
noise. For example, in ionisation detectors, many atoms or molecules of the
sensitive matter are turned into electron-ion pairs. If the energy of
${\mathcal S}$ is much higher than the energy of one ionisation, say about 10
eV, then many electron-ion pairs are produced and the positive as well as the
negative total charge is collected at electrodes \cite{leo}.

In the so-called cryogenic detectors \cite{stefan}, ${\mathcal S}$ interacts,
e.g., with superheated superconducting granules by scattering off a nucleus in
a granule and the phase transition from the superconducting into the normally
conducting phase of only one granule can lead to a perceptible electronic
signal. A detector can contain very many granules (typically $10^9$) in order
to enhance the probability of such scattering if the interaction between
${\mathcal S}$ and the nuclei is very weak (weakly interacting massive
particles, neutrino). Modern detectors are constructed so that their signal is
electronic. For example, to a scintillating matter, a photomultiplier is
attached, etc., see Ref.\ \cite{leo}.

In any case, in order to make a detector respond ${\mathcal S}$ must lose some
of its energy to the detector. The larger the loss, the better the
signal. Thus, most detectors are built in such a way that ${\mathcal S}$ loses
all its kinetic energy and is absorbed by the detector (in this way, also its
total momentum can be measured). Let us call such detectors {\em
absorbing}. If the bulk of the sensitive matter is not large enough ${\mathcal
S}$ can leave the detector after the interaction with it, in which case we
call the detector {\em non-absorbing}. Observe that a detector is absorbing
even if most copies of ${\mathcal S}$ leave the detector without causing a
response but cannot leave if there is a response (e.g., neutrino detectors).

Suppose that ${\mathcal S}$ is prepared in such a way that it must cross a
detector. Then, the probability of the detector response is generally $\eta <
1$. We call a detector {\em ideal}, if $\eta = 1$.

An important assumption, corroborated by all experiments, is that a real
detector either gives a signal or remains silent in each individual
registration. This corresponds here to the objectification requirement.

After these preparatory remarks, consider a kind of repeatable premeasurement
that is often described in textbooks (see, e.g., Ref.\ \cite{peres}, p. 27,
where it is called 'repeatable test'), for example a Stern-Gerlach-like
measurement of spin. Let coordinate system $\{x^1,x^2,x^3\}$ be chosen. Silver
atoms evaporate in an oven ${\mathcal O}$, form a beam $B_0$ along $x^2$-axis
passing through a velocity selector ${\mathcal V}$, and then through an
inhomogeneous magnetic field produced by device ${\mathcal M}_1$. ${\mathcal
M}_1$ splits $B_0$ into two beams, $B_{1+}$ and $B_{1-}$, of which $B_{1+}$ is
associated with positive and $B_{1-}$ with negative spin $x^1$-component, the
corresponding vector states being denoted by $|1+\rangle$ and
$|1-\rangle$. Beam $B_{1-}$ is blocked off by a screen. This is the
preparatory part of the experiment.

Next, beam $B_{1+}$ runs through another magnetic device, ${\mathcal
M}_3^{(1)}$ with centre at $\vec{x}_{(1)}$ and finally strike an array of
ideal detectors $\{{\mathcal A}^{(1)}_k\}$ placed and oriented suitably with
respect to ${\mathcal M}_3^{(1)}$. Two detectors of array $\{{\mathcal
A}^{(1)}_k\}$ respond, let us denote them by ${\mathcal A}_+$ and ${\mathcal
A}_-$, revealing the split of $B_{1+}$ into two beams, $B_{3+}$ and $B_{3-}$,
caused by ${\mathcal M}_3^{(1)}$. Let the orientation of ${\mathcal
M}_3^{(1)}$ be such that $B_{3+}$ corresponds to positive and $B_{3-}$ to
negative spin $x^3$-component, the associated states of silver atoms being
$|3+\rangle$ or $|3-\rangle$. The beams $B_{3+}$ and $B_{3-}$ are spatially
sufficiently separated when they arrive at the detectors so that they hit
different detectors ${\mathcal A}_+$ and ${\mathcal A}_-$. In this way, the
values of the quantum mechanical centres of mass of silver atoms at the moment
of their interaction with the detectors are measured. These centres of mass
denoted by $\vec{x}_{3+}$ and $\vec{x}_{3-}$ are registered by the detectors
in a rather coarse-grained way, which is, however, perfectly sufficient to
determine the spins of the atoms. Pointer values are the electronic responses
of detectors ${\mathcal A}_+$ and ${\mathcal A}_-$. Let us call experiment I
what is performed by ${\mathcal O}$, ${\mathcal V}$, ${\mathcal M}_1$,
${\mathcal M}_3^{(1)}$ and $\{{\mathcal A}^{(1)}_k\}$.

Let us now remove $\{{\mathcal A}^{(1)}_k\}$, place device ${\mathcal
M}_3^{(2)}$ of the same macroscopic structure and orientation as ${\mathcal
M}_3^{(1)}$ with centre position $\vec{x}_{(2)}$ in the way of $B_{3+}$ so
that $B_{3-}$ passes by and arrange array $\{{\mathcal A}^{(2)}_k\}$ so that
it has the same relative position with respect to ${\mathcal M}_3^{(2)}$ as
$\{{\mathcal A}^{(1)}_k\}$ had with respect to ${\mathcal M}_3^{(1)}$. Now,
only one detector will respond, namely that which is approximately at the
position $\vec{x}_{3+} - \vec{x}_{(1)} + \vec{x}_{(2)}$. Let us call
experiment II what is performed by ${\mathcal O}$, ${\mathcal V}$, ${\mathcal
M}_1$, ${\mathcal M}_3^{(1)}$, ${\mathcal M}_3^{(2)}$ and $\{{\mathcal
A}^{(2)}_k\}$. The result of experiment II is described in Ref.\ \cite{peres}
as 'two consecutive identical tests following each other with a negligible
time interval between them ... yield identical outcomes'.

Clearly, experiment II does not consist of two copies of experiment I
performed after each other. The only repetition is that device ${\mathcal
M}_3^{(2)}$ is placed after ${\mathcal M}_3^{(1)}$ and has the same structure
and orientation with respect to its incoming beam $B_{3+}$ as ${\mathcal
M}_3^{(1)}$ has with respect to $B_{1+}$. Device ${\mathcal M}_3^{(1)}$ splits
$B_{1+}$ into $B_{3+}$ and $B_{3-}$ but ${\mathcal M}_3^{(2)}$ does not split
$B_{3+}$. One may say that it leaves $B_{3+}$ unchanged. Let us define the
action of device ${\mathcal M}_3^{(k)}$ together with the {\em choice} of
($\pm$)-beam for each $k = 1,2$ as a test (in the sense of Ref.\ \cite{peres})
or a premeasurement. Let the outcomes be the thought responses of imaginary
detectors placed in the way of the chosen beam. Then the (counterfactual)
outcomes can be assumed to be identical indeed and we have an example of
repeatable premeasurement that satisfies Definition 22.

The procedures defined in this way are premeasurements that can be described
by von Neumann model. The value of the spin is transformed by the magnetic
field into the value $\vec{x}_{3+}$ of its centre-of mass-positions that can
be considered as the eigenvalues of the pointer observable associated with
effects $|3\pm\rangle\langle 3\pm|$. However, the premeasurement cannot be
considered as an instance of registration because it does not give us any
information about the silver atoms. Try to suppose, e.g., that the arrangement
measures effects $|3\pm\rangle\langle 3\pm|$ depending on which of the
outgoing beams is chosen. Now, how can we recognise whether the outcome is
'yes' or 'not'?  There is no change of a classical property of an apparatus
due to its interaction with a microsystem that would indicate which of the
values $\vec{x}_{3+}$ and $\vec{x}_{3-}$ results. But premeasurement is
allowed not to give definite responses by each individual action. To obtain
definite values, additional detectors are needed. Without the additional
detector, however, this real premeasurement is not a measurement.

The above observations about the role of detectors seem to be obvious. It is a
strange fact, however, that the detectors are neglected or deemed irrelevant
in old as well as in modern theoretical accounts of quantum measurement. For
example, Ref.\ \cite{pauli}, p. 64, describes a measurement of energy
eigenvalues with the help of scattering similar to Stern-Gerlach experiment,
and it explicitly states:
\begin{quote} We can consider the centre of mass as a 'special' measuring
apparatus...
\end{quote} Similarly, Ref.\ \cite{peres}, p. 17 describes Stern-Gerlach
experiment:
\begin{quote} The microscopic object under investigation is the magnetic
moment $\mathbf \mu$ of an atom.... The macroscopic degree of freedom to which
it is coupled in this model is the centre of mass position $\mathbf r$... I
call this degree of freedom {\em macroscopic} because different final values
of $\mathbf r$ can be directly distinguished by macroscopic means, such as the
detector... From here on, the situation is simple and unambiguous, because we
have entered the macroscopic world: The type of detectors and the detail of
their functioning are deemed irrelevant.
\end{quote} In our opinion, this notion of measurement is erroneous and leads
directly to von Neumann model and to the problem of quantum measurement.

Suppose next that there are non-absorbing ideal detectors that do not disturb
the spin state of the atom. This might work, at least approximately. Let
experiment I' be the same as I with the only change that the array
$\{{\mathcal A}^{(1)}_k\}$ is replaced by $\{{\mathcal A}^{p(1)}_k\}$
containing the non-absorbing detectors. Let experiment II' starts as I' and
proceeds as II but with $\{{\mathcal A}^{(2)}_k\}$ replaced by $\{{\mathcal
A}^{p(2)}_k\}$ made from the non-absorbing detectors. Clearly, the action of
$\bigl({\mathcal M}_3^{(j)} + \{{\mathcal A}^{p(j)}_k\}\bigr)$ for each
$j=1,2$ is a repeatable premeasurement according to Definition 22, and it is
even a repeatable measurement because of the responses of the real detectors,
but it definitely cannot be described by a von Neumann theoretical model. For
the detectors to respond, some part of the energy of the atoms is needed, so
that condition (\ref{vNeum}) is not satisfied.

An interesting difference emerges here between what we can say about the
system (silver atom) on the one hand and about states on the other in their
relation to the beams $B_{3+}$ and $B_{3-}$. Whereas $B_{3+}$ is associated
with $|3+\rangle$ and $B_{3-}$ with $|3-\rangle$, each atom is in a linear
superposition of the two states $|3+\rangle$ and $|3-\rangle$ that equals to
the prepared state $|1+\rangle$. One cannot even say that all atoms in beam
$B_{3+}$ are in state $|3+\rangle$ because no atom is just in $B_{3+}$. Unlike
the states, the atoms are not divided between the beams. Indeed, the two beams
could be guided so that no detectors are in their two ways and that they meet
each other again. Then, they would interfere and if the two ways are of equal
length, so that no relative phase shift results, the original state
$|1+\rangle$ would result. This would happen even if the beams are very thin,
containing always at most one silver atom. Hence, each atom in its wave aspect
had to go both ways simultaneously.

Let us however observe that each of the beams $B_{3+}$ and $B_{3-}$ by itself
behaves as if it were a prepared beam of silver atoms in a known state, which
are $|3+\rangle$ and $|3-\rangle$, respectively. The fact that we place some
arrangement ${\mathcal A}$ of devices that do not contain any detector in the
way of beam $B_{3+}$ and leave $B_{3-}$ alone justifies our use of state in
$|3+\rangle$ in all calculations of what will be the outcome after arrangement
${\mathcal A}$ is passed. The voluntary element of beam choice in this
experiment can be interpreted as a preparation or as a reselection of
ensemble, but not as a collapse of the wave function. Indeed, the {\em whole}
outcome will be a linear superposition of states in each of the two beams at
the time the upper beam passes ${\mathcal A}$. Only if we put any detector
after ${\mathcal A}$ or, for that matter, a detector or just a screen into the
way of $B_{3-}$, then something like a collapse of the wave function can
happen.

As yet, we have considered only the so-called direct registrations. A complete
theory of registration must distinguish two kinds of registration.

A {\em direct registration} first manipulates ${\mathcal S}$ by classical
fields and screens so that the prepared beam is split into spatially separated
beams, each of which associated with one eigenvalue of the registered
observable. Then, there is a set of detectors each of which can be hit by only
one beam.

An {\em indirect registration}, such as scattering or QND measurement (see
e.g.\ \cite{bragin} and \cite{peres} p.\ 400), lets ${\mathcal S}$ interact
with an auxiliary microsystem ${\mathcal S}'_1$ and it is only ${\mathcal
S}'_1$ that is then subject to a direct registration. For the measurement to
be QND, several further conditions must be satisfied, but this does not
concern us here. After a QND procedure, ${\mathcal S}$ remains available to
another one: another system ${\mathcal S}'_2$ of the same type as ${\mathcal
S}'_1$ interacts with ${\mathcal S}$ and is then directly registered
etc. Information given by the detectors of the direct registrations reveals
also something about ${\mathcal S}$. Thus, detectors are necessary for
indirect registrations, too.

The analysis of the present section motivates the following
generalisation. First, an arrangement of devices that acts in agreement with
von Neumann model of direct premeasurement is not a registration
apparatus. Second:
\begin{assump} Any registration apparatus for microsystems must contain at
least one detector and every reading of a pointer value is a signal from a
detector.
\end{assump} Assumption 6 seems to be obvious but it has far reaching
consequences.

As explained in Sec.\ 2.2.4, a preparation of a quantum object ${\mathcal S}$
of type $\tau$ must separate it from other systems of type $\tau$ in a space
domain $D$. Assumption 6 implies, however, that this need not be sufficient
for ${\mathcal S}$ to be an object on which registrations can be
performed. Indeed, on the one hand, any direct registration first manipulates
${\mathcal S}$ individually by fields and screens and then detect it by a
detector. For example, the scattering of electrons on a crystal of graphite
can be described as manipulating each electron by an effective potential
(=field). Also, ${\mathcal S}$ must have sufficient kinetic energy to excite
the detector. On the other hand, an indirect registration on ${\mathcal S}$ is
only possible if ${\mathcal S}$ is sufficiently isolated so that the auxiliary
microsystem must interact with ${\mathcal S}$ rather than with any other
microsystem.

It follows: For ${\mathcal S}$ to have any observable, it must be prepared
\begin{itemize}
\item in a domain $D$ that separates ${\mathcal S}$ from other microsystems
not necessarily of the same type
\item so that either ${\mathcal S}$ is available to manipulation by fields and
screens\footnote{A screen can be theoretically described as a boundary
condition that absorbs everything and releases nothing.} and has sufficient
kinetic energy to excite a detector, in the case of a direct registration,
\item or ${\mathcal S}$ is the only microsystem that can interact with the
auxiliary microsystem in the case of an indirect one.
\end{itemize}

Definition 16 and Assumption 6 imply that both preparations and registrations
of microsystem ${\mathcal S}$ are processes during which the separation status
of ${\mathcal S}$ changes. In particular, in order that observable $\mathsf E$
of any microsystem ${\mathcal S}$ be registered, ${\mathcal S}$ must enter the
sensitive matter part ${\mathcal M}$ of the corresponding detector. ${\mathcal
M}$ contains subsystems of the same type as those that are also contained in
${\mathcal S}$, either because they were there already before the detection
started or because ${\mathcal M}$ becomes polluted by them during the
experiment. Moreover, ${\mathcal M}$ hinders individual manipulation by fields
and screens that would be necessary for each given observable of ${\mathcal
S}$ to be registered.

Assumption 6 also leads to restrictions on observables of macroscopic
systems. In general, a macroscopic system ${\mathcal A}$ is a composite
quantum system with very many different microsystem constituents. One can
subdivide these microsystems into type classes. Consider first observables
that concern properties of microscopic subsystem ${\mathcal S}$ of type
$\tau$. If we apply the basic rules of observable construction for systems of
identical microsystems (Sec.\ 2.2.1), then e.g.\ the position and momentum of
any individual microsystem ${\mathcal S}$ are not observables of ${\mathcal
A}$ but 'collective' operators such as
\begin{equation}\label{coll} \sum_k a(\vec{x}_k;\vec{x}'_k)\ ,
\end{equation} where $a(\vec{x}_k;\vec{x}'_k)$ is the kernel of an operator in
$Q$-representation acting on $k$-th subsystem of type $\tau$, could
be. Suppose that there is an apparatus ${\mathcal B}$ suitable to measure
$a(\vec{x}_k;\vec{x}'_k)$ on an individually prepared system ${\mathcal
S}$. One can imagine that applying ${\mathcal B}$ to ${\mathcal A}$ in some
way similar to that described in Sec.\ 2.2.2 would measure ${\mathsf
a}_{\text{col}}$ because all subsystems of type ${\mathcal S}$ would
automatically contribute to the result of the measurement. However, in Sec.\
2.2.2, the registration apparatus was applied to individually prepared
subsystems. It follows from Assumption 6 that the apparatus ${\mathcal B}$
cannot be applied to ${\mathcal A}$ in this way because none of the identical
subsystems of type $\tau$ is prepared individually. Readings of ${\mathcal B}$
are signals of its detector and for any detector to register ${\mathcal S}$,
${\mathcal S}$ must be isolated to be manipulable, have sufficient kinetic
energy, etc. Hence, to measure collective observable (\ref{coll}), we need a
method that makes measurements directly on ${\mathcal A}$.

This holds for direct registrations. As an example of an indirect one consider
scattering $X$-rays off a crystal ${\mathcal A}$. By it, relative positions of
its nuclei can be recognised (an indirect registration). But rather than a
position of an individual nucleus it is a space dependence of the average
nuclear density due to all nuclei. Such an average nuclear density could be
obtained with the help of an operator of the symmetrised form (\ref{coll}). In
general, scattering of a microsystem ${\mathcal S}'$ off a macrosystem
${\mathcal A}$ can be determined in terms of a potential $V_k(\vec{x},\xi,
\vec{x}_k,\xi_k)$ that describes the interaction between ${\mathcal S}'$ and
one of the microscopic subsystems of ${\mathcal A}$. The whole interaction
Hamiltonian is then the sum analogous to (\ref{coll}),
$$
\sum_k V_k(\vec{x},\xi, \vec{x}_k,\xi_k)\ ,
$$
which however extends over all microscopic subsystems of ${\mathcal A}$ that
can interact with ${\mathcal S}'$.  It is also important to realise that there
are very few interactions that can supply potentials useful for practical
experiment.

Other examples have to do with additive quantities, such as momentum and
angular momentum. Total values of these quantities can be measured and they
are of the form (\ref{coll}), only the sum must extend to all microscopic
subsystems of ${\mathcal A}$ independently of type.

We notice, first, that an observable $a(\vec{x}_k;\vec{x}'_k)$ of a
microsystem ${\mathcal S}$ can be promoted to a collective observable
(\ref{coll}) of ${\mathcal A}$ if ${\mathcal A}$ itself admits a direct or
indirect registration of it, which happens only in special and seldom
cases. Second, such a collective observable is still too 'sharp', because only
some averages with rather large variances can be observed. It is impossible to
obtain its single eigenvalues as results of registration (for an example, see
Ref.\ \cite{peres}, p.181).

A different case is the kinetic energy of ${\mathcal S}$. Again, it cannot be
measured by the method kinetic energy is measured on individual systems of
type ${\mathcal S}$. In the special case that ${\mathcal A}$ is in
thermodynamic equilibrium, the average of the kinetic energy of ${\mathcal S}$
would have the meaning of $1/k_B$ times the temperature of ${\mathcal
A}$. Hence, a viable method to measure the average is to measure the
temperature of ${\mathcal A}$. Again, this is a very special case that works
only under specific conditions.

Thus, one of the consequences of Assumption 6 are principal and severe limits
on promotion of POV measures to observables.

\section{A simple modification of BCL model} Sec.\ 6.2 motivated the first
idea relevant for the present section, namely that the reading of a
registration apparatus for microsystems is in fact an electronic signal from a
detector. This gives us much clearer notion of registration apparatus. The
second relevant idea comes from Sec.\ 2.2.4, namely that fixed-status quantum
mechanics (FSQM) has its limits. This consequence of general rules of standard
quantum mechanics about indistinguishable microsystems leads to a significant
modification of quantum theory of measurement. The necessary changes are:
\begin{enumerate}
\item Each preparation of microsystem ${\mathcal S}$ must give ${\mathcal S}$
a non-trivial separation status $D$, $D \neq \emptyset$. Thus, a quantum
object comes into being only in preparation\footnote{There is no problem for
realism here. Even in classical physics, we often experiment with objects that
are set up in the laboratory and need not exist before the experiment.}.
\item Object ${\mathcal S}$ must be prepared so that it can then be
manipulated and controlled by devices within $D$ such as electric and magnetic
fields. This can be viewed as additional condition on separation status, see
Ref.\ \cite{hajicek3}.
\item Macrosystem ${\mathcal A}$ such as a blocking screen, a scattering
target or a detector that contains microsystems indistinguishable from
${\mathcal S}$ or from some subsystems of ${\mathcal S}$ must lie behind the
boundary of $D$. Other microsystems of the same type as ${\mathcal S}$ that
may be in $D$ due, e.g., to imperfect vacuum must not disturb registration of
${\mathcal S}$. Corrections to FSQM description of the behaviour of the
composed system ${\mathcal S} + {\mathcal A}$ due to a necassary
separation-status change of ${\mathcal S}$ must be carefully chosen.
\end{enumerate} The usual method of FSQM is to specify initial states of both
${\mathcal S}$ and ${\mathcal A}$ before their interaction, choose some
appropriate interaction Hamiltonian and calculate the corresponding unitary
evolution of the composed system ${\mathcal S} + {\mathcal A}$ ignoring the
problem with separation-status change. As shown in Sec.\ 6.1, the results are
wrong for registration apparatuses. The purpose of this section is to make
some corrections. First, we must modify and extend BCL model.

Let system ${\mathcal S}$ of type $\tau$ be prepared as a quantum object of
separation status $D$ in a domain $D$. Let ${\mathsf O}$ be the observable to
be registered with eigenvalues $o_k$ and eigenvectors $\phi_{kl}$ as in Sec.\
6.1. Let ${\mathcal S}$ be prepared in state $\phi$ and manipulated so that
its component eigenstates of ${\mathsf O}$ become spatially separated.

Next, we construct a sufficiently detailed model of detector ${\mathcal
A}$. Let ${\mathcal A}$ be an array of $N$ ideal monoatomic-gas ionisation
detectors ${\mathcal A}_k$ similar to that of Sec.\ 6.2. Let index $k$
enumerate the detectors and let the sensitive matter of each detector have
separation status $D_k$ where $D_k \cap D_l = \emptyset$ for all $k \neq l$
and $ D_k \cap D = \emptyset$ for all $k$. Let the sensitive matter be a
system of identical atoms. Let each atom be modelled by a particle with mass
$\mu$, spin zero and a further degree of freedom, ionisation, with two values,
'non-ionised' and 'ionised'. We simplify the model further by assuming that
the ionisation and translation degrees of freedom define two different formal
subsystems, ${\mathcal A}_{\text{ion}}$ and ${\mathcal A}_{\text{tra}}$ which
can be decomposed in subsystems ${\mathcal A}^k_{\text{ion}}$ and ${\mathcal
A}^k_{\text{tra}}$ of ${\mathcal A}_k$. Let $\chi_{kn}$ be the state
describing $n$ ionised atoms in $k$th detector. The states
$$
\prod_k \otimes \chi_{kn(k)}
$$
for all $n(k)$'s form a basis of the Hilbert space of ${\mathcal
A}_{\text{ion}}$, where $n(k)$ is a map from $\{1,\cdots,N\}$ into
non-negative integers. Let us assume that the initial state of ${\mathcal
A}_{\text{ion}}$ is
\begin{equation}\label{beginA} \psi = \prod_k \otimes \chi_{k0}\ ,
\end{equation} the perfectly non-ionised state. We can further assume that the
initial state ${\mathsf T}_{\text{tra}}$ of ${\mathcal A}_{\text{tra}}$ is
close to a state of maximum entropy with sufficiently low temperature so that
ionisations due to atomic collisions have a very low probability.

Let each ${\mathcal A}_k$ further contains family ${\mathcal S}'_k$ of systems
of type $\tau$ identical with ${\mathcal S}$ and let the number of such
subsystems in ${\mathcal S}'_k$ be $M_k$. Thus, ${\mathcal A}_k = {\mathcal
A}^k_{\text{ion}} + {\mathcal A}^k_{\text{tra}} + {\mathcal S}'_k$ Let the
subsystems of ${\mathcal S}'_k$ be distributed over the sensitive matter of
the $k$-th detector so that the separation status of ${\mathcal S}'_k$ is
$D_k$. We assume that the subsystems of ${\mathcal S}'_k$ do not interact with
each other, with the rest of the detector and with ${\mathcal S}$. Let the
state of ${\mathcal S}'_k$ be ${\mathsf T}_k$. Hence, ${\mathsf T}_k$
satisfies
$$
{\mathsf T}_k = {\mathsf P}^{(M_k)}_\tau{\mathsf T}_k{\mathsf P}^{(M_k)}_\tau\
,
$$
where ${\mathsf P}^{(M_k)}_\tau$ is the projection to the symmetric or
antisymmetric subspace of ${\mathbf H}_{{\mathcal R}(\tau)}^{M_k}$, depending
on $\tau$, see Sec.\ 2.2.1.

Let the measurement coupling be a coupling between ${\mathcal S}$ and the
ionisation degree of freedom of each atom in the sensitive matter of the whole
array. That is, ${\mathcal S}$ interacts directly only with ${\mathcal
A}_{\text{ion}}$. In a single detector, after the ionisation of the first
atom, all subsequent ionisations lie along a ray track inside the same
detector. An explanation of the fact that e.g.\ a spherical wave can produce a
straight track is given in Ref.\ \cite{mott}, where it is shown that the
position of the track head, the first ionisation of the track, determines the
track. This can be considered as a necessary property of every measurement
coupling that is possible in the case considered here.

Let our first attempt at the new rule be based on the assumption that the
dynamical rule of the standard quantum mechanics holds. Under this assumption,
let the measurement coupling fulfils the conditions of the BCL model, so that,
if systems ${\mathcal S}'_k$ did not exist, we would be lead to Eq.\
(\ref{unitar}), where $\psi$ is given by Eq.\ (\ref{beginA}),
$$
\psi_k = \left(\prod_{j=1}^{k-1} \otimes \chi_{j0}\right)\otimes \left(\sum_n
a_n\chi_{kn}\right)\otimes \left(\prod_{j=k+1}^N \otimes \chi_{j0}\right)\ ,
$$
and $a_n$ are coefficients independent of $k$ satisfying $\sum_n |a_n|^2 =
1$. This is again a simplifying assumption: each ${\mathcal S}$ creates always
the same ionisation state in any detector. Then Eqs.\ (\ref{finalSA}) and
(\ref{Phik}) are valid. Now, we can define the observable ${\mathsf A}$ by its
eigenstates (\ref{beginA}) with eigenvalue 0 and $\psi_k$'s with eigenvalues
$o_k$ assuming $o_k \neq 0$ for all $k$.

In Sec.\ 6.1, states $\psi_k$ were called 'end states' of ${\mathcal A}$ and
${\mathsf A}$ was called 'pointer observable'.  Here, we prefer to call
${\mathsf A}$ {\em trigger observable} and $\psi_k$ {\em trigger states}
because there is a further evolution of ${\mathcal A}$ independent of
${\mathcal S}$ that leads from $\psi_k$ to the concentration of charges at the
electrodes and an electronic signal of $k$-th detector. This is due to a
coupling between ${\mathcal A}_{\text{ion}}$ and ${\mathcal A}_{\text{tra}}$
mediated by the electrostatic field of the electrodes: ionised atoms move in a
different way than the non-ionised ones. This motion leads to atom collisions
and further ionisation in a complicated irreversible process. Only then, the
true end states with true pointer values are achieved. The pointer values are
some averages with some variances, in agreement with the expectation of Refs.\
\cite{PHJT,hajicek}. However, what is important for us happened already at the
trigger stage and we can ignore the evolution from a trigger state to a
detector signal.

Let further the support of trigger states $\varphi_{kl}$ of ${\mathcal S}$ be
$D_k$ for all $k,l$. It follows then that,
\begin{equation}\label{endorth} \langle \varphi_{kl}|\varphi_{mn}\rangle =
\delta_{km}\delta_{ln}\ .
\end{equation} Thus, we assume a special case of the BCL dynamics. Then, that
Eq.\ (\ref{gemengA}) would hold.

The existence of systems ${\mathcal S}'_k$ leads however to change of the
separation status of ${\mathcal S}$ after ${\mathcal S}$ enters the sensitive
matter of a detector. We can take this into account similarly as in Sec.\
2.2.4 and replace state $\Phi_{\text{end}}$ of system ${\mathcal S} + {\mathsf
A}_{\text{ion}}$ by a suitably entangled state ${\mathsf T}_{\text{trig}}$ of
system ${\mathcal S} + \sum_k {\mathcal S}'_k + {\mathcal
A}_{\text{ion}}$. However, many choices are possible, for example:
$$
{\mathsf T}_{\text{trig1}} = \sum_{kl} \sqrt{p_k p_l}\ \nu^2 {\mathsf P}^{(1 +
\sum_k M_k)}_\tau (|\varphi^1_k\rangle \langle\varphi^1_l| \otimes {\mathsf
T}_1 \otimes \cdots \otimes {\mathsf T}_N) {\mathsf P}^{(1 + \sum_k M_k)}_\tau
\otimes |\psi_k\rangle \langle \psi_l|\ ,
$$
where $\nu$ is a suitable normalisation factor (projections do not preserve
norm). State ${\mathsf T}_{\text{trig1}}$ entails a maximal entanglement of
identical systems within ${\mathcal S} + \sum_k {\mathcal S}'_k$. This is not
in agreement with Rule 14 because for each $k$, the separation status of
system ${\mathcal S}'_k$ is $D_k$ as is the support of the state
$|\varphi^1_k\rangle$.

An attempt that entails a minimal entanglement of identical systems that is
still compatible with our ideas on separation status is:
\begin{multline}\label{dynam1} {\mathsf T}_{\text{trig2}} = \sum_{kl}
\sqrt{p_k p_l}\ \nu_k \nu_l {\mathsf W}_{kl} \otimes {\mathsf T}_1 \otimes
\cdots \otimes {\mathsf T}_{k-1} \otimes {\mathsf T}_{k+1} \otimes \\ \cdots
\otimes {\mathsf T}_{l-1} \otimes {\mathsf T}_{l+1}\otimes\cdots \otimes
{\mathsf T}_N \otimes |\psi_k\rangle \langle \psi_l|\ .
\end{multline} The form of the summands is valid just for non-diagonal terms
$k < l$, but the sum must be understood as consisting of all non-diagonal
terms, $k \neq l$, with operators
$$
{\mathsf W}_{kl} = {\mathsf P}^{(M_k+1)}_\tau ({\mathsf T}_k \otimes
|\varphi^1_k\rangle)(\langle\varphi^1_l| \otimes {\mathsf T}_l){\mathsf
P}^{(M_l+1)}_\tau
$$
on the extended Hilbert space ${\mathbf H}_\tau^{1 + M_k + M_l}$ of system
${\mathcal S} + {\mathcal S}'_k + {\mathcal S}'_l$, as well as the diagonal
ones with the operator
$$
{\mathsf W}_{kk} = {\mathsf P}^{(M_k+1)}_\tau (|\varphi^1_k\rangle \langle
\varphi^1_k|\otimes {\mathsf T}_k){\mathsf P}^{(M_k+1)}_\tau
$$
on the Hilbert space of system ${\mathcal S} + {\mathcal S}'_k$.
Normalisation factor $\nu_k$ is determined by the condition
$$
\nu_k = \frac{1}{\sqrt{tr[{\mathsf W}_{kk}]}}\ .
$$
Then, ${\mathsf T}_{\text{trig2}}$ satisfies the normalisation condition of a
state. Indeed, the trace of a tensor product of operators is a product of the
traces of the operator factors, see Proposition\ 24. As $tr[|\psi_k\rangle
\langle \psi_l|] = \delta_{kl}$, only the diagonal elements of the sum
contribute. We have
\begin{prop} The trace of operator ${\mathsf W}_{kl}$ for $k \neq l$ is zero.
\end{prop} For proof, see next section.

Proposition 22 implies immediately that the trigger state of ${\mathcal
A}_{\text{ion}}$, obtained from Eq.\ (\ref{dynam1}) and (\ref{endorth}), is
given by Eq.\ (\ref{gemengA}). However, the right-hand side of Eq.\
(\ref{gemengA}) is not the gemenge structure of the state because state
${\mathsf T}_{\text{trig2}}$ does not satisfy the conditions of Rule
11. Hence, our first choice of dynamics is not supported by experimental
evidence.

Another attempt is to take just the diagonal elements of ${\mathsf
T}_{\text{trig2}}$:
\begin{equation}\label{dynam2} {\mathsf T}_{\text{trig3}} = \sum_k p_k \nu^2_k
{\mathsf T}_1 \otimes \cdots \otimes {\mathsf T}_{k-1} \otimes {\mathsf
W}_{kk} \otimes {\mathsf T}_{k+1} \otimes \cdots \otimes {\mathsf T}_N \otimes
|\psi_k\rangle \langle \psi_k |
\end{equation}

State ${\mathsf T}_{\text{trig2}}$ of ${\mathcal S} + {\mathcal
A}_{\text{ion}}$ contains many correlations between systems ${\mathcal
A}_{\text{ion}}$ and ${\mathcal S} + \sum_k {\mathcal S}'_k$. The only
correlations that remain in ${\mathsf T}_{\text{trig3}}$ are, for each $k$,
those between the trigger states $\psi_k$ of ${\mathcal A}_{\text{ion}}$ and
states $\nu^2_k{\mathsf W}_{kk}$ of system ${\mathcal S} + {\mathcal S}'_k$.

States ${\mathsf T}_{\text{trig1}}$, ${\mathsf T}_{\text{trig2}}$ or ${\mathsf
T}_{\text{trig3}}$ describe ${\mathcal S}$ being inside ${\mathcal A}$ and the
registration is running. A part of this registration is the registration of
${\mathsf A}$. Hence the observables that can be measured at this stage of the
experiment must commute with ${\mathsf 1} \otimes {\mathsf A}$ to be jointly
measurable with it. However, we have:
\begin{prop} Let ${\mathsf B}$ be a sharp observable of system of system
${\mathcal S} + \sum_k {\mathcal S}_k + {\mathcal A}_{\text{ion}}$ that
commutes with ${\mathsf 1}_{{\mathcal S} + \sum_k {\mathcal S}_k} \otimes
{\mathsf A}$. Then
$$
tr[{\mathsf B}{\mathsf T}_{\text{trig2}}] = tr[{\mathsf B}{\mathsf
T}_{\text{trig3}}]\ .
$$
\end{prop} The proof of Proposition 23 will be given in the next
section. Proposition 23 means that the states ${\mathsf T}_{\text{trig2}}$ and
${\mathsf T}_{\text{trig3}}$ cannot be distinguished from each other at this
stage of the experiment. If the detector is non-absorbing, and system
${\mathcal S}$ is released from it, then the existing experimental evidence
supports the notion that the state after the releasing is ${\mathsf
T}_{\text{trig3}}$ with the gemenge structure given by the right-hand side of
Eq.\ (\ref{dynam2}) rather than ${\mathsf T}_{\text{trig2}}$.

This motivates the following assumption:
\begin{rl} Let a microsystem ${\mathcal S}$ be detected by a detector
${\mathcal A}_k$ and let the measurement coupling modified as above satisfy
Eq.\ (\ref{endorth}). Then the state of ${\mathcal S} + \sum_k{\mathcal S}'_k
+ {\mathcal A}_{\text{ion}}$ is
\begin{multline}\label{trig} {\mathsf T}_{\text{trig}} = \left(\sum_k\right)
p_k (|\psi_{10}\rangle \langle \psi_{10}| \otimes {\mathsf T}_1) \otimes
\cdots \otimes (|\psi_{k-10}\rangle \langle \psi_{k-10}| \otimes{\mathsf
T}_{k-1}) \\ \otimes \left|\sum_n a_n\chi_{kn}\right\rangle \left\langle
\sum_n a_n\chi_{kn}\right| \otimes \nu^2_k {\mathsf P}^{(M_k+1)}_\tau
(|\varphi^1_k\rangle \langle \varphi^1_k|\otimes {\mathsf T}_k){\mathsf
P}^{(M_k+1)}_\tau \\ \otimes (|\psi_{k+10}\rangle \langle \psi_{k+10}| \otimes
{\mathsf T}_{k+1}) \otimes \cdots \otimes (|\psi_{N0}\rangle \langle
\psi_{N0}| \otimes {\mathsf T}_N)\ .
\end{multline}
\end{rl} State ${\mathsf T}_{\text{trig}}$ satisfies two conditions:
\begin{enumerate}
\item The trigger state of ${\mathcal A}_{\text{ion}}$ has gemenge structure
(\ref{gemengA}). That ${\mathsf T}_{\text{trig}}$ satisfies this condition
follows from Rule 11.
\item It contains the same measurable correlation between systems ${\mathcal
A}_{\text{ion}}$ and ${\mathcal S} + {\mathcal S}'_k$ as ${\mathsf
T}_{\text{trig2}}$. That ${\mathsf T}_{\text{trig}}$ satisfies this condition
follows from Proposition 23.
\end{enumerate} All other correlations are erased during the change of
separation status of ${\mathcal S} + {\mathcal A}$. What survives and what is
erased is uniquely determined in this case by the modified BCL model and Rule
17. In particular, the probability reproducibility condition determines states
$\varphi_{kl}$ from the initial state $\psi$ of ${\mathcal A}_{\text{ion}}$
uniquely and the initial state $\phi$ of ${\mathcal S}$ determines states
$\varphi^1_k$ uniquely. Thus, the evolution from state $|\phi\rangle \langle
\phi| \otimes {\mathsf T}_1 \otimes \cdots \otimes {\mathsf T}_N) \otimes
|\psi_0\rangle \langle \psi_0|$ to state ${\mathsf T}_{\text{trig}}$ is
non-unitary but still deterministic. Rule 17 is a new general rule which has
to be added to quantum mechanics. To choose such a rule, we have looked at
observations and experiments.

The correct interpretation of Rule 17 distinguishes two cases. If the
detectors are absorbing, then state ${\mathsf T}_{\text{trig}}$ evolves with
${\mathcal S}$ staying inside ${\mathcal A}$ and being lost always in one of
the detectors. If they are non-absorbing, then state ${\mathsf
T}_{\text{trig}}$ evolves to
\begin{equation}\label{truend} {\mathsf T}_{\text{truend}} =
\left(\sum_k\right) p_k|\varphi^1_k\rangle \langle\varphi^1_k| \otimes
{\mathsf T}'_k\ ,
\end{equation} where ${\mathsf T}'_k$ are the end states of ${\mathcal A}_k$
including the detector response. In this case, states $\nu^2_k{\mathsf
W}_{kk}$ evolved to the release of ${\mathcal S}$ in state $\varphi^1_k$ that
is correlated with detector signals. Each release is understood as an instance
of preparation and the whole procedure is a random mixture of these single
preparations. In both cases, the end state of ${\mathcal A}_{\text{ion}}$ has
then gemenge structure (\ref{gemengA}).

As an example of a system of non-absorbing detectors, the MWPC telescope for
particle tracking can be mentioned \cite{leo}. It is a stack of the so-called
multiwire proportional chambers (MWPC), which is arranged so that a particle
runs through, exciting each of them. The resulting system of electronic
signals contains the information about the particle track.

A registration by a non-absorbing detector is similar to a scattering of a
microsystem by a macroscopic target. First, let us consider no-entanglement
processes such as the scattering of electrons on a crystal of graphite with a
resulting interference pattern \cite{DG} or the splitting of a laser beam by a
down-conversion process in a crystal of KNbO$_3$ (see, e.g., Ref.\
\cite{MW}). No-entanglement processes can be described by the following model.
Let the initial state of the target ${\mathcal A}$ be $\mathsf T$ and that of
the microsystem be $\phi$. We assume that the end state of the target is
${\mathsf T}'$ and the end-state of the microsystem is $\varphi$ and that we
have a unitary evolution:
$$
|\phi\rangle\langle\phi| \otimes {\mathsf T} \mapsto
|\varphi\rangle\langle\varphi| \otimes {\mathsf T}'\ .
$$
The two systems are not entangled by their interaction, hence there is no
necessity to divide the resulting correlations between ${\mathcal S}$ and
${\mathcal A}$ in what survives and what is erased. The end state is already
of the form (\ref{truend}) and it has a trivial gemenge structure. In this
way, our corrections to FSQM become trivial.

A more complicated case is an entanglement scattering. Let microsystem
${\mathcal S}$ in initial state $\phi$ be scattered by a macrosystem
${\mathcal A}$ in initial state ${\mathsf T}$ and let this lead to excitation
of different microscopic subsystems of ${\mathcal A}$.  Scattering of neutrons
on spin waves in ferromagnets, transmutation of nuclei inside ${\mathcal A}$
or, for that matter, ionising an atom in a gas detector are examples. We have,
therefore, a more general situation than that in which Rule 17 gives a unique
result. It seems that the change of status must lead to survival of some
correlations between ${\mathcal S}$ and ${\mathcal A}$ while some are
erased. However, in this situation it has still to be investigated which is
which. Clearly, the definitive general rule will depend on the two interacting
systems and on the interaction Hamiltonian. More theoretical and experimental
work is necessary to guess the general rule.

One can wonder whether a more detailed quantum mechanical model can be
constructed of what happens during a change of separation status. The reason
why this cannot be done within FSQM is that FSQM is not applicable to changes
of separation status. Hence, just a new law added to FSQM is needed.

\subsection{Proofs of Propositions 22 and 23} For the proofs, we shall need
some simple properties of trace.
\begin{prop}
\begin{enumerate}
\item Let operator ${\mathsf B}$ has a form ${\mathsf B} =
\sum_{mn}B_{mn}|m\rangle \langle n|$, where $\{|n\rangle\}$ is some family of
states. Then,
$$
tr[{\mathsf B}] = \sum_{mn}B_{mn}\langle m|n\rangle\ .
$$
\item Let operator ${\mathsf B}$ on Hilbert space ${\mathbf H}$ leave a
subspace ${\mathbf H}_1 \in {\mathbf H}$ invariant and acts trivially on
${\mathbf H}_1^\bot$. Then
$$
tr_{\mathbf H}[{\mathsf B}] = tr_{{\mathbf H}_1}[{\mathsf B}]\ .
$$
\item Let three trace-class operator ${\mathsf B}$, ${\mathsf B}_1$ and
${\mathsf B}_2$ satisfy ${\mathsf B} = {\mathsf B}_1 \otimes {\mathsf
B}_2$. Then
$$
tr[{\mathsf B}] = tr[{\mathsf B}_1] tr[{\mathsf B}_2]\ .
$$
\end{enumerate}
\end{prop}
\par \vspace{.5cm} \noindent {\bf Proof of Proposition 22} Our method is to
expand operator ${\mathsf W}_{kl}$ in a basis and to use Proposition 24. The
basis can be chosen as follows. In our model, states $\varphi^1_k$ and
${\mathsf T}_k$ have supports inside $D_k$. Hence, they can be expanded into a
basis $\{\phi^k_m\}, m =1, 2, \cdots $ of ${\mathbf H}_{D_k}$. We can choose
$\phi^k_1 = \varphi^1_k$. Then, the family of vectors $\{\phi^k_m\}\cup
\{\phi^l_n\}$ is a basis for the extended Hilbert space $({\mathbf H}_\tau)^{1
+ M_k + M_l}$ of system ${\mathcal S} + {\mathcal S}'_k + {\mathcal
S}'_l$. Any term in expansion of operator ${\mathsf W}_{kl}$ into basis
$\{\phi^k_m\}\cup \{\phi^l_n\}$ will contain composite vectors
$$
\text{Perm}[1 + M_k](|\phi^k_1\rangle \otimes |\phi^k_{m(1)}\rangle) \otimes
\cdots \otimes |\phi^k_{m(M_k)}\rangle \otimes
\text{Perm}[M_l](|\phi^l_{n(1)}\rangle \otimes \cdots \otimes
|\phi^l_{n(M_l)}\rangle)
$$
left and
$$
\text{Perm}[M_k](\langle \phi^k_{r(1)}| \otimes \cdots \otimes \langle
\phi^k_{r(M_k)})| \otimes \text{Perm}[1 + M_l](\langle \phi^l_1| \otimes
\langle \phi^l_{s(1)}|\otimes \cdots \otimes \langle \phi^l_{s(M_l)}|)
$$
right, where $\text{Perm}[N]$ is an arbitrary but fixed order permutation of
$N$ states and the state choices $m(\cdot)$, $n(\cdot)$, $r(\cdot)$ and
$s(\cdot)$ are mappings from the first $M_k$ or $M_l$ positive integers to
positive integers (they need not be injections). Clearly, the two composite
vectors are orthogonal for any permutations and state choices because there is
one more element of $\{\phi^k_m\}$ in the product defining the first one than
in the second one, and one more element of $\{\phi^l_m\}$ in the product
defining the second one than in the first one. Hence, the trace is zero,
Q.E.D.
\par \vspace{.5cm} \noindent {\bf Proof of Proposition 23} According to Sec.\
6.1, $\{\psi_k\}$ together with $\psi$ is an orthonormal basis in a Hilbert
space on which ${\mathsf A}$ is an observable,
$$
{\mathsf A} = \sum_k o_k |\psi_k\rangle \langle \psi_k|
$$
and all $o_k$ are different from each other.

Let $\{\Phi_k\}$ be a basis for the Hilbert space of system ${\mathcal S} +
\sum_k {\mathcal S}'_k$. Then, any observable ${\mathsf B}$ can be written as
$$
{\mathsf B} = \sum_{klmn} B_{klmn}|\Phi_k\rangle \langle \Phi_m| \otimes
|\psi_l\rangle \langle \psi_n|\ .
$$
In order to commute with ${\mathsf 1} \otimes {\mathsf A}$, ${\mathsf B}$ has
to be diagonal in $l$ and $n$:
$$
{\mathsf B} = \sum_{klm} B_{klm}|\Phi_k\rangle \langle \Phi_m| \otimes
|\psi_l\rangle \langle \psi_l|\ .
$$
Then, we easily obtain
$$
tr[{\mathsf T}_{\text{trig2}}{\mathsf B}] = \sum_{kml} B_{kml} p_m \nu^2_m
\langle \Phi_l| {\mathsf T}_1 \otimes \cdots \otimes {\mathsf T}_{m-1} \otimes
{\mathsf W}_{mm} \otimes {\mathsf T}_{m+1} \otimes \cdots \otimes {\mathsf
T}_N |\Phi_l\rangle
$$
and the same result for $tr[{\mathsf T}_{\text{trig3}}{\mathsf O'}]$, which
proves Proposition 23.

\chapter{Conclusion} Our careful study of quantum-mechanical practice, both
experimental and theoretical, has lead us to quite a new picture of quantum
mechanics. It differs from most current approaches in several points.

First, we do not consider a value of an observable of system ${\mathcal S}$ as
a property of ${\mathcal S}$ but only as an indirect piece of information on
such properties. Each value of observable is determined by both objective
properties of ${\mathcal S}$ and the registration apparatus. Thus, we reject
attempts to view all observables of ${\mathcal S}$, or at least some subsets
of them, as objective properties of ${\mathcal S}$.

Second, we find a qualitative difference between quantum microsystems and
large composite systems. In nature, microsystems occur in families of a fixed
type, the elements of which are identical, and do not possess any
individuality in strict sense. Sufficiently large composite systems are
different for the following reason. The larger a composite system is, the less
probable the existence of an identical system somewhere in the universe is.

Third, classical properties are objective quantum properties of macrosystems
associated with states that are close to maximum entropy. We have shown how
this principle is applied to classical mechanics by introducing a new class of
states, the maximum-entropy packets.

Fourth, we give preparation a different and much greater significance than is
usually assumed. In a preparation procedure, the prepared system gains its
objective quantum properties such as states, gemenge structures, averages and
variances of observables etc. so that it is justified to speak of a physical
object. This is what we call quantum object. Next, a preparation must separate
a microsystem from the set of identical microsystems, at least approximately
so that it obtains a non-trivial separation status. Only then, it can be
viewed as an individual system and a particular set of observables depending
on the status becomes measurable on it. In this way, the standard rule for
composition of identical microsystems can be weakened. This is justified by
the idea of cluster separability. Finally, a preparation also separates the
microsystem from all othe microsystems so that it can be individually
manipulated by external fields or matter shields and registered by detectors.

Fifth, macrosystems, as well as generally large composite systems, have much
less observables than one would expect according to standard quantum
mechanics. There are two reasons why only very few observables concerning
single constituents of such a system can be measured. On the one hand, the
constituents may be elements of a large family of identical systems from which
they are not separated by preparation and do not, therefore, possess any
measurable observables of their own. On the other, they are not separated from
other microsystems to be individually manipulable by fields and shields and
registrable by detectors. The differences between macroscopic and microscopic
systems that are the contents of the second, third and fifth points are not
due to inapplicability of quantum mechanics to macroscopic systems. Just the
opposite is true: they result from strict and careful application of standard
quantum mechanics to macroscopic systems.

Sixth, we give registration a more specific form than is usually assumed. Any
apparatus that is to register a microsystem directly contains a detector and
the 'pointer' value that is read off the apparatus is a signal from the
detector. We assume that each detector contains a macroscopic piece of
sensitive matter with which the registered microsystem becomes unified and
looses its object status. The post-Everett trend in the theory of quantum
measurement is to avoid collapse of the wave function. In a sense, the present
paper is heading in the opposite direction. We even replace the collapse by a
more radical transformation, a change in microsystem kinematic
description. This change is, in plain words, a kind of disappearance of a
registered object during its registration. Nevertheless, our result for
non-absorbing detectors and the collapse idea by von Neumann have some
features in common.

Finally, standard quantum mechanics does not provide information about
processes, in which the separation status of microsystems
changes. Preparations and registrations belong to such processes. Unjustified
application of standard rules to such processes leads to contradictions with
experimental evidence. However, one can add new rules to quantum mechanics
governing such processes without violating its logic.

The main purpose of Chap.\ 5 is to show how the new ideas on measurement
problem work by studying well-understood, restricted class of physical
conditions in which the following assumptions are a good approximation: (a)
non-relativistic quantum mechanics, (b) measurement performed directly on
microsystems, (c) a simple modification of BCL model of measurement and (d)
pointer readings being signals from detectors.

In our future work, we have gradually to remove the restrictions. First, we
must turn to other models of measurement, for example to different (non-ideal)
kinds of detectors or to the more realistic unitary premeasurements for which
the no-go theorems such as Theorem 6.2.1 in Ref.\ \cite{BLM}, p. 76, hold. The
main point is again that the state resulting from the unitary evolution
contains information about properties of the composite system ${\mathcal S} +
{\mathcal A}$ that could be measured only if more observables than ${\mathsf
O}$ of ${\mathcal S}$ existed. Thus, a change of this illusory state analogous
to that given by Rule 17 could be justified, it could again remove the
objectification problem and the no-go theorems could be avoided. The exact
division line between correlations that survive and those that are erased
might again be determined by a careful analysis of observational facts.

Next, relativistic corrections have been neglected so that all notions and
rules of non-relativistic quantum mechanics could be used. An extension of the
present results to relativistic fields seems to be a realistic project.

Let us finish by giving a summary of how our theory changes quantum mechanics.
Standard quantum mechanics has the so-called minimal interpretation (see,
e.g., Ref.\ \cite{BLM}, p.\ 9). As measurement is concerned, it just
postulates the probability reproducibility and objectification avoiding any
deeper theory. It does not worry about origin of classical properties and
about philosophy. Moreover, common practice of quantum mechanics does not take
the symmetrisation or anti-symmetrisation over all identical systems seriously
ignoring all unknown identical systems. This {\em Minimal Quantum Mechanics}
has been in general use and it has been formidably successful. Our theory
takes over Minimal Quantum Mechanics just giving a new foundation to the
identical systems that justifies the common practice. To this, it adds a
quantum theory of measurement that is motivated by the new ideas on identical
systems and satisfies probability reproducibility and objectification
requirements. Finally, it adds a realist interpretation and a theory of
classical systems that do not contradict Minimal Quantum Mechanics.

\subsection*{Acknowledgements}The authors are indebted to \v{S}tefan
J\'{a}no\v{s} for help with experimental physics and to Heinrich Leutwyler for
useful discussions. J. T. acknowledges partial support of the Ministry of
Education of Czech Republic, research projects MSM6840770039 and LC06002.

\end{document}